%% file: paperCDS.tex
\documentclass[ALICE,manyauthors]{cernphprep}

\include{setup}

\usepackage{capt-of}
\usepackage{epstopdf}

\usepackage[T1]{fontenc} 
\usepackage{setspace}
\begin{document}%
%
%
\begin{titlepage}
\PHyear{2014}
\PHnumber{104}                 
\PHdate{15 May}              
%
%
\title{Elliptic flow of identified hadrons in Pb--Pb collisions at $\mathbf{\sqrt{\textit{s}_{\rm NN}}}$=2.76~TeV}
\ShortTitle{Elliptic flow of identified hadrons}   
%
\Collaboration{ALICE Collaboration%
         \thanks{See Appendix~\ref{app:collab} for the list of collaboration
                      members}}
\ShortAuthor{ALICE Collaboration}      
\begin{abstract}
The elliptic flow coefficient ($v_2$) of identified particles in Pb--Pb collisions at $\sqrt{s_\mathrm{{NN}}} = 2.76$~TeV was measured with the 
ALICE detector at the LHC. The results were obtained with the Scalar Product method, a two-particle correlation technique, using a pseudo-rapidity 
gap of $|\Delta\eta| > 0.9$ between the identified hadron under study and the reference particles. The $v_2$ is reported for $\pi^{\pm}$, $\mathrm{K}^{\pm}$, $\mathrm{K}^0_\mathrm{S}$, p+$\overline{\mathrm{p}}$, 
$\mathrm{\phi}$, $\Lambda$+$\overline{\mathrm{\Lambda}}$, $\Xi^-$+$\overline{\Xi}^+$ and $\Omega^-$+$\overline{\Omega}^+$ in several collision 
centralities.  In the low transverse momentum ($\pt$) region, $\pt <2$~\gevc, $v_2(p_\mathrm{T})$ exhibits a particle mass dependence consistent with elliptic flow accompanied by the transverse radial expansion of the system with a common velocity field. The experimental data for $\pi^{\pm}$ and $\mathrm{K}$ are described fairly well by hydrodynamical calculations coupled to a hadronic cascade model (VISHNU) for central collisions. However, the same calculations fail to reproduce the $v_2(p_\mathrm{T})$ for p+$\overline{\mathrm{p}}$, $\mathrm{\phi}$, $\Lambda$+$\overline{\mathrm{\Lambda}}$ and $\Xi^-$+$\overline{\Xi}^+$.  For transverse momentum values larger than about 3~GeV/$c$, 
particles tend to group according to their type, i.e.~mesons and baryons. However, the experimental data at the LHC exhibit deviations from the number of constituent quark (NCQ) scaling at the level of $\pm$20$\%$ for $p_{\mathrm{T}} > 3$~GeV/$c$.
\end{abstract}
\end{titlepage}
\setcounter{page}{2}


\input{section/introduction.tex}
\input{section/experimentalSetup.tex}

\input{section/identificationOfHadrons.tex}

\input{section/flowMethods.tex}

\input{section/systematics.tex}

\input{section/discussion.tex}

\input{section/conclusion.tex}
%
%
\newenvironment{acknowledgement}{\relax}{\relax}
\begin{acknowledgement}
\section*{Acknowledgements}
\input{section/acknowledgements.tex}    
\end{acknowledgement}
%
%
%
\input{section/references.tex}

\newpage
\appendix
\input{section/appendixB.tex}
%
%
\newpage
\section{The ALICE Collaboration}
\label{app:collab}
\input{AliceAuthorlistCERNPREP.tex}  
\end{document}

%% file: setup.tex
\usepackage{hyperref}
\usepackage{graphicx}
\usepackage{dcolumn}
\usepackage{bm}
\usepackage[normalem]{ulem}
\usepackage{lineno}
\usepackage{color}
\definecolor{dgreen}{cmyk}{1.,0.,1.,0.2}
\definecolor{orange}{cmyk}{0.,0.353,1.,0.}

\newcommand{\ITS}{\rm{ITS}}
\newcommand{\SPD}{\rm{SPD}}
\newcommand{\SDD}{\rm{SDD}}
\newcommand{\SSD}{\rm{SSD}}
\newcommand{\TPC}{\rm{TPC}}

\newcommand{\TOF}{\rm{TOF}}
\newcommand{\VZERO}{\rm{VZERO}}

\newcommand{\pt}{{p_{\mathrm T}}}
\newcommand{\gevc}{GeV/$c$ }

\newcounter{vers}

\newcommand{\vtwo}{${v}_2$}

\newcommand{\kzero}{K$^0_\mathrm{s}$}
\newcommand{\lambdas}{$\Lambda$($\bar{\Lambda}$)}

%% file: section/introduction.tex
\section{Introduction}
\label{Sec:Introduction}
Lattice quantum chromodynamics calculations predict a transition from ordinary nuclear matter to the Quark-Gluon Plasma (QGP) \cite{Satz:2000bn,Bass:1998vz,Shuryak:1984nq,Cleymans:1985wb}, in which the constituents, the 
quarks and the gluons, are deconfined. At low values of the baryochemical potential, a crossover transition is expected to take place at a temperature of about 150~MeV and at an energy density of about 0.5~GeV/fm$^3$~\cite{Borsanyi:2010cj,Bhattacharya:2014ara}. These conditions are 
accessible in the laboratory by colliding heavy ions at ultra--relativistic energies. The study of the properties of this deconfined matter is the main goal of the heavy-ion collision program at the Large Hadron Collider (LHC). The existence of the QGP has been stipulated by observations at the Relativistic Heavy Ion Collider (RHIC) \cite{Arsene:2004fa,Adcox:2004mh,Back:2004je,Adams:2005dq}. The first experimental 
results from the heavy-ion program at the LHC \cite{Aamodt:2010pb,Aamodt:2010cz,Chatrchyan:2011pb,Aamodt:2011mr,Aamodt:2010pa,ATLAS:2011ah,ATLAS:2012at,Chatrchyan:2012wg,Chatrchyan:2012ta,ALICE:2011ab, Aad:2013xma,Chatrchyan:2013kba,Aamodt:2010jd,Aad:2010bu,Chatrchyan:2011sx} have also provided evidence of the existence of the QGP in this new energy regime. 

Anisotropic flow, which characterises the momentum anisotropy of the final state particles, can probe the properties, such as the ratio of shear viscosity to entropy density ($\eta/s$), of the system created in heavy-ion interactions.
In nuclear collisions, the impact parameter vector and the beam axis define the reaction plane. It was recently realized that the overlap region of the colliding nuclei exhibits an irregular shape driven by the initial density profile of nucleons participating in the collision which is different from one event to the other. The symmetry plane of this irregular shape fluctuates around the reaction plane in every event. This spatial anisotropy of the overlap region is transformed into an anisotropy 
in momentum space through interactions between partons and at a later stage between the produced particles. 
The resulting anisotropy is usually characterised by the Fourier
coefficients \cite{Voloshin:1994mz,Poskanzer:1998yz} according to

\begin{equation}
v_n = \langle \cos \big [n(\varphi - \Psi_{n}) \big] \rangle,
\label{Eq:Fourier}
\end{equation}

\noindent where $\varphi$ is the azimuthal angle of particles, $n$ is the order of the flow harmonic and $\Psi_n$ is the angle of the spatial plane of symmetry of harmonic $n$ \cite{Bhalerao:2006tp,Alver:2008zza,Ollitrault:2009ie,Alver:2010gr,Qiu:2011iv}. The second Fourier coefficient, \vtwo, measures the 
azimuthal momentum space anisotropy of particle emission relative to the second harmonic symmetry plane and is known as elliptic flow.

The study of anisotropic flow in nucleus-nucleus collisions at RHIC~\cite{Arsene:2004fa,Adcox:2004mh,Back:2004je,Adams:2005dq} contributed significantly in establishing that the produced system is described as a strongly coupled Quark-Gluon Plasma (sQGP) with a small value of the ratio of shear viscosity to entropy density ($\eta/s$), very close to the conjectured lower limit of $\hbar/4\pi k_{\mathrm{B}}$, where $\hbar$ and $k_{\mathrm{B}}$ are the reduced Planck and Boltzmann constants, respectively \cite{Kovtun:2004de}. 
Recent anisotropic flow measurements for charged particles at the LHC \cite{Aamodt:2010pa,ATLAS:2011ah,ATLAS:2012at,Chatrchyan:2012wg,Chatrchyan:2012ta,ALICE:2011ab,Aad:2013xma,Chatrchyan:2013kba} indicate that the system created in Pb--Pb collisions at $\sqrt{s_\mathrm{{NN}}} = 2.76$~TeV also behaves as a strongly interacting liquid. 
An additional constraint on the value of $\eta/s$ can be obtained by studying the flow coefficients of Eq.~\ref{Eq:Fourier} as a function 
of collision centrality and transverse momentum for different particle species~\cite{Arsene:2004fa,Adcox:2004mh,Back:2004je,Adams:2005dq}. An interplay of radial flow (i.e. azimuthally symmetric) and anisotropic flow leads to a characteristic 
mass dependence of $v_n(\pt)$~\cite{Voloshin:1996nv,Huovinen:2001cy,Teaney:2000cw,Shen:2011eg}, first observed by the 
E877 Collaboration at the AGS for the case of directed flow ($v_1$)~\cite{Barrette:1997pt,Barrette:1998bz} and by the NA49 Collaboration 
at SPS \cite{Appelshauser:1997dg,Alt:2003ab}. This interplay was then studied in detail for $v_2$ at RHIC, where the characteristic mass ordering of the $v_2(p_{\mathrm{T}})$ (the $\pt$-differential \vtwo) for $p_{\mathrm{T}} < 2$~GeV/$c$ was reported~\cite{Adams:2003am,Abelev:2007qg,Adams:2004bi,Adler:2003kt,Afanasiev:2007tv,Adare:2006ti,Adare:2012vq}.

The comparison of $v_2(\pt)$ measurements to hydrodynamic calculations in the low transverse momentum region has established that elliptic flow is built up mainly during the early, 
partonic stages of the system and is thus governed by the evolution of the QGP medium \cite{Arsene:2004fa,Adcox:2004mh,Back:2004je,Adams:2005dq}. However, the hadronic rescattering that follows the QGP phase could also contribute to the development of $v_2$~\cite{Hirano:2005xf}. The development of anisotropic flow at the partonic stage may be probed by studying particles with a small hadronic cross section, which are expected to be less affected by the hadronic stage and thus more sensitive to the early (partonic) stages of the collision. The $\phi$, $\Xi^-$+$\overline{\Xi}^+$ and $\Omega^-$+$\overline{\Omega}^+$ are argued to be such weakly coupled probes \cite{Biagi:1980ar,Bass:1999tu,Dumitru:1999sf,Bass:2000ib,Shor:1984ui}.

In addition, at RHIC energies, in the intermediate $\pt$ region~($2 < \pt < 6$~GeV/$c$) the $v_2(\pt)$ of baryons is larger than 
that of mesons. In~\cite{Voloshin:2002wa}, it was suggested that this phenomenon can find an explanation in a picture where flow 
develops at the partonic level and quarks coalesce into hadrons during the hadronization. The proposed mechanism was argued to lead 
to the observed hierarchy in the values of $v_2(\pt)$, referred to as number of constituent quarks (NCQ) scaling, in the intermediate $p_{\mathrm{T}}$ region where hydrodynamic flow might no longer be dominant and may compete with the corresponding contribution from jet fragmentation. The expectation was investigated by several studies of the quark coalescence picture both experimentally~\cite{Adams:2003am,Abelev:2007qg,Adams:2004bi,Adler:2003kt,Afanasiev:2007tv,Adare:2006ti,Adare:2012vq} and theoretically~\cite{Molnar:2003ff,Greco:2003mm,Fries:2003kq,Hwa:2003ic}. 

In \cite{Abelev:2012di}, we presented the first measurements of $v_2(\pt)$ for identified $\pi^{\pm}$, p and $\overline{\mathrm{p}}$ at the LHC in the range $3 < p_{\mathrm{T}} < 20$~GeV/$c$. In the present article, the $v_2(\pt)$ of identified particles is reported for $0.2 < p_{\mathrm{T}} < 6.0$~GeV/$c$ measured in Pb--Pb collisions at the centre of mass 
energy per nucleon pair $\sqrt{s_\mathrm{{NN}}} = 2.76$~TeV with the ALICE detector \cite{Carminati:2004fp,Alessandro:2006yt,Aamodt:2008zz} 
at the LHC. Results on $v_2(p_\mathrm{T})$ for identified mesons ($\pi^{\pm}$, $\mathrm{K}^{\pm}$, $\mathrm{K}^0_\mathrm{S}$, 
$\phi$) and baryons (p, $\Lambda$, $\Xi^-$, $\Omega^-$, and their antiparticles), measured in $|y| < 0.5$ (where $y$ is the rapidity of each particle) are presented. The $v_2(\pt)$ values of particles and antiparticles were measured separately and were found to be compatible within the statistical uncertainties. Thus, in this article the $v_2(\pt)$ for the sum of particles and antiparticles is reported.  For the reconstruction of the decaying particles presented in Section~\ref{Sec:AnalysisDetails}, the following channels were 
used: $\mathrm{K}^0_\mathrm{S} \rightarrow \pi^+ + \pi^-$, $\phi \rightarrow \mathrm{K}^+ + \mathrm{K}^-$, $\Lambda \rightarrow \mathrm{p} + \pi^-$ ($\overline{\Lambda} \rightarrow \overline{\mathrm{p}} + \pi^+$), $\Xi^- \rightarrow \Lambda + \pi^-$ ($\overline{\Xi}^+ \rightarrow \overline{\Lambda} + \pi^+$), and $\Omega^- \rightarrow \Lambda + \mathrm{K}^-$ ($\overline{\Omega}^+ \rightarrow \overline{\Lambda} + \mathrm{K}^+$). The results are obtained with the Scalar Product method described briefly in Section~\ref{Sec:FlowMethods}, and in detail in~\cite{Adler:2002pu,Ref:EventPlane}, using a pseudo-rapidity gap of $|\Delta\eta| > 0.9$ between the identified hadrons under study and the charged reference particles (see Section~\ref{Sec:FlowMethods} for details). This method suppresses  the contribution to $v_2(\pt)$
from correlations not related to the symmetry plane, i.e. non-flow effects, such as correlations arising from jets and resonance decays. The $v_2(\pt)$ is reported 
for different centralities of Pb--Pb collisions, which span the range 0--60$\%$ of the inelastic cross section \cite{Abelev:2013qoq}, where the contribution from non-flow effects is small as compared to the collective flow signal.

%% file: section/experimentalSetup.tex
\section{Experimental setup}
\label{Sec:ExpSetup}
ALICE \cite{Aamodt:2008zz} is one of the four major experiments at the LHC. It is particularly designed to cope with the large 
charged--particle densities present in central Pb--Pb collisions \cite{Aamodt:2010pb}. ALICE uses a right-handed Cartesian system with its origin at the second LHC Interaction Point (IP2). The beam direction defines the $z$-axis, the $x$-axis is horizontal and points towards the centre of the LHC, and the $y$-axis is vertical and points upwards. The apparatus 
consists of a set of detectors located in the central barrel positioned inside a solenoidal magnet which generates a $0.5$~T field parallel to the beam direction, and a set of forward detectors. The central 
detector systems allow for full azimuthal coverage for track reconstruction within a pseudo-rapidity window of $|\eta| < 0.9$. 
The experimental setup provides momentum resolution of about 1 to 1.5~$\%$ for the momentum range covered in this article, and 
particle identification (PID) over a broad momentum range. 

For this analysis, the charged particles were reconstructed using the Time Projection Chamber (\TPC) 
\cite{Alme:2010ke} or the combination of the \TPC~and the Inner Tracking System (\ITS) \cite{Aamodt:2008zz}. The \TPC~is the main 
tracking detector of the central barrel. The detector provides full azimuthal coverage in the pseudo-rapidity range $|\eta| < 0.9$. 
The \ITS~consists of six layers of silicon detectors employing three different technologies. The two innermost 
layers, positioned at $r = 3.9$~cm and 7.6~cm,  are Silicon Pixel Detectors (\SPD), followed by two layers of Silicon Drift Detectors (\SDD) ($r = 15$~cm and 23.9~cm). 
Finally the two outermost layers are double--sided Silicon Strip Detectors (\SSD) at $r = 38$~cm and 43~cm.

Charged particles were identified using the information from the \TPC~and the Time of Flight (\TOF) 
detector \cite{Aamodt:2008zz}. The \TPC~provides a simultaneous measurement of the momentum of a particle and 
its specific ionisation energy loss ($\mathrm{d}E/\mathrm{d}x$) in the gas. The detector provides a sufficient separation (i.e.~better than 2 standard deviations) for the hadron species at $p_{\rm{T}} < 0.7$~GeV/$c$ and the possibility to identify particles on a statistical basis in the relativistic rise region of $\mathrm{d}E/\mathrm{d}x$ (i.e.~$2 < p_{\rm{T}} < 20$~GeV/$c$) 
\cite{Abelev:2014ffa}. The $\mathrm{d}E/\mathrm{d}x$ resolution for the 5$\%$ most central Pb-Pb collisions is 6.5$\%$ and improves for peripheral collisions. The \TOF~detector surrounds the \TPC~and provides a $3\sigma$ separation between $\pi$--K and K--$\rm{p}$ 
up to $p_{\rm T} = $ 2.5~GeV/$c$ and $p_{\rm T} = 4$~GeV/$c$, respectively \cite{Abelev:2014ffa}. This is done by measuring the arrival 
time of particles with a resolution of about $80$~ps. The start time for the \TOF~measurement is provided by the T0 detectors, two arrays of Cherenkov counters positioned at opposite sides of the interaction points covering $4.6 < \eta < 4.9$ (T0-A) and $-3.3 < \eta < -3.0$ (T0-C). The start time is also determined using a combinatorial algorithm that compares the timestamps of particle hits measured by the TOF to the expected times of the tracks, assuming a common event time $t_{ev}$ \cite{Abelev:2014ffa,Akindinov:2013tea}. Both methods of estimating the start time are fully efficient for the 60$\%$ most central Pb--Pb collisions.

A set of forward detectors, the \VZERO~scintillator arrays~\cite{Abbas:2013taa}, were used in the trigger logic and for the centrality 
and reference flow particle determination for the Scalar Product method described in Section~\ref{Sec:FlowMethods}. The \VZERO~consists of two systems, the \VZERO-A 
and the \VZERO-C, positioned on each side of the interaction point, and cover the pseudo-rapidity ranges 
of $2.8 < \eta < 5.1$ and $-3.7 < \eta < -1.7$ for \VZERO-A and \VZERO-C, respectively.
For more details on the ALICE experimental setup, see \cite{Aamodt:2008zz}.

%% file: section/identificationOfHadrons.tex
\section{Event sample, track selection and identification}
\label{Sec:AnalysisDetails}

\subsection{Trigger selection and data sample}

In this analysis approximately 15~$\times$~10$^6$ Pb--Pb events were used. The sample was recorded during the first LHC 
heavy-ion data taking period in 2010 at $\sqrt{s_\mathrm{{NN}}} = 2.76$~TeV. Minimum bias Pb--Pb events were triggered by the coincidence of signals from the two \VZERO~detectors. An 
offline event selection exploiting the signal arrival time in \VZERO-A and \VZERO-C, with a 1~ns resolution, was used to discriminate background (e.g.~beam-gas) from collision events. This reduced the background events in the analysed sample to a negligible fraction ($< 0.1 \%$).  All events retained in the analysis have a reconstructed primary vertex 
position along the beam axis ($V_z$) within 10~cm from the centre of the detector. The vertex was estimated using either tracks reconstructed by the \TPC~or by the global tracking, i.e. combining information from all tracking detectors (the \TPC~and the \ITS).

The data were grouped according to fractions of the inelastic cross section and correspond to the 60$\%$ most central Pb--Pb collisions. The 0--5$\%$ interval corresponds 
to the most central (i.e.~small impact parameter) and the 50--60$\%$ interval to the most peripheral (i.e.~large impact parameter) collisions in the analysed sample. The centrality of the collision was estimated using the distribution of signal amplitudes from the \VZERO~scintillator detectors (default analysis). The systematic uncertainty due to estimating the centrality of the collision is determined using the charged particle multiplicity distribution of \TPC~tracks, and the number of \ITS~clusters. Details on the centrality determination can be found in \cite{Abelev:2013qoq}.

\subsection{Selection of $\pi^{\pm}$, $\mathrm{K}^{\pm}$ and $\mathrm{p}$+$\overline{\mathrm{p}}$}
\label{SubSec:PiKP}

Primary charged pions, kaons and (anti-)protons were required to have at least 70 reconstructed space points out 
of the maximum of 159 in the \TPC. The average $\chi^2$ of the track fit per TPC space point per degree of freedom (see \cite{Abelev:2014ffa} for details) was required to be below $2$. These selections reduce the contribution from short tracks, which are unlikely to originate from the primary vertex, to the analysed sample. To further reduce the contamination from secondary tracks 
(i.e.~particles originating either from weak decays or from the interaction of other particles with the material), only particles within a maximum distance of closest approach (DCA) between the tracks and the primary vertex in both the $xy$-plane ($d_{xy} < 2.4$~cm) and the $z$ coordinate ($d_{z} < 3.0$~cm) were analysed. The selection leads to 
an efficiency of about $80\%$ for primary tracks at $p_\mathrm{T} > 0.6$~GeV/$c$ and a contamination from secondaries of about $5\%$ at $p_{\rm{T}} = 1$~GeV/$c$ \cite{Abelev:2013vea}. These values depend strongly on particle species and transverse momentum \cite{Abelev:2013vea}. The $v_2(\pt)$ results are reported 
for $|y| < 0.5$ and $0.2 < p_{\rm{T}} < 6.0$~GeV/$c$ for $\pi^{\pm}$, $0.3 < p_{\rm{T}} < 4.0$~GeV/$c$ 
for $\mathrm{K}^{\pm}$ and $0.3 < p_{\rm{T}} < 6.0$~GeV/$c$ for p+$\overline{\mathrm{p}}$.

\begin{figure}[tbh!f]
  \includegraphics[width=\textwidth]{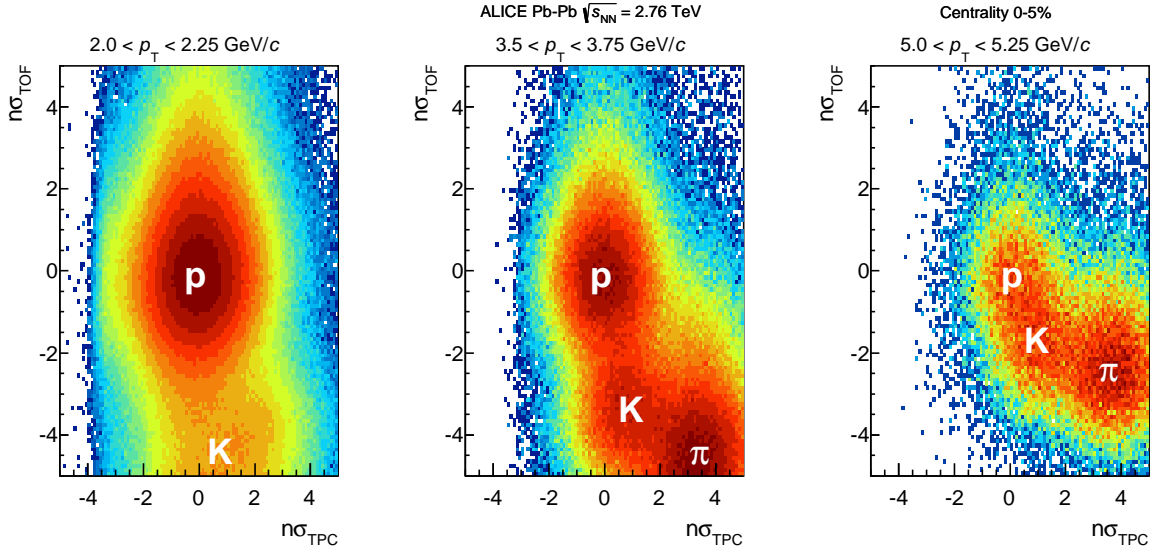}
  \caption{\label{fig.pid} The correlation between the number of standard deviations from the expected signal of the \TPC~($\sigma_{\mathrm{TPC}})$ and the \TOF~($\sigma_{\mathrm{TOF}})$ detectors using the proton mass hypothesis for three different transverse momentum intervals in the 5$\%$ most central Pb--Pb collisions.}
\end{figure}

For the identification of $\pi^{\pm}$, $\mathrm{K}^{\pm}$ and p+$\overline{\mathrm{p}}$ over the wide 
$p_{\rm{T}}$ range, the combination of information from the TPC and the 
TOF detectors was used. In particular, the identification was based on a two-dimensional correlation between the response of the \TPC~and the \TOF.
The particles were selected by requiring their signal to lie within three standard deviations (3$\sigma$) of both the dE/dx ($\sigma_{\mathrm{TPC}}$) and TOF ($\sigma_{\mathrm{TOF}}$) resolutions. For some particles (particularly kaons) with $p_{\mathrm{T}} > 3$~GeV/$c$ where the relevant bands for different particle species start to overlap, the requirement was changed to 2$\sigma$. This identification strategy results in a purer sample as compared to previous analyses reported by ALICE (see e.g. \cite{Abelev:2013vea}). It is adopted since it reduces the need for potential corrections due to particle misidentification that could introduce additional uncertainties to the measurement of $v_2$.
An example of a correlation plot between the number of standard deviation from the expected signal of the \TPC~and the \TOF~detectors for three different transverse momentum intervals in the 5$\%$ most central Pb--Pb collisions is presented in Fig.~\ref{fig.pid}. The resulting purity, estimated using Monte-Carlos (MC) simulations but also data-driven methods (e.g. selecting pions and (anti)protons from $\mathrm{K}^{0}_{s}$ and $\mathrm{\Lambda}$($\mathrm{\overline{\Lambda}}$) decays) was larger than 90$\%$ for $\pi^{\pm}$, $\mathrm{K}^{\pm}$ and p+$\overline{\mathrm{p}}$ throughout the analysed transverse momentum range.

Finally, since the contamination from secondary protons created through the interaction of particles with the detector material can reach values larger than 10$\%$ for $p_{\rm{T}} < 1$~GeV/$c$, only $\overline{\mathrm{p}}$ were considered for $p_{\rm{T}} < 3$~GeV/$c$, while for higher values of $p_{\rm{T}}$ a combined measurement of p and $\overline{\mathrm{p}}$ was used.

\subsection {Reconstruction of $\mathrm{K}^0_\mathrm{S}$ and $\mathrm{\Lambda}$+$\overline{\mathrm{\Lambda}}$}

The measurement of $\mathrm{K}^0_\mathrm{S}$, $\mathrm{\Lambda}$ and $\overline{\mathrm{\Lambda}}$ was performed using their 
weak decays in the following channels: $\mathrm{K}^0_\mathrm{S} \rightarrow \pi^+ + \pi^-$ (branching ratio 
69.2$\%$) and $\mathrm{\Lambda} \rightarrow \textrm{p} + \pi^{-}$, $\overline{\mathrm{\Lambda}} \rightarrow \overline{\textrm{p}} + \pi^{+}$ 
(branching ratio 63.9$\%$) \cite{Beringer:1900zz}.  The identification of these particles is based on the reconstruction of the secondary vertex exhibiting a characteristic V-shape, called V0, defined by the trajectories of the decay products. 

For all three particles, the decay products of the V0 candidates were required to have a minimum $p_\mathrm{T}$ of 0.1~GeV/$c$, while the criteria on the number of \TPC~space points and on the $\chi^2$ per \TPC~space point per degree of freedom were identical to those applied for primary particles. In addition, a selection of secondary particles based on a minimum 
DCA to the primary vertex of 0.1~cm was applied. Furthermore, a maximum 
DCA of 0.5~cm between the decay products at the point of the V0 decay was required to ensure that they are products of 
the same decay. The decay tracks were reconstructed within \mbox{$|\eta| < 0.8$}. Finally, for the $\Lambda$+$\overline{\Lambda}$ 
candidates with low values of transverse momentum, a particle identification cut to select their 
p+$\overline{\mathrm{p}}$ decay products was applied that relied on a 3-$\sigma$ band around the expected energy loss in the TPC, defined by a parameterization of the Bethe-Bloch curve. The selection parameters are summarised in Table~\ref{tab.v0daughtercuts}.

\begin{table}[h!bt]
\begin{center}
    \centering
    \begin{tabular}{@{}ll@{}}
      \multicolumn{2}{c}{\kzero{} and \lambdas{} decay products} \\\hline\hline
      TPC space points              & $\geq 70$ \\
      $\chi^2$ per TPC space point per d.o.f. & $\leq 2$\\
      DCA to primary vertex         & $\geq 0.1$~cm\\
      DCA between decay products        & $\leq 0.5$~cm\\
      $p_\mathrm{T}$                  & $\geq$ 0.1~GeV/$c$\\
      $|\eta|$                  & $<$ 0.8\\
      TPC PID compatibility selection for p+$\overline{\mathrm{p}}$ decay products of $\Lambda$+$\overline{\Lambda}$              & $\leq 3\sigma$ \\\hline
    \end{tabular}
   \caption{\label{tab.v0daughtercuts} Selection criteria for the decay products of the V0 candidates.}
\end{center}
 \end{table}

To reduce the contamination from secondary and background particles, mainly from other strange baryons affecting 
$\mathrm{\Lambda}$ and $\overline{\mathrm{\Lambda}}$, a minimum value of the cosine of the pointing angle ($\cos{\theta_\mathrm{p}} \geq 0.998$) was 
required. The pointing angle is defined as the angle between the momentum vector of the V0 candidate and the 
vector from the primary to the reconstructed V0 vertex \cite{Abelev:2013xaa}. To further suppress the background, only V0 candidates whose decay length was within three times the $c\tau$ value of 2.68~cm for $\mathrm{K}^0_\mathrm{S}$ 
and 7.89~cm for $\Lambda$ ($\overline{\Lambda}$) \cite{Abelev:2013xaa} were analysed. In addition, the radial position of 
the secondary vertex reconstruction was required to be more than 5 cm away from the primary vertex in the transverse plane (i.e. larger than the radius of the first SPD layer) in order to minimise possible biases introduced by the high occupancy in 
the first layers of the \ITS. Furthermore, the analysed V0 candidates were reconstructed within $|y| < 0.5$, to 
suppress any effects originating from the lower reconstruction efficiency close to the edges of the detector acceptance. Finally, an additional selection in the Armenteros-Podolanski variables\footnote{The Armenteros-Podolanski variables are the projection of the decay charged-track momentum on the plane perpendicular to the V0 momentum ($q_{\mathrm{T}}$) and the decay asymmetry parameter defined as $\alpha = (p_{\mathrm{L}}^+ - p_{\mathrm{L}}^-)/(p_{\mathrm{L}}^+ + p_{\mathrm{L}}^-)$, where $p_{\mathrm{L}}$ is the projection of the decay charged-track momentum on the momentum of the V0.} \cite{Ref:Armenteros} was applied for $\mathrm{K}^0_\mathrm{S}$ candidates, 
accepting particles with $q_{\mathrm{T}} \geq 0.2|\alpha|$. This was done to reduce the contamination from reconstructed V0 candidates originating from $\gamma$ conversion in the detector material and $\mathrm{\Lambda}$ and $\overline{\mathrm{\Lambda}}$ in the $\mathrm{K}^0_\mathrm{S}$ mass region. These selection parameters are summarised in Table~\ref{tab.v0cuts}.
  
\begin{table}[h!bt]
\begin{center}
    \begin{tabular}{@{}ll@{}}
      \multicolumn{2}{c}{\kzero{} and \lambdas{} candidates} \\\hline\hline
      Decay length                 & $\leq 3 c\tau$ \\
      $\cos{\theta_\mathrm{p}}$    & $\geq$ 0.998 \\
      Decay radius & $\geq 5$~cm \\
      $|y|$                     & $\leq$ 0.5\\
      $q_{\mathrm{T}}$ (\kzero{} only)  & $\geq 0.2|\alpha|$ \\
\hline
    \end{tabular}
   \caption{\label{tab.v0cuts} Topological selections for the \kzero{} and
    \lambdas{}}
\end{center}
\end{table}

\begin{figure}[tbh!f]
  \includegraphics[width=\textwidth]{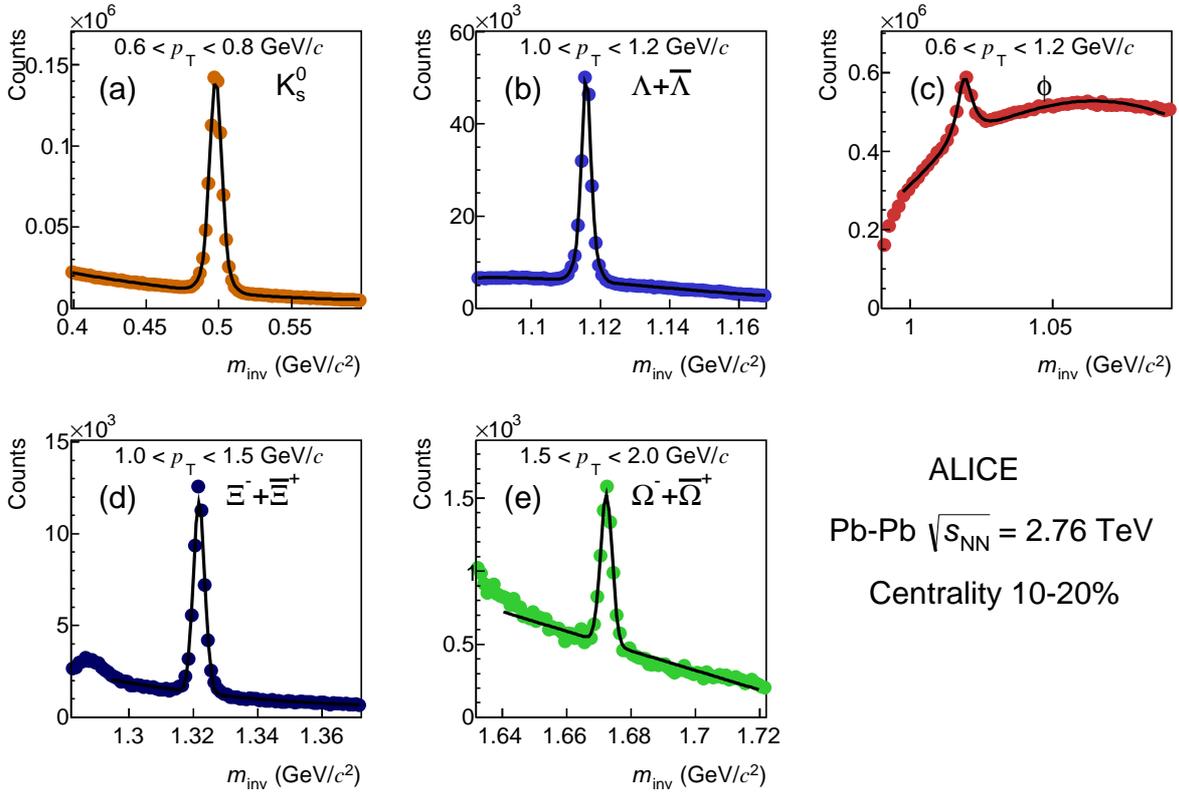}
  \caption{\label{fig.massplots} Invariant mass distributions in
  the 10--20$\%$ centrality interval of Pb--Pb collisions for reconstructed
    decaying particles: (a) $\mathrm{K}^0_\mathrm{S}$, (b) $\Lambda$+$\overline{\mathrm{\Lambda}}$, (c) $\phi$, (d) $\Xi^-$($\overline{\mathrm{\Xi}}^+$), and (e) $\Omega^-$($\overline{\mathrm{\Omega}}^+$).}
\end{figure}

Charged pions and pion-(anti-)proton pairs were then combined to obtain the invariant mass ($m_{\mathrm{inv}}$) for $\mathrm{K}^0_\mathrm{S}$ and $\mathrm{\Lambda}$ ($\overline{\mathrm{\Lambda}}$), respectively. Examples of such distributions 
for two of the lowest transverse momentum intervals used in this analysis for the 10--20$\%$ centrality of Pb--Pb 
collisions at 
$\sqrt{s_\mathrm{{NN}}} = 2.76$~TeV are shown in Fig.~\ref{fig.massplots} (a) and (b) for $\mathrm{K}^0_\mathrm{S}$ and $\mathrm{\Lambda}$, 
respectively. These distributions are fitted with a sum of a Gaussian function and a third-order polynomial to estimate the signal and the background in the mass peak. The signal to background ratio in the mass peak depends on the transverse momentum and on the event centrality and is better than 5 for both particles. The $v_2(\pt)$ results are reported 
for $|y| < 0.5$ and $0.4 < p_{\rm{T}} < 6.0$~GeV/$c$ for $\mathrm{K}^0_\mathrm{S}$ and $0.6 < p_{\rm{T}} < 6.0$~GeV/$c$ for $\mathrm{\Lambda}$ and $\overline{\mathrm{\Lambda}}$.

\subsection{Reconstruction of $\phi$}

The $\phi$-meson was reconstructed via its hadronic decay channel: $\phi \rightarrow \mathrm{K}^+ + \mathrm{K}^-$ (branching 
ratio 48.9$\%$) \cite{Beringer:1900zz}.
The selections applied for the decay products 
were identical to those of primary $\mathrm{K}^{\pm}$, described in Section~\ref{SubSec:PiKP}. The $\phi$-meson yield was extracted from the invariant mass ($m_{\mathrm{inv}}$) reconstructed from the unlike-sign kaon pairs. 

The combinatorial background was evaluated using the like-sign kaon pairs in each $p_{\rm{T}}$ and centrality interval. The 
like-sign background $m_{\mathrm{inv}}$ distribution is normalised to the corresponding distribution of unlike-sign pairs in the 
region above the $\phi$-meson mass ($1.04 < m_{\mathrm{inv}} < 1.09$~GeV/$c^{2}$).  An example of an invariant mass distribution before the like-sign subtraction for $0.6 < p_{\rm{T}} < 1.2$~GeV/$c$ is given in Fig.~\ref{fig.massplots} (c) for the 10--20$\%$ centrality interval of  Pb--Pb collisions. The remaining background was estimated using a third-order polynomial.

These invariant mass distributions were then fitted with a relativistic Breit-Wigner distribution, describing the signal in 
the mass peak. The $v_2(\pt)$ results for the $\phi$-meson are reported for $|y| < 0.5$ and $0.6 < p_{\rm{T}} < 6.0$~GeV/$c$ for the centrality intervals covering the 10--60$\%$ of the inelastic cross section. For the 10$\%$ most central Pb--Pb collisions, the extraction of the signal over the large combinatorial background resulted into large uncertainties using the currently analysed data sample.

\subsection{Reconstruction of $\mathrm{\Xi^-}$+$\overline{\mathrm{\Xi}}^+$ and $\mathrm{\Omega}^-$+$\overline{\mathrm{\Omega}}^+$}

The measurement of $\mathrm{\Xi^-}$+$\overline{\mathrm{\Xi}}^+$ and $\mathrm{\Omega}^-$+$\overline{\mathrm{\Omega}}^+$ was performed using the following decay channels: $\rm \Xi^{-} \rightarrow \rm \Lambda + \pi^{-}$, $\overline{\rm \Xi}^{+} \rightarrow \overline{\rm \Lambda} + \pi^{+}$
(branching ratio 99.9$\%$) and $\rm \Omega^{-} \rightarrow \rm \Lambda + \rm K^{-}$, $\overline{\rm \Omega}^{+} \rightarrow \overline{\rm \Lambda} + \rm K^{+}$ (branching ratio 67.8$\%$) \cite{Beringer:1900zz}. The reconstruction of $\mathrm{\Xi^-}$+$\overline{\mathrm{\Xi}}^+$ and $\mathrm{\Omega}^-$+$\overline{\mathrm{\Omega}}^+$ is performed based on the cascade topology of the decay, 
consisting of the V-shape structure of the $\rm{\Lambda}$-decay and a charged particle not associated to the V0, referred to as bachelor track i.e.~$\pi^{\pm}$ and $\mathrm{K^{\pm}}$ for the case of $\mathrm{\Xi^-}$+$\overline{\mathrm{\Xi}}^+$ and $\mathrm{\Omega}^-$+$\overline{\mathrm{\Omega}}^+$, respectively.

To reconstruct $\mathrm{\Xi^-}$+$\overline{\mathrm{\Xi}}^+$ and $\mathrm{\Omega}^-$+$\overline{\mathrm{\Omega}}^+$ candidates, topological and kinematic criteria were applied to first select 
the V0 decay products and then to match them with the secondary, bachelor track. The track selection criteria, summarised in Tables~\ref{tab:cutsCascade}-\ref{tab:trackSelection}, for the reconstruction of $\mathrm{\Xi^-}$+$\overline{\mathrm{\Xi}}^+$ and $\mathrm{\Omega}^-$+$\overline{\mathrm{\Omega}}^+$ follow the procedure described in \cite{ABELEV:2013zaa}. 
The cuts that contributed significantly to the reduction of the combinatorial background were the predefined window around 
the $\Lambda$+$\overline{\Lambda}$ mass, the DCA cut between the V0 and the bachelor track, and the V0 and cascade pointing angle with respect to the primary vertex position. Finally, for all three 
decay tracks, a particle identification cut for the pion, kaon or proton hypotheses was applied using their 
energy loss in the TPC. This was done by selecting particles within three standard deviations from the Bethe-Bloch curve for each mass hypothesis.

Examples of invariant mass distributions for two of the lowest transverse momentum intervals used in this analysis before the background subtraction for $\mathrm{\Xi^-}$+$\overline{\mathrm{\Xi}}^+$ and $\mathrm{\Omega}^-$+$\overline{\mathrm{\Omega}}^+$ for the 10--20$\%$ 
centrality class of Pb--Pb collisions can be seen in Fig.~\ref{fig.massplots} (d) and (e). These distributions are fitted with a sum of a Gaussian function and a third-order polynomial to estimate the signal and the background in the mass peak. The signal to background ratio in the mass peak varies from about 2 (central events) to larger than 10 (peripheral events) for  $\mathrm{\Xi^-}$+$\overline{\mathrm{\Xi}}^+$, while for $\mathrm{\Omega}^-$+$\overline{\mathrm{\Omega}}^+$ it is between 
1 (central events) and larger than 4 (peripheral events). The $v_2(\pt)$ results are reported 
for $|y| < 0.5$ and $1.0 < p_{\rm{T}} < 6.0$~GeV/$c$ for $\mathrm{\Xi^-}$+$\overline{\mathrm{\Xi}}^+$ and $1.5 < p_{\rm{T}} < 6.0$~GeV/$c$ for $\mathrm{\Omega}^-$+$\overline{\mathrm{\Omega}}^+$.

\begin{table}[h!bt]
\begin{center}
    \begin{tabular}{@{}ll@{}}
    \multicolumn{2}{c}{$\mathrm{\Xi^-}$+$\overline{\mathrm{\Xi}}^+$ and $\mathrm{\Omega}^-$+$\overline{\mathrm{\Omega}}^+$ candidates} \\\hline\hline
  DCA between V0 and bachelor track &$\leq 0.3$~cm\\
  $\cos{\theta_\mathrm{p}}$ & $\geq 0.999$\\
Decay radius & $0.9 \leq r \leq 100$~cm\\
      $|y|$                     & $\leq$ 0.5\\
  \hline
  \end{tabular}
  \caption{\label{tab:cutsCascade} Topological selections for $\mathrm{\Xi^-}$,$\overline{\mathrm{\Xi}}^+$, $\mathrm{\Omega}^-$ and $\overline{\mathrm{\Omega}}^+$ candidates.}  
\end{center}
\end{table}

\begin{table}[h!tb]
\begin{center}
     \begin{tabular}{@{}ll@{}}
    \multicolumn{2}{c}{$\Lambda$ ($\overline{\Lambda}$) decay products} \\\hline\hline
  V0 invariant mass & $1.108 \leq m_{\mathrm{inv}} \leq 1.124$~GeV/$c^2$ \\
  DCA of V0 to primary vertex & $\geq 0.05$~cm\\
  DCA of decay tracks to primary vertex & $\geq 0.1$~cm\\
  DCA between decay tracks & $\leq 1.0$~cm\\
  $\cos{\theta_\mathrm{p}}$ & $\geq 0.98$\\
Radius & $0.9 \leq r \leq 100$~cm\\
\hline
  \end{tabular}
  \caption{\label{tab:cutsV0} Topological selections for the $\Lambda$+$\overline{\Lambda}$ decay product of $\mathrm{\Xi^-}$,$\overline{\mathrm{\Xi}}^+$, $\mathrm{\Omega}^-$ and $\overline{\mathrm{\Omega}}^+$.}
\end{center}
\end{table}

\begin{table}[h!bt]
\begin{center}
    \begin{tabular}{@{}ll@{}}
    \multicolumn{2}{c}{Selection of bachelor tracks} \\\hline\hline
  DCA of bachelor track to primary vertex  & $ \geq 0.03$ cm\\
  $p_{\mathrm{T}}$ & $\geq 0.15$ GeV/$c$\\
  Pseudo-rapidity & $|\eta| < 0.8$\\
  Number of TPC space points & $ \geq 70$\\
  $\chi^2$ per TPC space point per d.o.f. & $ < 2$\\
  n$\sigma$ d$E$/d$x$ (TPC PID) & 3\\
  \hline
  \end{tabular}
  \caption{\label{tab:trackSelection} Selection criteria for secondary, bachelor tracks}  
\end{center}
\end{table}

%% file: section/flowMethods.tex
\section{Extraction of $v_2(\pt)$}
\label{Sec:FlowMethods}

The $v_2(\pt)$ was calculated with the Scalar Product (SP) \cite{Adler:2002pu,Ref:EventPlane}, a two-particle correlation method, using a pseudo-rapidity gap of $|\Delta\eta| > 0.9$ between the identified hadron under study and the reference flow particles. The applied gap reduces correlations 
not related to the symmetry plane $\Psi_n$, such as correlations due to resonance decays and jets, known as non--flow effects.

The SP method is based on the calculation of
the $Q$-vector \cite{Ref:EventPlane}, computed from a set of reference
flow particles (RFP) and defined as:

\begin{equation}
  \vec{Q}_n = \sum_{i \in \mathrm{RFP}} w_{i} e^{in\varphi_{i}},
\label{eq:qvector}
\end{equation}

\noindent where $\varphi_i$ is the azimuthal angle of the $i$-th reference flow particle, $n$ is the order of the harmonic and $w_i$ is a weight applied for every RFP.

The default results were obtained by dividing each event into three sub-events A, B and C using three different detectors. The reference flow particles were taken from sub-events A 
and C, using the \VZERO-A and \VZERO-C detectors, respectively. Each of the \VZERO~arrays consists of 32 channels and is segmented
in four rings in the radial direction, and each ring is divided in eight sectors in the azimuthal direction. They cover the pseudo-rapidity ranges 
of $2.8 < \eta < 5.1$ and $-3.7 < \eta < -1.7$ for \VZERO-A and \VZERO-C, respectively. Since these detectors do not provide tracking information, the 
amplitude of the signal from each cell, which is proportional to the number of particles that cause a hit, was used as a weight $w_i$. A calibration procedure~\cite{Abelev:2014ffa,Abbas:2013taa} was performed prior to the usage of these signals, to account for fluctuations induced by the performance of the hardware, and for different conditions of the LHC for each data taking period. The particles under study (i.e.~$\pi^{\pm}$, $\mathrm{K}^{\pm}$, 
$\mathrm{K}^0_s$, p, $\overline{\mathrm{p}}$, $\phi$, $\Lambda$, $\overline{\Lambda}$, $\Xi^-$, $\overline{\Xi}^+$, $\Omega^-$ and 
$\overline{\Omega}^+$) were taken from sub-event B within $|y| < 0.5$ as described in Section~\ref{Sec:AnalysisDetails}, using the region covered by the mid-rapidity detectors.  

The $v_2$ was then calculated using the unit flow vector 
$\vec{\mathrm{u}}_2^\mathrm{B} =  e^{2i\varphi^\mathrm{B}}$ measured in sub-event B according to

\begin{equation}
 v_2 = 
\sqrt{
  \frac{
  \big\langle \big\langle \vec{\mathrm{u}}_2^{\mathrm{B}} \cdot \frac{\vec{Q}_2^\mathrm{A*}}{M_\mathrm{A}}\big\rangle \big\rangle
  \big\langle \big\langle \vec{\mathrm{u}}_2^{\mathrm{B}} \cdot \frac{\vec{Q}_2^{\mathrm{C*}}}{M_{\mathrm{C}}}\big\rangle \big\rangle
  }{
  \big\langle \frac{\vec{Q}_2^\mathrm{A}}{M_\mathrm{A}} \cdot \frac{\vec{Q}_2^\mathrm{C*}}{M_\mathrm{C}} \big\rangle
  }
},
\label{eq:sp}
\end{equation}

\noindent where the two brackets in the numerator indicate an average over all  particles of interest and over all events, $M_{\mathrm{A}}$ and $M_{\mathrm{C}}$ are the estimates of multiplicity from the \VZERO-A and \VZERO-C detectors, and $\vec{Q}_2^\mathrm{A*}$, $\vec{Q}_2^\mathrm{C*}$ are the complex conjugates of the flow vector calculated in sub-event A and C, respectively. The non uniformity of the detector azimuthal efficiency is taken into account in the SP method by 
applying the inverse of the event-averaged signal as a weight for each of the \VZERO~segments~\cite{Abelev:2014ffa,Abbas:2013taa}, together with a recentring procedure (i.e. subtraction of the average centroid position of each sector)~\cite{Abelev:2014ffa}.

To investigate the dependence of the results on the applied pseudo-rapidity gap and the possible residual contribution from non-flow effects, the analysis was repeated taking the particles under study from $y > 0$ (or $y < 0$) and the reference particles from $-3.7 < \eta < -1.7$ i.e. \VZERO-C (or $2.8 < \eta < 5.1$ i.e. \VZERO-A). The results were consistent with the default ones within the uncertainties.

\subsection{Reconstruction of $v_2(\pt)$ with the invariant mass method}
\label{SubSec:V2Minv}

For the $v_2(\pt)$ measurement of $\mathrm{K}^0_s$, $\phi$, $\Lambda$ ($\overline{\Lambda}$), $\Xi^-$ ($\overline{\Xi}^+$), and $\Omega^-$ ($\overline{\Omega}^+$), the \vtwo~versus invariant mass 
($m_{\rm{inv}}$) method \cite{Borghini:2004ra,Abelev:2008ae} was used. The $v_2(\pt)$ of the particles of interest ($v_2^{\rm{Sgn}}(\pt)$) is extracted from the total 
$v_2^{\rm{Tot}}(\pt)$ of all pairs or triplets contributing to the invariant mass window and from background ($v_2^{\rm{Bg}}(\pt)$) contributions, measured with the SP method, weighted by their relative yields according to

\begin{equation}
v_2^{\rm{Tot}}(m_{\rm{inv}},\pt)
= v_2^{\mathrm{Sgn}}(\pt)\frac{N^{\rm{Sgn}}(m_{\rm{inv}},\pt)}{N^{\rm{Tot}}(m_{\rm{inv}},\pt)}
+ v_2^{\mathrm{Bg}}(m_{\rm{inv}},\pt)\frac{N^{\mathrm{Bg}}(m_{\rm{inv}},\pt)}{N^{\rm{Tot}}(m_{\rm{inv}},\pt)},
\label{eq:v2VsInvMass}
\end{equation}

\noindent where $N^{\rm{Tot}}$ is the total number of candidates, and $N^{\rm{Bg}}$ and $N^{\rm{Sgn}}$ are the yields of the background and signal respectively. The relative yields are determined from the fits to the invariant mass distributions shown in Fig.~\ref{fig.massplots} for each transverse momentum interval.

\begin{figure}[tbh!h]
\includegraphics[width=\textwidth]{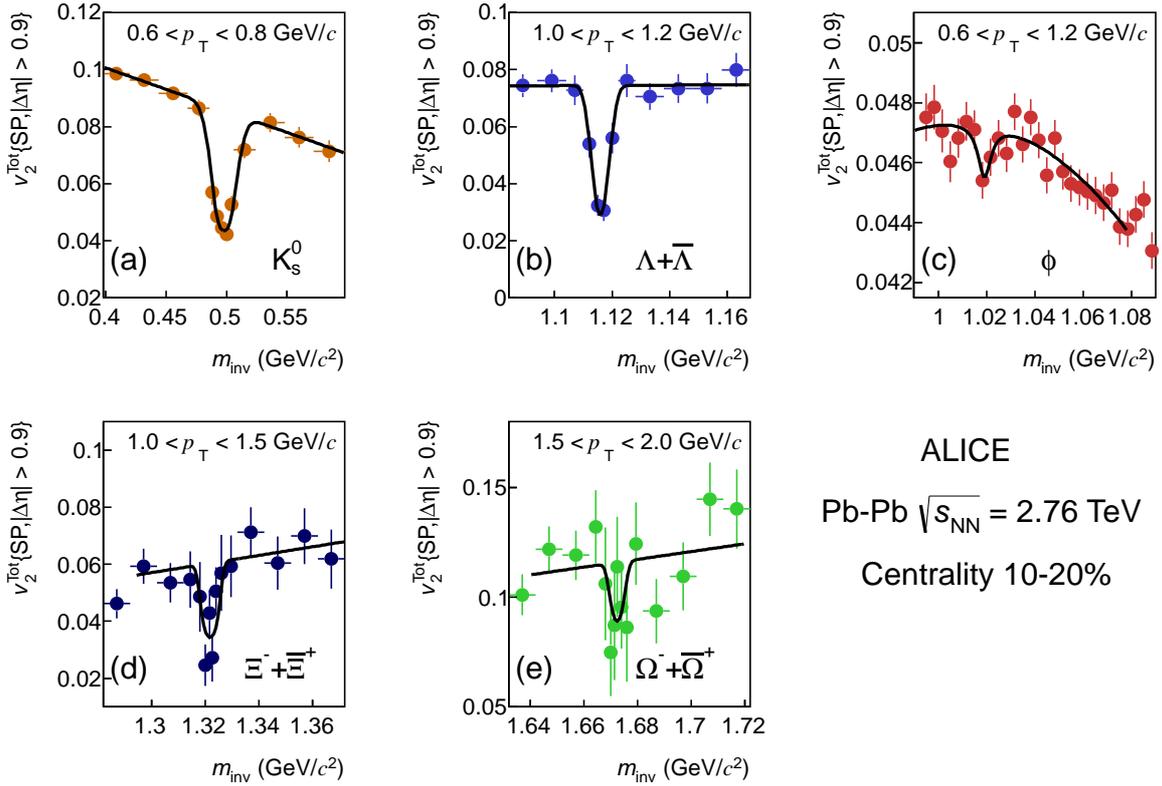}
\caption{\label{fig:v2mass} The measured value of $v_2^{\mathrm{Tot}}$ in
  the 10--20$\%$ centrality interval of Pb--Pb collisions as a function of the invariant mass
  for all decaying particles presented in this article.}
\end{figure}

For a given $p_T$, the observed $v_2^{\mathrm{Sgn}}$ is determined by fitting simultaneously the invariant mass distribution and the $v_2^{\rm{Tot}}(m_{\rm{inv}})$ according to Eq.~\ref{eq:v2VsInvMass}. The value of $v_2^{\mathrm{Bg}}$ in the peak region is obtained by interpolating the values from the two sideband regions. Figure~\ref{fig:v2mass} 
shows these fits for each decaying particles in a given characteristic $\pt$~range in the 10--20$\%$ centrality interval of Pb--Pb collisions.

%% file: section/systematics.tex
\section{Systematic uncertainties}
\label{Sec:Systematics}

The systematic uncertainties in all results were determined by varying the event and particle selections and by studying the detector response with Monte-Carlo (MC) simulations. The contributions from different sources, described below, were estimated for every particle species and centrality separately, as the maximum difference of $v_2(\pt)$ extracted from the variations of the cut values, relative to the main result extracted using the default selection criteria described in Section~\ref{Sec:AnalysisDetails}. 
The ranges of each individual contribution over all centralities, expressed in percentages of the measured values, are summarized in Table~\ref{tab:systematics1} for $\pi^\pm$, $\mathrm{K}^\pm$ and p+$\overline{\mathrm{p}}$ and Table~\ref{tab:systematics2} for the decaying particles.
The total systematic uncertainty was calculated as the quadratic sum of these individual contributions. 

\begin{table}[h!tb]
\begin{center}
  \begin{tabular*}{140mm}{@{\extracolsep{\fill}}lccc}
    Error source & $\pi^\pm$ & $\mathrm{K}^\pm$ & p+$\overline{\mathrm{p}}$  \\
		\hline
		\hline
		Vertex position &  \multicolumn{3}{c}{$\leq 0.1\%$}  \\
		\hline
		Centrality estimator &  \multicolumn{3}{c}{$\leq 0.1\%$}  \\
		\hline
		Magnetic field polarity &  \multicolumn{3}{c}{$\leq 0.1\%$}  \\
		\hline
		Number of TPC space points &  \multicolumn{3}{c}{$\leq 0.1\%$}  \\
		\hline
		$\chi^2$ per TPC space point &  \multicolumn{3}{c}{$\leq 0.1\%$}  \\
		\hline
		Particle identification & \multicolumn{3}{c}{$5-15\%$} \\
	         \hline
		Contamination & \multicolumn{3}{c}{$\leq 5\%$} \\
	         \hline
		Feed-down &\multicolumn{2}{c}{$\leq 0.1\%$} & $5\%$  \\
	         \hline
		Selection of reference particles &\multicolumn{3}{c}{$\leq 0.1\%$}  \\
	         \hline
		Local track density &\multicolumn{3}{c}{$\leq 15\%$}  \\
	         \hline
  \end{tabular*}
  \caption{\label{tab:systematics1} Summary of systematic errors for the $v_2(\pt)$ measurement for $\pi^\pm$, $\mathrm{K}^\pm$ and p+$\overline{\mathrm{p}}$. Percentages given are fractions of the measured values.}
\end{center}
\end{table}

\begin{table}[h!tb]
\begin{center}
  \begin{tabular*}{140mm}{@{\extracolsep{\fill}}lccccc}
    Error source & $\phi$ & $\mathrm{K}^0_\mathrm{S}$ & $\mathrm{\Lambda}$+$\overline{\mathrm{\Lambda}}$ & $\mathrm{\Xi^-}$+$\overline{\mathrm{\Xi}}^+$ & $\mathrm{\Omega}^-$+$\overline{\mathrm{\Omega}}^+$ \\
		\hline
		\hline
		Vertex position &  \multicolumn{5}{c}{$\leq 0.1\%$}  \\
		\hline
		Centrality estimator &  \multicolumn{5}{c}{$\leq 0.1\%$}  \\
		\hline
		Magnetic field polarity &  \multicolumn{5}{c}{$\leq 0.1\%$}  \\
		\hline
		Number of TPC space points &  \multicolumn{5}{c}{$\leq 0.1\%$}  \\
		\hline
		$\chi^2$ per TPC space point &  \multicolumn{5}{c}{$\leq 0.1\%$}  \\
		\hline
		Decay length & n/a & \multicolumn{4}{c}{$\leq 0.1\%$}  \\
		\hline
		Decay vertex (radial position) & n/a & \multicolumn{4}{c}{$\leq 0.1\%$}  \\
		\hline
		Armenteros-Podolanski variables & n/a & \multicolumn{2}{c}{$\leq 0.1\%$} & n/a & n/a  \\
		\hline
		DCA decay products to primary vertex & n/a & \multicolumn{4}{c}{$\leq 0.1\%$}  \\
		\hline
		DCA between decay products & n/a & \multicolumn{2}{c}{$\leq 10\%$} & n/a & n/a  \\
		\hline
		Pointing angle $\cos\theta_{\rm{p}}$ & n/a & \multicolumn{2}{c}{$\leq 10\%$} & n/a & n/a  \\
		\hline
		Particle identification & \multicolumn{5}{c}{$5-15\%$} \\
	         \hline
		Contamination & \multicolumn{5}{c}{$\leq 5\%$} \\
	         \hline
		Signal and background estimation & $5-10\%$ & $\leq 0.1\%$ & $\leq 0.1\%$ & \multicolumn{2}{c}{$5-10\%$} \\
	         \hline
		Feed-down & \multicolumn{5}{c}{$\leq 0.1\%$} \\
	         \hline
		Selection of reference particles & $5\%$ & $1-5\%$ & $\leq 0.1\%$ & \multicolumn{2}{c}{$1-5\%$} \\
	         \hline
		Local track density &\multicolumn{5}{c}{$\leq 0.1\%$}  \\
	         \hline
  \end{tabular*}
  \caption{\label{tab:systematics2} Summary of systematic errors for the $v_2(\pt)$ measurement for the decaying particles. Percentages given are fractions of the measured values (the notation n/a stands for non-applicable).}
\end{center}
\end{table}

The event sample was varied by (i) changing the cut on the position of the primary vertex along the beam axis ($V_z$) from $\pm10$~cm to $\pm7$~cm, (ii) changing the centrality selection criteria from the signal amplitudes of the \VZERO~scintillator detectors to the multiplicity of TPC tracks, and the number of ITS clusters. For all species and centralities, the resulting $v_2(p_\mathrm{T})$ was consistent with results obtained with the default cuts. Results from runs with different magnetic field polarities did not exhibit any systematic change in $v_2(\pt)$ for any particle species for any centrality.

In addition, the track selection criteria, such as the number of \TPC~space points and the $\chi^2$ per TPC space point per degree of freedom were varied, for both primary hadrons (i.e.~$\pi^{\pm}$, $\mathrm{K}^{\pm}$ and p+$\overline{\mathrm{p}}$) and the daughters of decaying particles. No systematic deviations in the values of $v_2(\pt)$ relative to the results obtained with the default selection were found. To estimate the uncertainties for the decaying particles, the ranges of the cuts for the decay length, the radial position of the decay vertex, the correlation between the Armenteros-Podolanski variables, and the DCA of the decay products to the primary vertex were varied by as much as three times the default values. These variations did not affect the measured result. Differences were observed for the cases of $\mathrm{K}^0_\mathrm{S}$ and $\mathrm{\Lambda}$($\mathrm{\overline{\Lambda}}$), when changing the requirement on the minimal distance between the two daughter tracks (DCA) and the pointing angle $\cos\theta_{\rm{p}}$. These differences resulted in systematic uncertainties on the measured $v_2(\pt)$ of $\leq 10\%$ for both $\mathrm{K}^0_\mathrm{S}$ and $\mathrm{\Lambda}$+$\overline{\mathrm{\Lambda}}$. 

Systematic uncertainties associated with the particle identification procedure were studied by varying the number of standard deviations (e.g.~between 1-4$\sigma$) around the expected energy loss in the TPC (similarly for the \TOF) for a given particle species. Furthermore, the contamination of the kaon and proton samples was studied in collision data by selecting pions and (anti)protons from $\mathrm{K}^{0}_{s}$ and $\mathrm{\Lambda}$($\mathrm{\overline{\Lambda}}$) decays, respectively, and then determining the number that passed the kaon selections. The resulting uncertainties related to the particle (mis-)identification on the extracted $v_2(\pt)$ values depend weakly on centrality, increase with transverse momentum and are in the range 5--15$\%$ for all particle species. 

The feed-down from weakly decaying particles was found to be a significant factor only for p+$\overline{\mathrm{p}}$. Its contribution was determined by selecting p($\overline{\mathrm{p}}$) from $\Lambda$($\overline{\rm \Lambda}$) decays and measuring their anisotropy with the SP method. It was found that the systematic uncertainty in the extracted $v_2(\pt)$ resulting from this source was at maximum 5$\%$ for all centralities. 

The systematic uncertainty originating from the signal extraction and the background description, used in the method described in Section~\ref{SubSec:V2Minv}, was studied by extracting the yields with a simple bin-counting method. The uncertainty was further investigated by using different functions to describe the signal (e.g.~Breit-Wigner, Gaussian and double Gaussian) and background (e.g.~polynomial of different orders) in the invariant mass distribution. In addition, for the case of the $\phi$-meson, a subtraction of the background estimated with the mixed events method was used. The mixed events were formed by combining tracks from separate events belonging to the same centrality interval, with a reconstructed primary vertex position along the beam axis within $\pm 2$~cm) from the value of the original event. The corresponding systematic uncertainties in the extracted $v_2(\pt)$ from the previous sources were below $0.1 \%$ for $\mathrm{K}^0_\mathrm{S}$ and $\Lambda$($\overline{\rm \Lambda}$). For the $\phi$-meson, $\mathrm{\Xi^-}$($\overline{\mathrm{\Xi}}^+$) and $\mathrm{\Omega}^-$($\overline{\mathrm{\Omega}}^+$) they were found to be in the range 5--10$\%$.

The systematic uncertainties originating from the selection of reference flow particles were extracted by measuring $v_2(\pt)$ with reference particles estimated either with the three sub-event method described in Section~\ref{Sec:FlowMethods}, or using two sub-events with either the \VZERO-A or the \VZERO-C detectors separately. This resulted in a systematic uncertainty in the extracted $v_2(\pt)$ up to 5$\%$ for the $\phi$-meson, $\mathrm{\Xi^-}$($\overline{\mathrm{\Xi}}^+$) and $\mathrm{\Omega}^-$($\overline{\mathrm{\Omega}}^+$).

Finally, due to the anisotropy of particle production there are more particles in the direction of the symmetry plane than in the direction perpendicular to the plane. Consequently, the detector occupancy varies as a function of the angle relative to the symmetry plane. The track finding and track reconstruction are known to depend slightly on the detector occupancy. A local track density dependent efficiency would reduce the reconstructed~{\vtwo} for all charged tracks proportional to the modulation of the efficiency. 
In order to investigate how a variation in occupancy affects the efficiency for track finding and track reconstruction, dedicated Monte-Carlo events using a generator without any physics input (i.e. a so-called toy-model) with the particle yields and ratios, momentum spectra, and flow coefficients (e.g.~$v_2(\pt)$, $v_3(\pt)$) measured in data for every centrality interval were generated. The ALICE detector response for these events was determined using a GEANT3~\cite{Brun:1994aa} simulation. The occupancy dependence of the tracking and matching between the \TPC~and the \TOF~contributed to the systematic uncertainty of $v_2(\pt)$ for $\pi^{\pm}$, $\mathrm{K}^{\pm}$ and p+$\overline{\mathrm{p}}$ with less than $10\%$, independent of momentum. An additional contribution of less than 6\% of the measured $v_2(\pt)$ for $p_{\mathrm{T}} > 2.5$~GeV/$c$ for the same particles resulted from the sensitivity of the \TPC~$\mathrm{d}E/\mathrm{d}x$ measurement to the local track density.
The analysis of the MC events did not indicate any additional systematic effect related to the detector occupancy for the other particle species and was in agreement with a numerical calculation of the particle reconstruction efficiency as a function of the total event multiplicity.

%% file: section/discussion.tex
\section{Results and discussion}
\label{Sec:Results}

\begin{figure}[tbh!f]
\includegraphics[width=\textwidth]{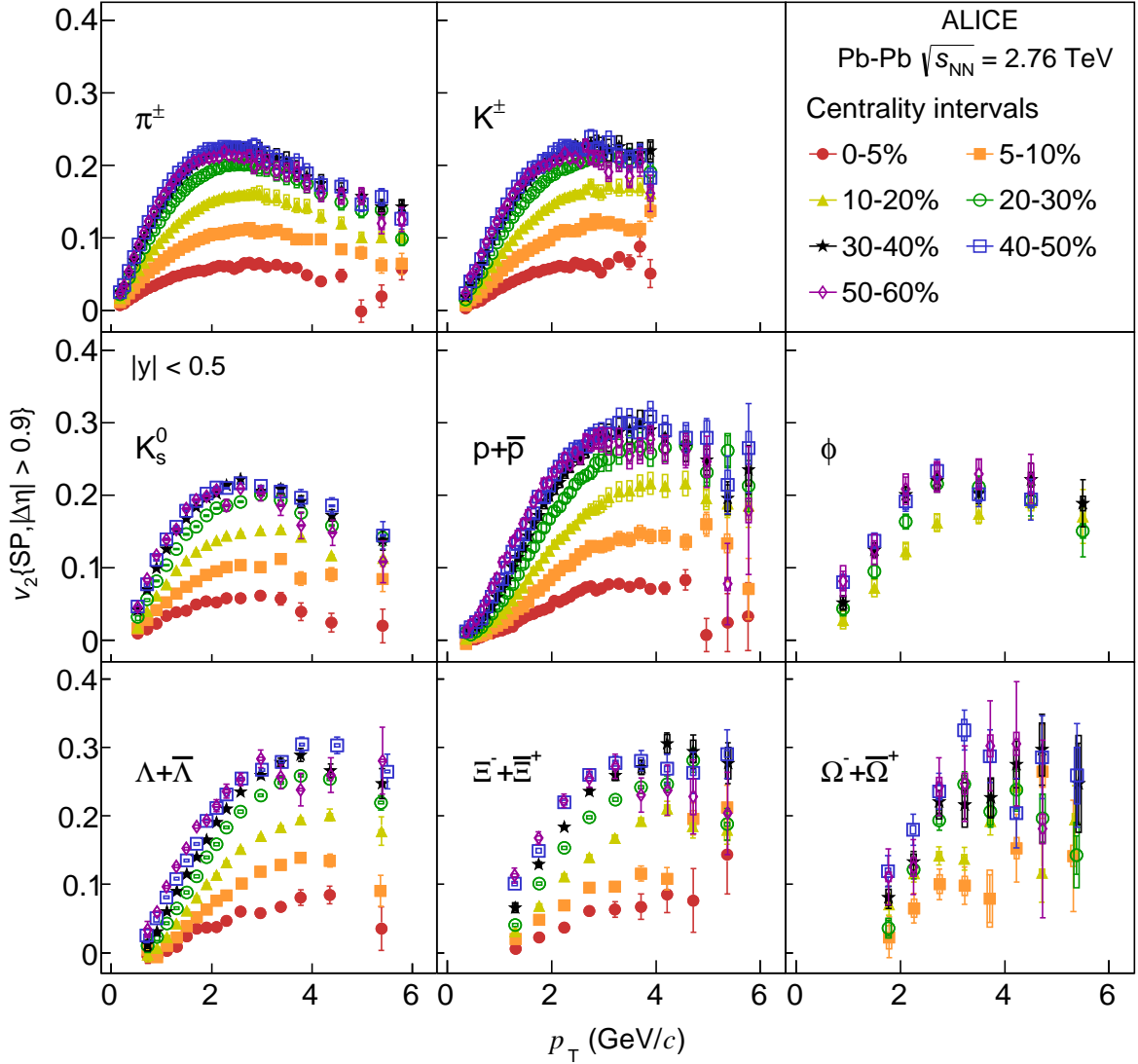}
\caption{The $p_{\rm{T}}$-differential \vtwo~for different centralities of Pb--Pb collisions at $\sqrt{s_\mathrm{{NN}}} = 2.76$~TeV grouped by particle species.}
\label{fig:pTDifferentialv2AllSpecies} 
\end{figure}

Figure~\ref{fig:pTDifferentialv2AllSpecies} presents the $p_{\rm{T}}$-differential \vtwo~for all identified particles 
measured in Pb--Pb collisions at $\sqrt{s_\mathrm{{NN}}} = 2.76$~TeV. To illustrate the development of \vtwo~as a function of centrality for $\pi^{\pm}$, $\mathrm{K}^{\pm}$, 
$\mathrm{K}^0_\mathrm{S}$, p+$\overline{\mathrm{p}}$, $\phi$, $\Lambda$+$\overline{\mathrm{\Lambda}}$, $\mathrm{\Xi^-}$+$\overline{\mathrm{\Xi}}^+$ and $\mathrm{\Omega}^-$+$\overline{\mathrm{\Omega}}^+$, the results are grouped by particle 
species in different panels. The error
bars correspond to statistical uncertainties, while the hollow boxes around each point indicate the systematic 
uncertainties. The same convention for these uncertainties is used for the rest of the figures in this article. The systematic uncertainties in many cases are smaller than the marker size.
 
The value of $v_{2}(p_{\rm{T}})$~progressively increases from central to peripheral collisions up to the 40--50$\%$ centrality 
interval for all particle species. This is consistent with the picture of the final state anisotropy driven by the geometry of the collision, as represented by the initial state eccentricity which increases for peripheral collisions. For more peripheral events (i.e.~50--60$\%$), the magnitude 
of \vtwo~does not change significantly within the systematic uncertainties compared to the previous centrality interval. According to~\cite{Song:2013qma}, this might originate from a convolution of different effects such as the smaller lifetime of the fireball in peripheral compared to more central collisions that does not allow $v_2$ to further develop. The authors also attributed this effect to the less significant (compared to more central events) contribution of eccentricity fluctuations and to final state hadronic effects.
The transverse momentum dependence of \vtwo~exhibits an almost linear 
increase up to about 3~GeV/$c$. This initial rise is followed by a saturation and then a decrease observed for all particles and centralities. The position of the maxima depends on the particle species and on the centrality interval.

\begin{figure}[tbh!f]
\includegraphics[width=\textwidth]{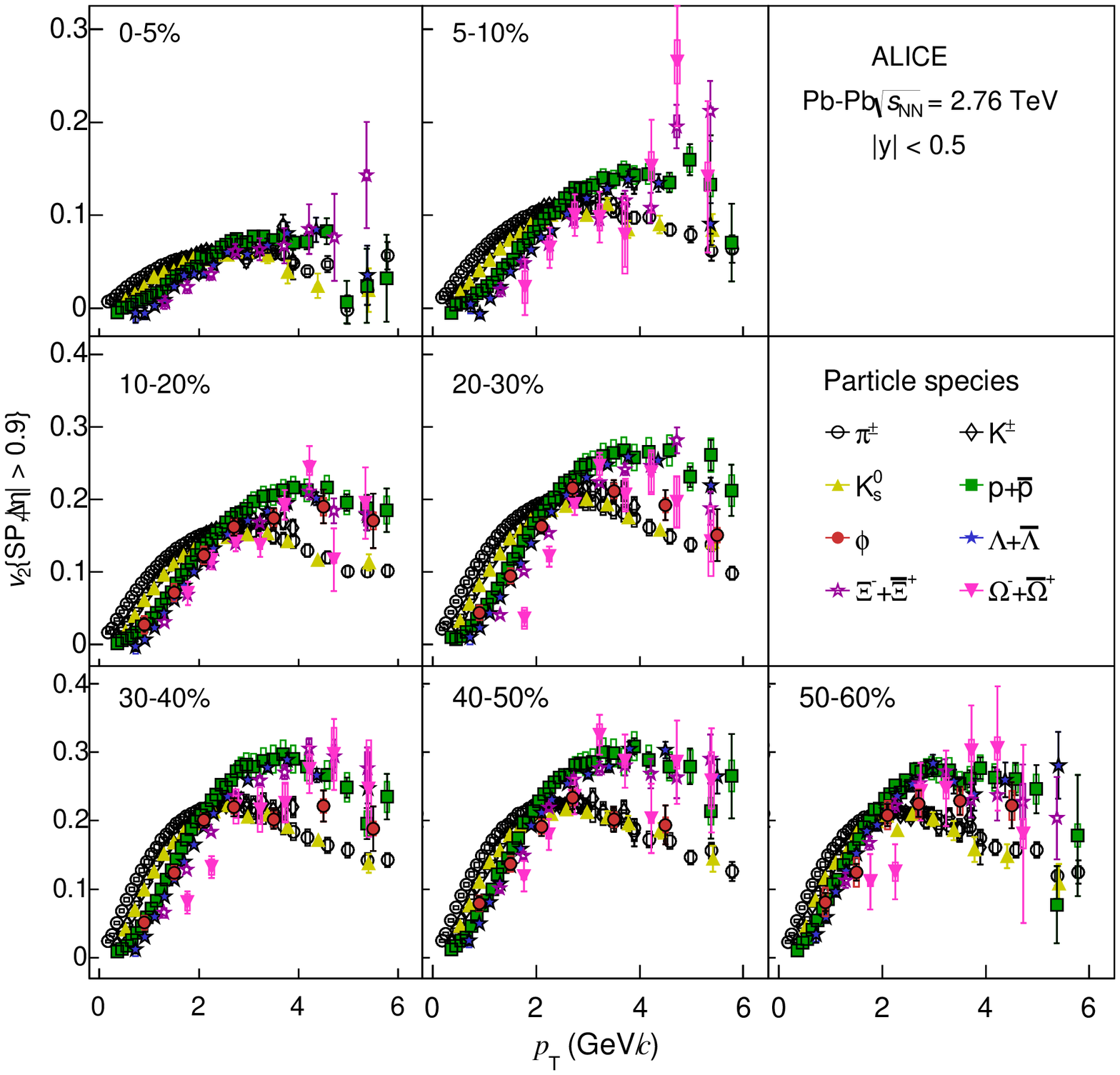}
\caption{The $p_{\rm{T}}$-differential \vtwo~for different particle species grouped by centrality class of Pb--Pb collisions at $\sqrt{s_\mathrm{{NN}}} = 2.76$~TeV.}
\label{fig:pTDifferentialv2AllSpecies2}
\end{figure}

Figure~\ref{fig:pTDifferentialv2AllSpecies2} presents the same data points shown in Fig.~\ref{fig:pTDifferentialv2AllSpecies}, arranged into panels of different event centrality selection, illustrating how $v_2(\pt)$ develops 
for different particle species within the same centrality interval. The panels are arranged by decreasing centrality from left to right and top to bottom. The top left plot presents results for the 5$\%$ most central Pb--Pb collisions, while the most
peripheral interval presented in this article, the 50--60$\%$ centrality, is shown in the bottom right plot.

A clear mass ordering is seen for all centralities in the low $p_{\rm{T}}$ region (i.e.~$p_{\rm{T}} \le 3$~GeV/$c$), 
attributed to the interplay between elliptic and radial flow~\cite{Huovinen:2001cy,Teaney:2000cw,Voloshin:1996nv,Shen:2011eg}. Radial flow tends to create a depletion in the particle $p_{\mathrm{T}}$ spectrum at low values, which increases with increasing particle mass and transverse velocity. When introduced in a system that exhibits azimuthal anisotropy, this depletion becomes larger in-plane than out-of-plane, thereby reducing $v_2$. The net result is that at a fixed value of $p_{\mathrm{T}}$, heavier particles have smaller $v_2$ value compared to lighter ones. In addition, a crossing between the \vtwo~values of baryons (i.e.~p, $\Lambda$, $\Xi$ and 
$\Omega$ and their antiparticles) and the corresponding values of pions and kaons is observed, that takes place between 2 and 3.5~GeV/$c$, 
depending on the particle species and centrality. It is seen that the crossing between e.g.~$\pi^{\pm}$ and p+$\overline{\mathrm{p}}$ happens at lower $p_{\rm{T}}$ for peripheral than for central collisions. For more central collisions, the crossing point moves to higher $p_{\rm{T}}$ values, since the common velocity
field, which exhibits a significant centrality dependence \cite{Abelev:2013vea}, affects heavy particles more. For higher 
values of $p_{\rm{T}}$ ($p_{\rm{T}} > 3$~GeV/$c$), particles tend to group according to their type, i.e.~mesons 
and baryons. This feature will be discussed in detail in Section~\ref{SubSec:NCQ}.

Figure~\ref{fig:pTDifferentialv2AllSpecies2} also shows how \vtwo~develops for $\mathrm{K}^{\pm}$ and  $\mathrm{K}^{0}_{s}$ as a function of transverse momentum for different centralities. A centrality and $p_{\rm{T}}$~dependent difference is observed in these two measurements. 
In particular, the $v_2(\pt)$ for neutral kaons is systematically lower than that of their charged counterparts. 
The difference between the two measurements reaches up to two standard deviations in central, and is on the level of one standard deviation in peripheral Pb--Pb collisions.
A number of cross checks performed using data (e.g.~calculating the $v_2(\pt)$ of kaons identified via the kink topology of their leptonic decay, studies of feed-down corrections) as well as analysis of the dedicated MC simulations described in Section \ref{Sec:Systematics} did not reveal an origin for the difference. 
Additionally, no physics mechanism (e.g.\ feed-down from $\phi$, larger mass for $\mathrm{K}^{0}_{s}$ than $\mathrm{K}^{\pm}$ by about 4~MeV/$c^2$) responsible for the difference could be found. 
Therefore, for the remaining figures of this article, the $v_2(\pt)$ results for $\mathrm{K}^{\pm}$ and $\mathrm{K}^0_\mathrm{S}$ were considered as two independent measurements of kaon flow. 
Following the description given by the PDG~(Section~5 of~\cite{Beringer:1900zz}) we averaged the two sets of data points in the overlapping $p_{\mathrm{T}}$ region (i.e.\ for $p_{\mathrm{T}}  < 4$~GeV/$c$) using the statistical and the total (uncorrelated) systematic uncertainties in every transverse momentum interval as a weight. 
The uncertainty on the average was obtained from the individual uncertainties added in quadrature and the differences between the two measurements assigned assymetrically. 
With this procedure, the averaged values for kaons are closer to the $\mathrm{K}^0_\mathrm{S}$, which have a smaller error (and hence larger weight), but the assigned asymmetric error covers both the original charged and neutral data points.
For $p_{\mathrm{T}} > 4$~GeV/$c$, only the $\mathrm{K}^0_\mathrm{S}$ data points were used and their uncertainty has not been modified.
Since our studies did not identify a common underlying effect (e.g.\ from charged particle tracking) as the source of the difference, the additional uncertainties were not propagated to other particles.

Among all particle species, the $\phi$-meson is of particular interest since its mass is close to that of p and 
$\Lambda$. It provides an excellent testing ground of both the mass ordering 
and the baryon-meson grouping at low and intermediate $p_{\rm{T}}$, respectively. The 
\vtwo~values of the $\phi$-meson in Fig.~\ref{fig:pTDifferentialv2AllSpecies2} indicate that for $p_{\rm{T}} < 3$~GeV/$c$ 
it follows the mass-ordered hierarchy. However, for higher $p_{\rm{T}}$ values the 
$\phi$ data points appear to follow the band of baryons for central events within uncertainties. For peripheral collisions though, the $v_2$ values of the $\phi$-meson shift progressively to the band of mesons. This is congruous with the observation that the $(\mathrm{p} + \overline{\mathrm{p}})/\phi$ ratio, calculated from the transverse momentum spectra, is almost constant as a function of transverse momentum in central Pb--Pb events, while for peripheral collisions the ratio decreases with increasing $p_{\mathrm{T}}$, as reported in~\cite{Abelev:2014uua}. 

Finally, the multi-strange baryons, i.e.~$\mathrm{\Xi^-}$+$\overline{\mathrm{\Xi}}^+$ and $\mathrm{\Omega}^-$+$\overline{\mathrm{\Omega}}^+$, provide another interesting test of both the mass ordering 
and the baryon-meson grouping.
Similar to all other particle species, a mass ordering is reported at low $p_{\rm{T}}$ values.  At intermediate 
$p_{\rm{T}}$ values, both particles seem to follow the band formed by the other baryons, within the statistical and systematic uncertainties. 

\subsection{Comparison with hydrodynamic calculations}
\label{SubSec:Hydro}

It has been established that hydrodynamic~\cite{Song:2007fn,Song:2007ux,Song:2008si} as well as hybrid models (hydrodynamic system evolution followed by a hadron cascade model)~\cite{Song:2010mg,Song:2011hk,Song:2013qma} describe the soft particle production at both RHIC and the LHC fairly well. 

\begin{figure}[tbh!f]
  \includegraphics[width=\textwidth,height=17cm]{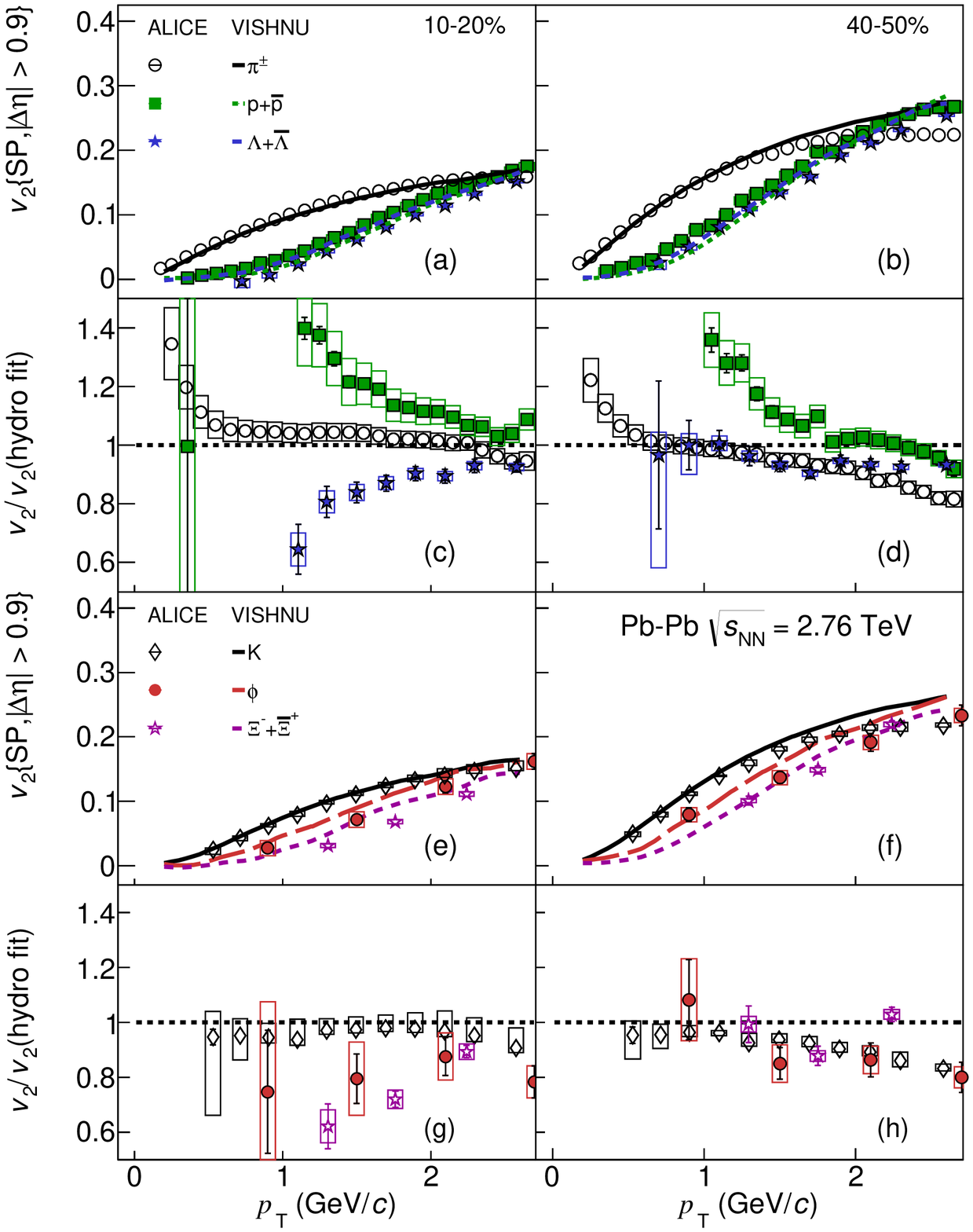}
  \caption{The $p_{\rm{T}}$-differential \vtwo~for different particle species in (a), (b), (e), (f), measured with the scalar product 
  method with a pseudo-rapidity gap $|\Delta\eta| > 0.9$ in Pb--Pb collisions at $\sqrt{s_\mathrm{{NN}}} = 2.76$~TeV, 
  compared to theoretical, hydrodynamic calculations coupled to a hadronic cascade model \cite{Song:2007fn,Song:2007ux,Song:2008si}. The panels (c), (d), (g) and (h), show the dependence of the ratio of the experimental points to a fit over the theoretical calculations as a function of $p_{\rm{T}}$. The left and right plots present the comparison for the 10--20$\%$ and 40--50$\%$ centrality intervals, respectively. The low transverse momentum points for p+$\overline{\mathrm{p}}$ are out of scale in panels (c) and (d).}
\label{fig:v2DataVsHydro}
\end{figure}

In Fig.~\ref{fig:v2DataVsHydro}, the $v_2$ measurements for two centrality intervals, the 10--20$\%$ in the left column and the 40--50$\%$ interval in the right column, are compared to hydrodynamic calculations coupled to a hadronic 
cascade model (VISHNU) \cite{Song:2010mg,Song:2011hk,Song:2013qma}. The usage of such a hybrid approach provides the possibility of investigating the influence of the hadronic stage on the development 
of elliptic flow for the different particle species. It also provides an excellent testing ground for the particles that 
are estimated to have small hadronic cross section ($\phi$, $\Xi$) and are thus expected not to be affected by 
this stage. VISHNU uses the MC-KLN model \cite{Drescher:2007ax} to describe the initial conditions, an initial time after which the hydrodynamic evolution begins at $\tau_0 = 0.9$~fm/$c$ and a value of $\eta/s = 0.16$, almost two times the lower bound of $1/4\pi$ (for $\hbar = k_{\mathrm{B}} = 1$). The transition from the hydrodynamic description to the microscopic evolution of the hadronic matter is done at a temperature of T = 165~MeV. More information about the hadronic 
cascade model can be found in \cite{Bass:1998ca,Bleicher:1999xi}. These theoretical calculations are represented in Fig.~\ref{fig:v2DataVsHydro} by the different curves with the line colour matching that of 
the experimental measurement for each species.

Figures~\ref{fig:v2DataVsHydro}-(a), (b), (e), (f), present the $p_{\rm{T}}$-differential \vtwo~for different particle species, while Fig.~\ref{fig:v2DataVsHydro}-(c), (d), (g) and (h) show the ratio of the measurement to a fit to the theoretical calculations as a function of $p_{\rm{T}}$.
It is seen that VISHNU gives a qualitatively similar picture with a similar mass ordering to that seen experimentally for most particle species.

For more central collisions the measured $v_2(\pt)$ for the $\pi^{\pm}$ is systematically above the theoretical calculations for $p_{\rm{T}} < 2$~GeV/$c$, whereas the kaon measurement is described fairly well for the same range. In addition, the model calculations appear to underestimate significantly the elliptic flow for protons, but
overestimate \vtwo~of $\Xi^{-}$+$\overline{\Xi}^+$. This multi-strange baryon is estimated to have small hadronic cross sections and thus could be unaffected from the hadronic rescattering in the later stages of the collision \cite{Biagi:1980ar,Bass:1999tu,Dumitru:1999sf,Bass:2000ib}. Furthermore, for $\Lambda$+$\overline{\Lambda}$, the model does not preserve the mass ordering observed in the experiment and overestimates the \vtwo. This could indicate that the implementation of the hadronic cascade phase and the hadronic cross-sections within the model need further improvements.

Finally, the $\phi$-meson was argued to reflect the properties of the early partonic stages in the 
evolution of the system, being less affected by the hadronic interactions. The latter is suggested by 
phenomenological calculations to stem from the small hadronic interaction cross section of the $\phi$-meson \cite{Shor:1984ui}. It is seen that VISHNU systematically overestimates $v_2(\pt)$ and expects that the measurement does not follow the mass 
ordering for $p_{\rm{T}} < 2$~GeV/$c$. This might be an indication that the $\phi$-meson's hadronic cross section 
is underestimated in these calculations.

For peripheral collisions, the model calculations agree better with the results for $\pi^{\pm}$, $\mathrm{K}$ and $\Lambda$+$\overline{\Lambda}$. However, VISHNU under-predicts the $v_2(\pt)$ values of p+$\overline{\mathrm{p}}$ and over-predicts the values for $\mathrm{K}$, $\phi$ and $\Xi^{-}$+$\overline{\Xi}^+$. 

\subsection{Comparison with RHIC results at $\sqrt{s_\mathrm{{NN}}} = 0.2$~TeV}
\label{SubSec:Rhic}

\begin{figure}[tbh!f]
  \includegraphics[width=\textwidth]{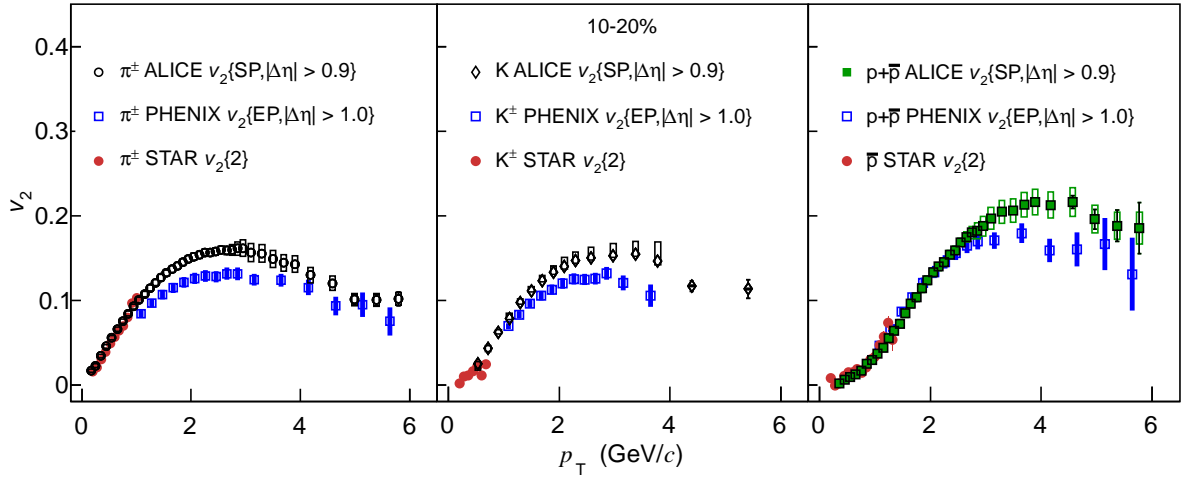}
  \caption{The comparison of the $p_{\rm{T}}$-differential \vtwo~for $\pi^{\pm}$, $\mathrm{K}$ and p+$\overline{\mathrm{p}}$ 
  for the 10--20$\%$ centrality class of Pb--Pb and Au-Au collisions at the LHC and RHIC, respectively. The RHIC 
  points are extracted from \cite{Adams:2004bi} (STAR) and \cite{Adare:2012vq} (PHENIX).}
\label{fig:v2DataVsRHIC}
\end{figure}

The mass ordering in the $p_{\rm{T}}$-differential \vtwo~and the qualitative agreement with 
hydrodynamic calculations were first reported in Au-Au collisions at RHIC energies by both STAR
\cite{Adams:2003am,Abelev:2007qg,Adams:2004bi} and PHENIX experiments \cite{Adler:2003kt,Afanasiev:2007tv,Adare:2006ti,Adare:2012vq}. In addition, one of the first experimental observations 
at the LHC \cite{Aamodt:2010pa} was that the $p_{\rm{T}}$-differential 
\vtwo~for inclusive charged particles remains almost unchanged between RHIC and LHC for several centrality intervals. On the other hand, the integrated $v_2$ values at the LHC were about 30$\%$ higher compared with RHIC. The comparison of the $v_2(\pt)$ values for different particle species in these two different energy
regimes could provide additional insight into the dynamics of anisotropic flow and the effect of radial expansion of the system.

Figure~\ref{fig:v2DataVsRHIC} presents the comparison between the measurements for $\pi^{\pm}$, $\mathrm{K}$ and p+$\overline{\mathrm{p}}$ performed at the LHC and the results from Au-Au collisions at $\sqrt{s_\mathrm{{NN}}} = 200$~GeV from STAR \cite{Adams:2004bi} and from PHENIX \cite{Adare:2012vq}. The comparison is based on the 10--20$\%$ 
centrality interval, one of the most central classes, where the values of the transverse expansion velocity extracted 
from a blast-wave fit to the identified particle spectra are $0.57 \pm 0.01 (\mathrm{stat.})$ at RHIC \cite{Adams:2005dq} and $0.639 \pm 0.004 (\mathrm{stat.}) \pm 0.022 (\mathrm{syst.})$ at the LHC \cite{Abelev:2013vea}. The $v_2(\pt)$ from STAR 
is calculated using the two particle cumulant analysis (i.e.~$v_2\{2\}(\pt)$) \cite{Adams:2004bi}, while PHENIX reconstructed $v_2(\pt)$ using the event plane method with a pseudo-rapidity gap of $|\Delta\eta| > 1.0$ 
\cite{Adare:2012vq}. These two measurements have different sensitivity to 
non-flow effects, which makes a quantitative comparison difficult.

At low values of transverse momentum ($p_{\rm{T}} < 1.5$~GeV/$c$) the $v_2(\pt)$ reported from STAR and ALICE exhibits qualitatively similar behavior. On the 
other hand, for $p_{\rm{T}} > 1.5$~GeV/$c$ for $\pi^{\pm}$ and $\mathrm{K^{\pm}}$ and for $p_{\rm{T}} > 2.5$~GeV/$c$ for p+$\overline{\mathrm{p}}$, the \vtwo~measurements at the LHC are significantly higher than 
those at the lower energies. Although this direct quantitative comparison might be subject to e.g.~different non-flow contributions, spectra, radial flow, the qualitative picture that emerges from the $p_{\mathrm{T}}$-differential $v_2$ appears similar at the LHC and RHIC.

\begin{figure}[tbh!f]
   \includegraphics[width=\textwidth]{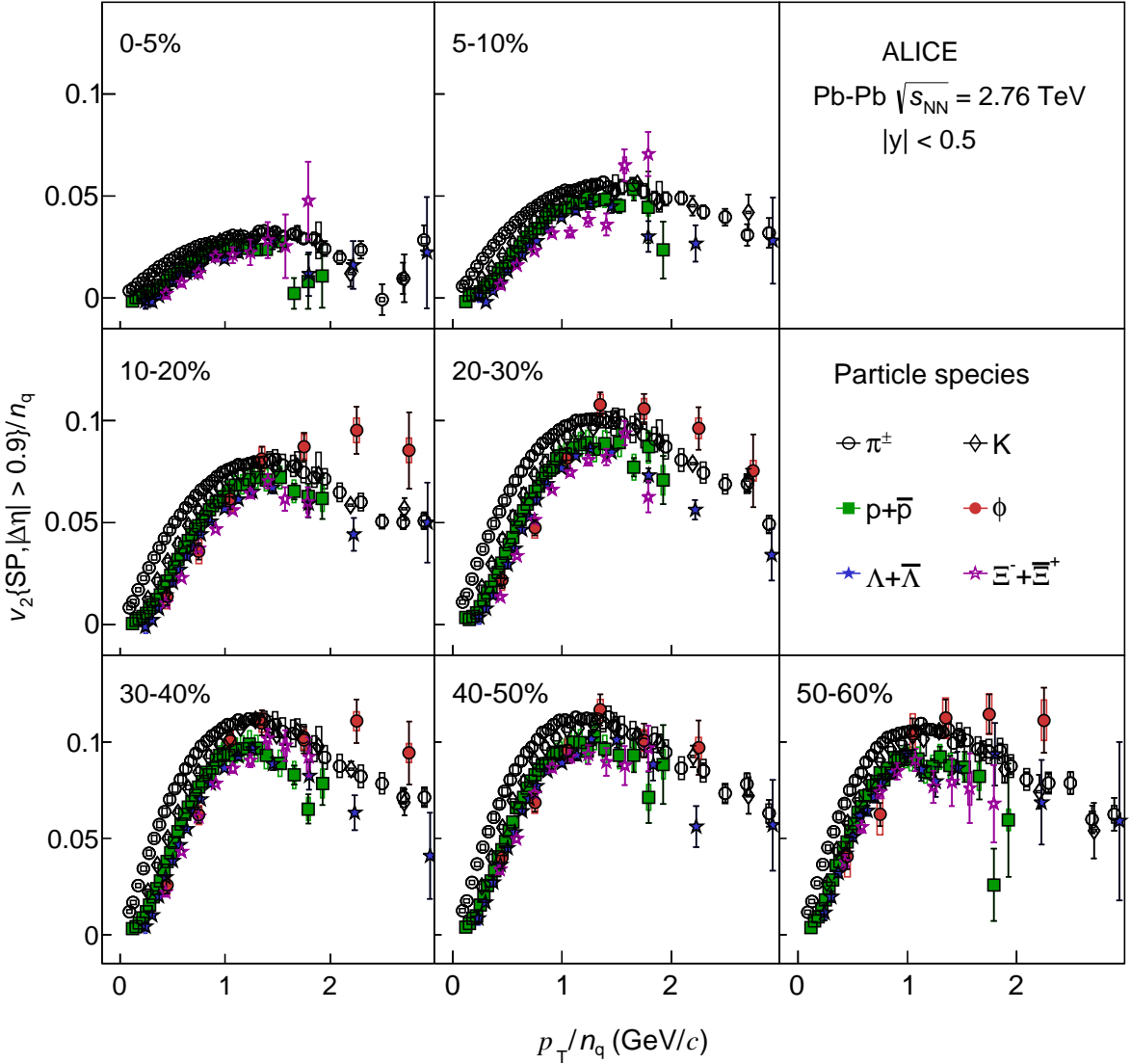}
  \caption{The $p_{\rm{T}}/n_q$ dependence of $v_2/n_q$ for $\pi^{\pm}$, $\mathrm{K}$, p+$\overline{\mathrm{p}}$, 
  $\phi$, $\Lambda$+$\overline{\mathrm{\Lambda}}$, and $\mathrm{\Xi^-}$+$\overline{\mathrm{\Xi}}^+$ for Pb--Pb collisions in various centrality intervals at $\sqrt{s_\mathrm{{NN}}} = 2.76$~TeV.}
\label{fig:pTScaling}
\end{figure}

\subsection{Test of scaling properties}
\label{SubSec:NCQ}

\begin{figure}[tbh!f]
  \includegraphics[width=\textwidth]{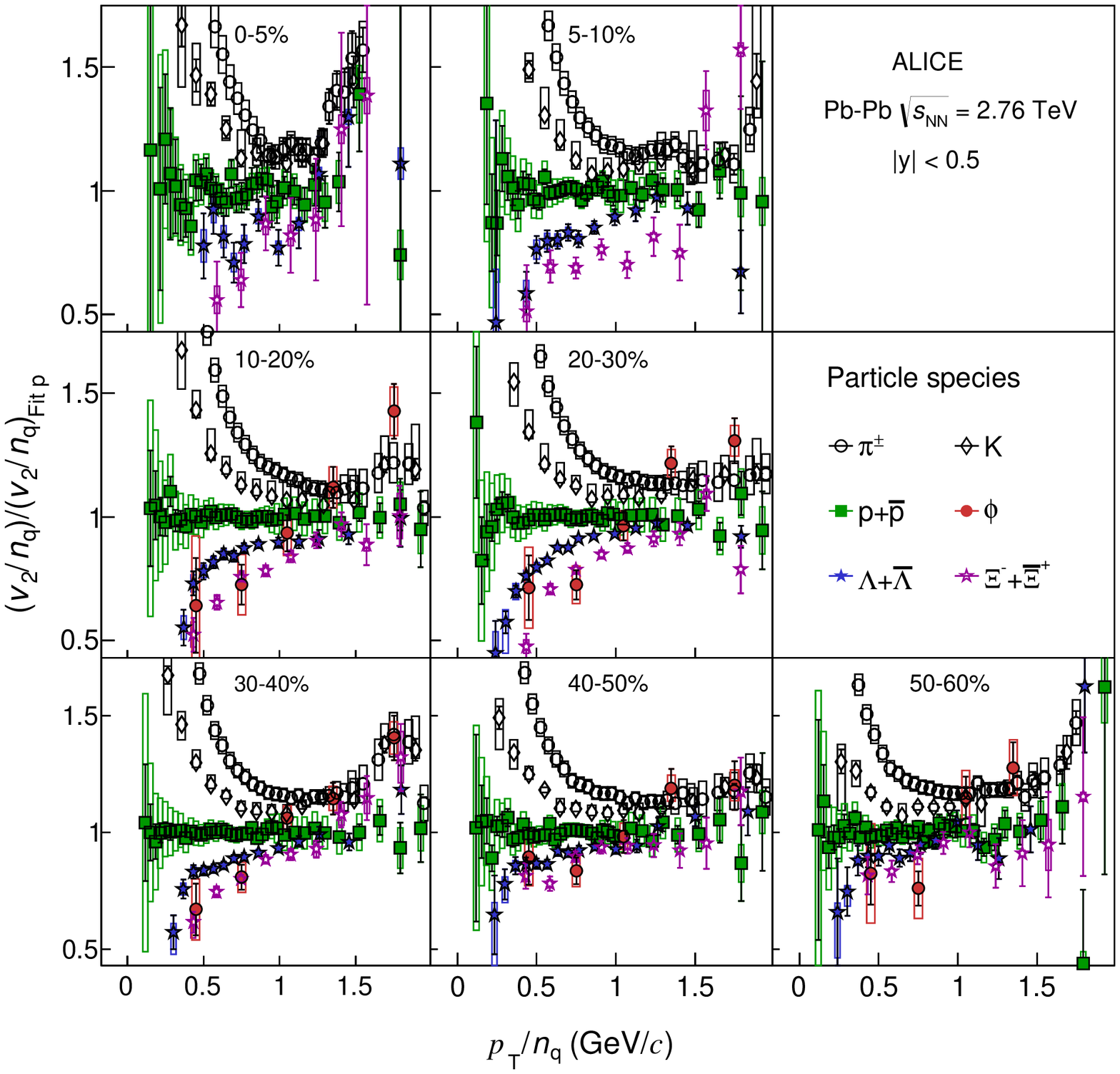}
  \caption{The $p_{\rm{T}}/n_q$ dependence of the double ratio of $v_2/n_q$ for every particle 
  species relative to a fit to $v_2/n_q$ of p and $\overline{\mathrm{p}}$ (see text for details) for Pb--Pb collisions at $\sqrt{s_\mathrm{{NN}}} = 2.76$~TeV. }
\label{fig:pTScaling2}
\end{figure}

\begin{figure}[tbh!f]
  \includegraphics[width=\textwidth]{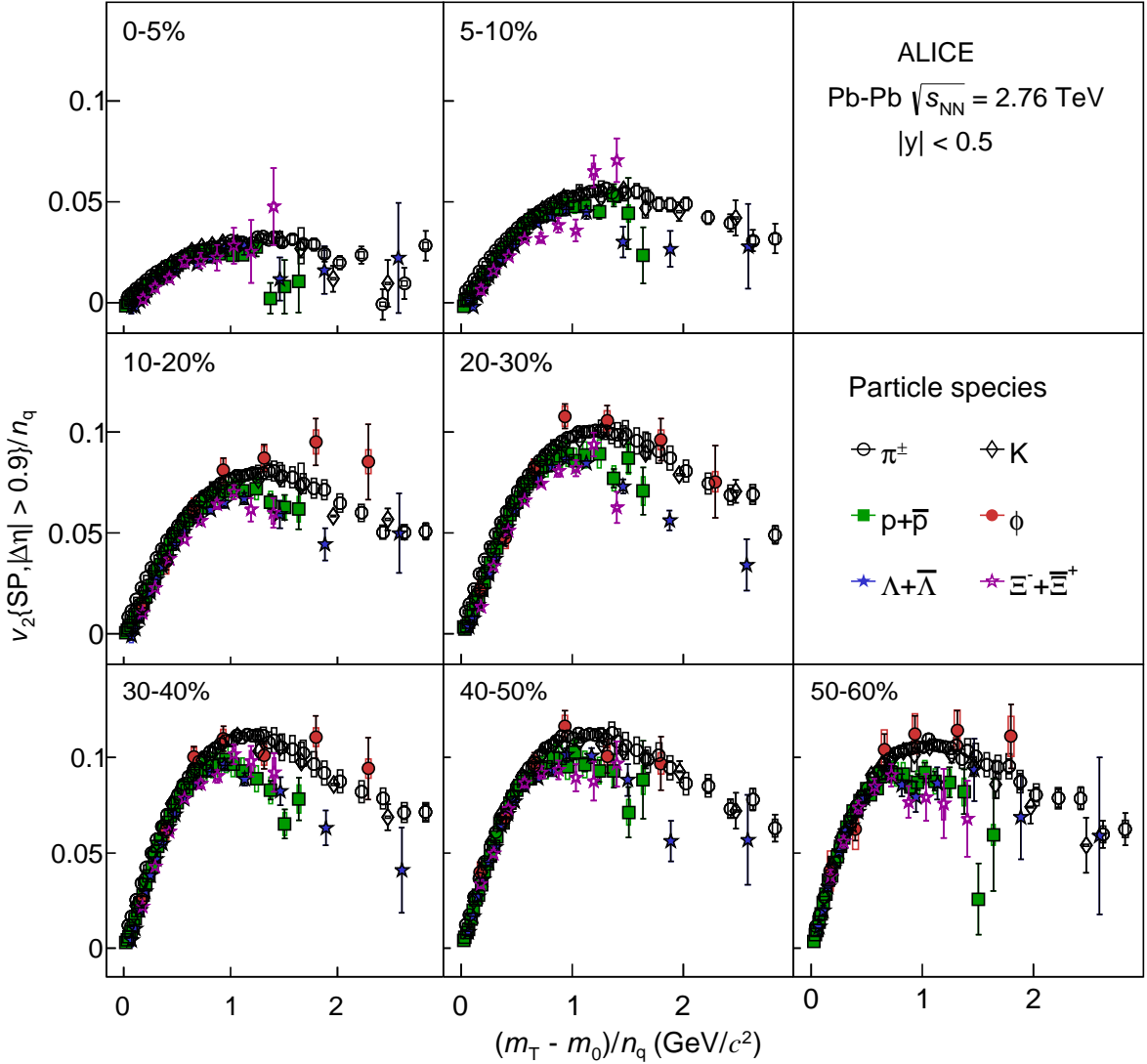}
  \caption{The $(m_{\rm{T}} - m_0)/n_q$ dependence of $v_2/n_q$ for
    $\pi^{\pm}$, $\mathrm{K}$, p+$\overline{\mathrm{p}}$, 
  $\phi$, $\Lambda$+$\overline{\mathrm{\Lambda}}$, and $\mathrm{\Xi^-}$+$\overline{\mathrm{\Xi}}^+$ for
     Pb--Pb collisions in various centrality intervals at $\sqrt{s_\mathrm{{NN}}} = 2.76$~TeV.}
\label{fig:mTScaling}
\end{figure}

\begin{figure}[tbh!f]
  \includegraphics[width=\textwidth]{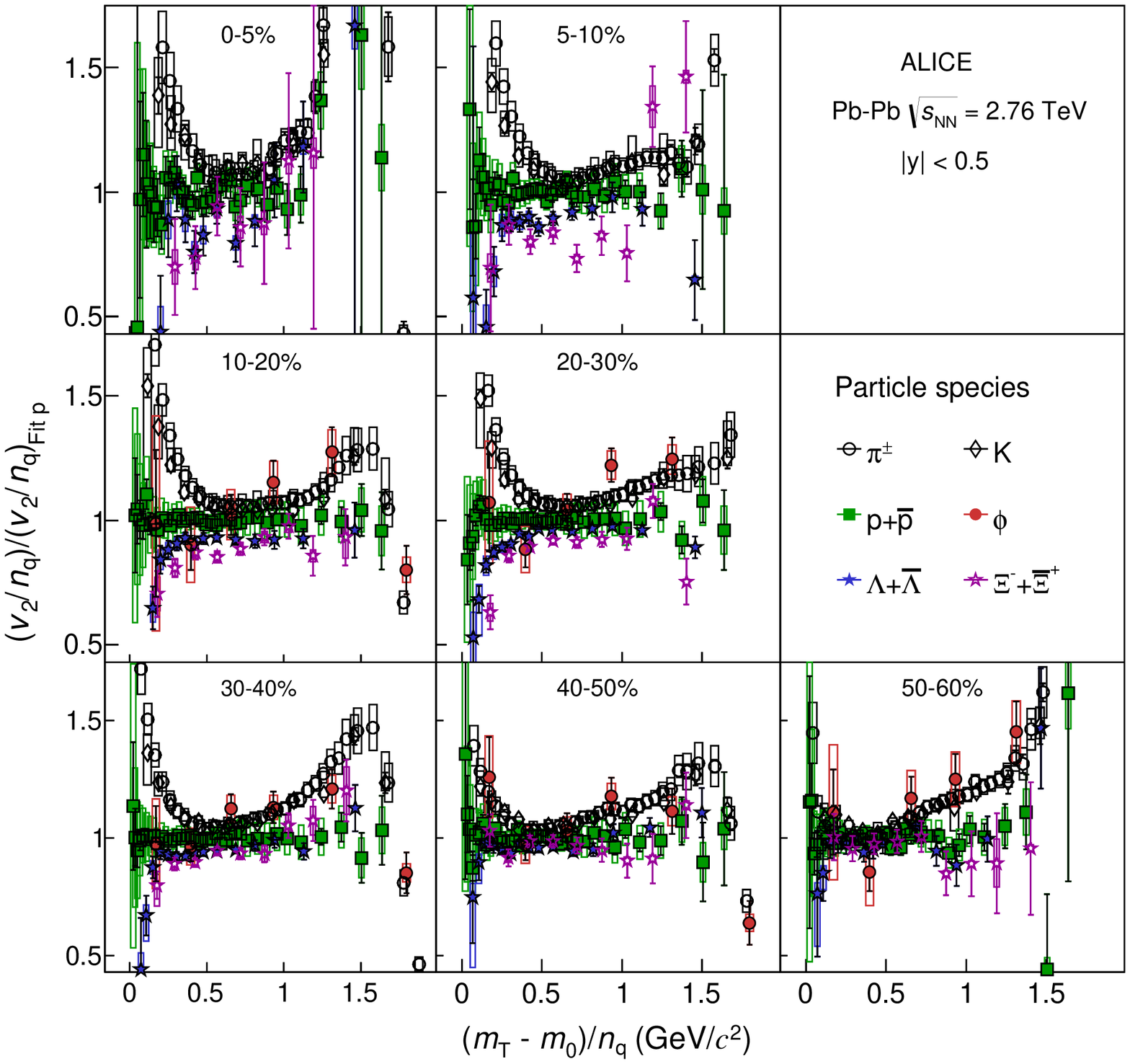}
  \caption{The $(m_{\rm{T}} - m_0)/n_q$ dependence of the double ratio of $v_2/n_q$ for every particle species
    relative to a fit to $v_2/n_q$ of p and $\overline{\mathrm{p}}$ (see text for
    details) for
    Pb--Pb collisions at $\sqrt{s_\mathrm{{NN}}} = 2.76$~TeV.}
\label{fig:mTScaling2}
\end{figure}

One of the experimental observations reported at RHIC was that at intermediate values of transverse 
momentum, particles tend to group based on their hadron type 
\cite{Adams:2003am,Abelev:2007qg,Adler:2003kt,Afanasiev:2007tv,Adare:2006ti} i.e. baryons and mesons. 
It was also reported that if both \vtwo~and $p_{\rm{T}}$ are scaled by the number of constituent quarks ($n_q$), the 
various identified hadron species approximately follow a common behaviour
\cite{Adams:2003am,Abelev:2007qg,Adler:2003kt,Afanasiev:2007tv,Adare:2006ti}. 
The PHENIX Collaboration suggested extending the scaling to the lower $\pt$ region by plotting elliptic 
flow as a function of the transverse kinetic energy defined as $KE_{\rm{T}} = m_{\rm{T}} - m_0$, where $m_{\mathrm{T}} = \sqrt{p_{\mathrm{T}}^2 + m_0^2}$ 
is the transverse mass \cite{Adler:2003kt,Afanasiev:2007tv,Adare:2006ti}.
Initially, this representation was observed to work well at RHIC energies. However, recent publications report deviations from this scaling for Au-Au collisions \cite{Adare:2012vq}. Such baryon versus 
meson grouping triggered significant theoretical debate over its origin. The effect was successfully reproduced by 
models invoking quark coalescence as the dominant hadronization mechanism in this momentum range
\cite{Molnar:2003ff,Greco:2003mm,Fries:2003kq,Hwa:2003ic}. Thus, the number of constituent quark (NCQ) scaling of
\vtwo~has been interpreted as evidence that quark degrees of freedom dominate in the early stages of heavy-ion 
collisions when collective flow develops \cite{Molnar:2003ff,Greco:2003mm,Fries:2003kq,Hwa:2003ic}.

To test the scaling properties of $v_2$, $v_2/n_q$ is plotted as a function of $p_{\rm{T}}/n_q$ in Fig.~\ref{fig:pTScaling} for $\pi^{\pm}$, $\mathrm{K}$, p+$\overline{\mathrm{p}}$, 
  $\phi$, $\Lambda$+$\overline{\mathrm{\Lambda}}$, and $\mathrm{\Xi^-}$+$\overline{\mathrm{\Xi}}^+$. In the intermediate transverse momentum region (i.e.~$3 < p_{\rm{T}} < 6$~GeV/$c$ or for $p_{\rm{T}}/n_q > 1$~GeV/$c$), where the coalescence mechanism is argued to be dominant \cite{Voloshin:2002wa,Adams:2003am,Abelev:2007qg,Adams:2004bi,Adler:2003kt,Afanasiev:2007tv,Adare:2006ti,Adare:2012vq,Molnar:2003ff,Greco:2003mm,Fries:2003kq,Hwa:2003ic}, the measurements at the LHC indicate that the scaling is only approximate. The magnitude of the observed deviations seems to be similar for all centrality intervals.

To quantify the deviation, the $p_{\rm{T}}/n_q$ dependence of $v_2/n_q$ for p and $\overline{\mathrm{p}}$ is fitted with a seventh order polynomial function and the ratio of $(v_2/n_q)/(v_2/n_q)_{\mathrm{Fit p}}$ for each particle species is calculated. The corresponding $p_{\rm{T}}/n_q$ dependence of this double ratio is presented in Fig.~\ref{fig:pTScaling2} for the various centrality intervals. Figure~\ref{fig:pTScaling2} illustrates that for $p_{\rm{T}}/n_q > 1$~GeV/$c$ the data points exhibit deviations from an exact scaling at the level of $\pm$20$\%$ with respect to the reference ratio for all centrality intervals.

Figure~\ref{fig:mTScaling} presents the $(m_{\rm{T}} - m_0)/n_q$ dependence of $v_2/n_q$. In this representation, introduced to extend the scaling to low values of transverse momentum, the data points illustrate significant deviations for $(m_{\rm{T}} - m_0)/n_q < 0.6-0.8$~GeV/$c^2$. For the intermediate region the scaling, if any, is approximate for all centrality intervals. To quantify these deviations, in Fig.~\ref{fig:mTScaling2} the $(m_{\rm{T}} - m_0)/n_q$ dependence of $v_2/n_q$ for p and $\overline{\mathrm{p}}$ are fitted with a seventh order polynomial function and the double ratio of $(v_2/n_q)/(v_2/n_q)_{\mathrm{Fit p}}$ for each particle species is then formed. It is seen that there is no scaling for $(m_{\rm{T}} - m_0)/n_q < 0.6-0.8$~GeV/$c^2$, while for higher values there are deviations at the level of $\pm$20$\%$ with respect to the reference ratio for all centrality intervals. 

\begin{figure}[tbh!f]
  \includegraphics[width=\textwidth]{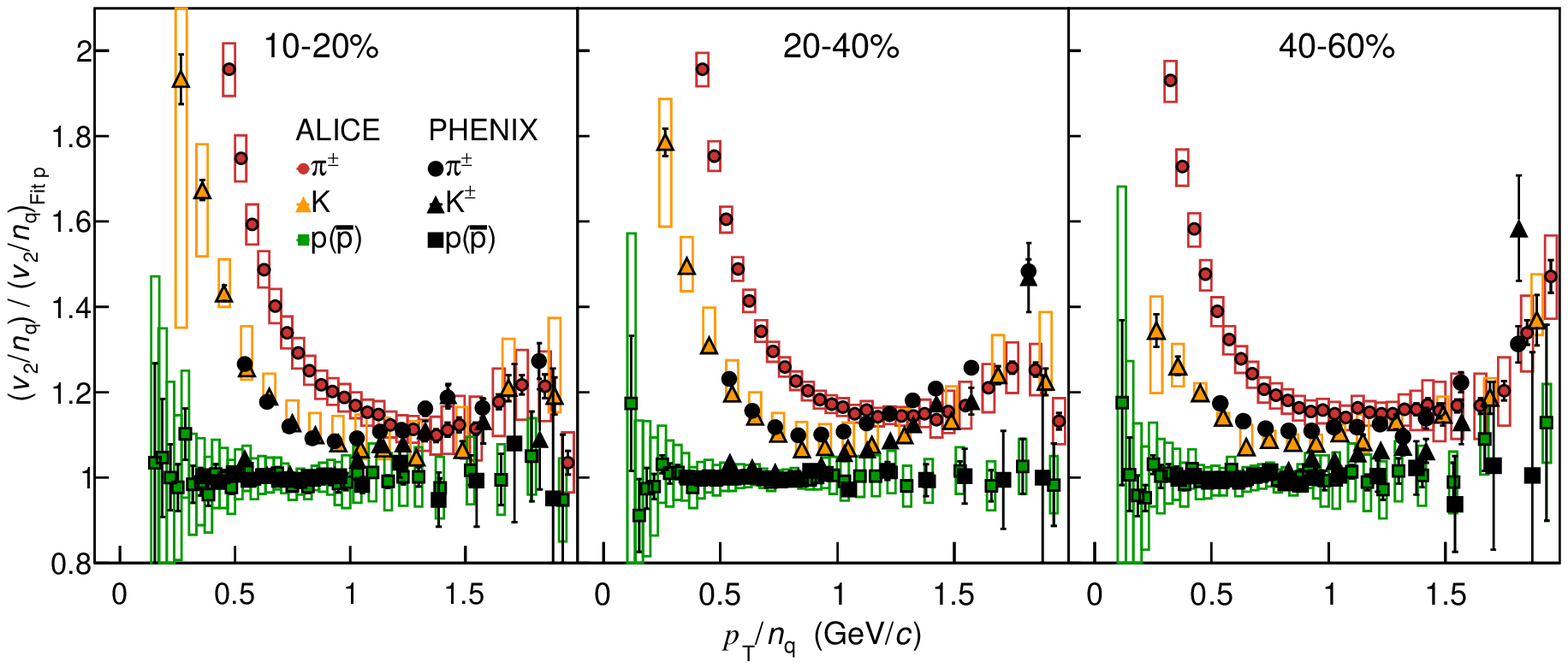}
    \caption{The $p_{\rm{T}}/n_q$ dependence of the double ratio of $v_2/n_q$ for $\pi^{\pm}$, $\mathrm{K}$ relative to a fit to $v_2/n_q$ of p and $\overline{\mathrm{p}}$ (see text for details) in Pb--Pb collisions at $\sqrt{s_\mathrm{{NN}}} = 2.76$~TeV. The LHC points are compared with the results from Au--Au collisions at $\sqrt{s_\mathrm{{NN}}} = 0.2$~TeV from \cite{Adare:2012vq}.}
\label{fig:pTScalingComparisonToRhic}
\end{figure}

Figure~\ref{fig:pTScalingComparisonToRhic} presents the comparison of the $p_{\rm{T}}/n_q$ dependence of the double ratio of $v_2/n_q$ for $\pi^{\pm}$, $\mathrm{K}$ relative to a fit to $v_2/n_q$ of p and $\overline{\mathrm{p}}$ for both the LHC and RHIC energies. The RHIC data points are extracted from \cite{Adare:2012vq}. It is seen that the deviations at intermediate values of transverse momentum are qualitatively similar at the two energy regimes. However, there are differences in the $p_{\rm{T}}/n_q$ evolution of this double ratio for $\pi^{\pm}$ and $\mathrm{K}$ between ALICE and PHENIX.

\begin{figure}[tbh!f]
  \includegraphics[width=\textwidth]{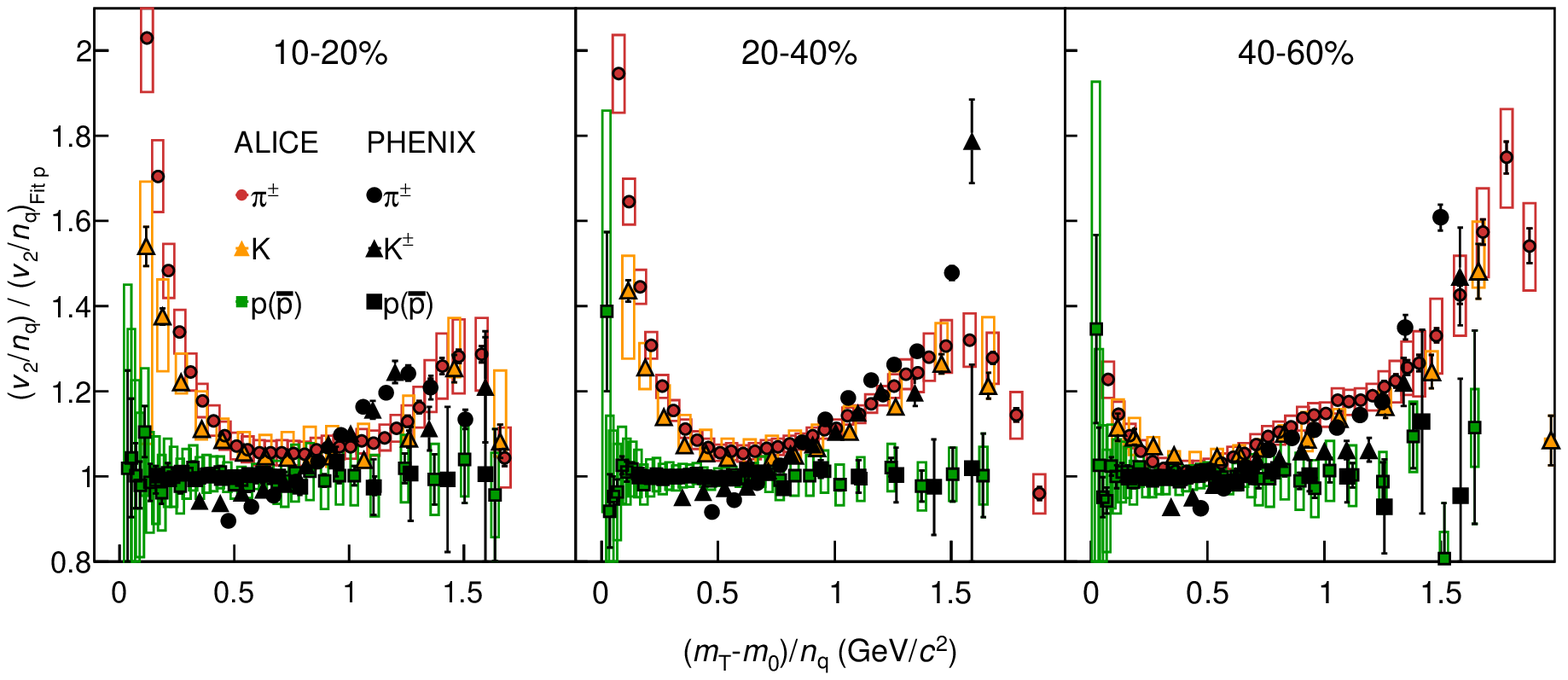}
  \caption{The $(m_{\rm{T}} - m_0)/n_q$ dependence of the double ratio of $v_2/n_q$ for $\pi^{\pm}$, $\mathrm{K}$ relative to a fit to $v_2/n_q$ of p and $\overline{\mathrm{p}}$ (see text for details) in Pb--Pb collisions at $\sqrt{s_\mathrm{{NN}}} = 2.76$~TeV. The LHC points are compared with the results from Au--Au collisions at $\sqrt{s_\mathrm{{NN}}} = 0.2$~TeV from \cite{Adare:2012vq}.}
\label{fig:mTScalingComparisonToRhic}
\end{figure}

Figure~\ref{fig:mTScalingComparisonToRhic} presents the comparison of the $(m_{\rm{T}} - m_0)/n_q$ dependence of the double ratio of $v_2/n_q$ for $\pi^{\pm}$, $\mathrm{K}$ relative to a fit to $v_2/n_q$ of p and $\overline{\mathrm{p}}$ between ALICE and PHENIX~\cite{Adare:2012vq}. As in Fig.~\ref{fig:pTScalingComparisonToRhic}, the deviations are qualitatively similar at the two energy regimes but the $(m_{\rm{T}} - m_0)/n_q$ evolution of the double ratio is different for $\pi^{\pm}$ and $\mathrm{K}$ at the LHC and RHIC.

%% file: section/conclusion.tex
\section{Conclusions}
\label{Sec:Conclusions}

In summary, the first measurements of \vtwo~as a function of transverse momentum for 
$\pi^{\pm}$, $\mathrm{K}^{\pm}$, $\mathrm{K}^0_\mathrm{S}$, p+$\overline{\mathrm{p}}$, $\phi$, $\Lambda$+$\overline{\mathrm{\Lambda}}$,
$\Xi^-$+$\overline{\Xi}^+$ and $\Omega^-$+$\overline{\Omega}^+$ for various centralities of Pb--Pb collisions at 
$\sqrt{s_{\mathrm{NN}}} = 2.76$~TeV were reported. The second Fourier coefficient was calculated with the Scalar Product method, 
using a pseudo-rapidity gap of $|\Delta\eta| > 0.9$ between the identified hadron under study and each of the reference flow particles. 
A distinct mass 
ordering was found for all centralities in the low transverse momentum region i.e.~for $p_{\rm{T}} < 3$~GeV/$c$, 
which is attributed to the interplay between elliptic and radial flow that modifies the $v_2(p_{\rm{T}})$ according to particle mass. The $v_2(p_{\rm{T}})$ for heavy particles appears to be shifted to higher $p_{\rm{T}}$ with respect to the $v_2(p_{\rm{T}})$ values of light particles. In this 
transverse momentum range, the experimental points for $\pi^{\pm}$ and $\mathrm{K}$ are described fairly well for peripheral collisions by hydrodynamic calculations coupled to a hadronic cascade model (VISHNU) indicating that a small value of $\eta/s$ (close to the lower bound) is favoured. However, 
for central collisions and for heavy particles, the same theoretical calculations tend to overestimate (i.e.~$\Lambda$, $\Xi$) or underestimate (i.e.~p) the measured \vtwo. VISHNU fails to describe the measured $v_2$ of $\phi$, which could be an indication that this particle has a larger hadronic cross section than its current theoretical estimate.
In the intermediate transverse momentum region (i.e.~$3 < p_{\rm{T}} < 6$~GeV/$c$), where at RHIC 
there was evidence that coalescence is the dominant hadronization
mechanism, our data exhibit deviations from the number of constituent quark (NCQ) scaling at the level of $\pm$20$\%$.

%% file: section/acknowledgements.tex
The ALICE Collaboration would like to thank all its engineers and technicians for their invaluable contributions to the construction of the experiment and the CERN accelerator teams for the outstanding performance of the LHC complex.
The ALICE Collaboration gratefully acknowledges the resources and support provided by all Grid centres and the Worldwide LHC Computing Grid (WLCG) collaboration.
The ALICE Collaboration acknowledges the following funding agencies for their support in building and
running the ALICE detector:
State Committee of Science,  World Federation of Scientists (WFS)
and Swiss Fonds Kidagan, Armenia,
Conselho Nacional de Desenvolvimento Cient\'{\i}fico e Tecnol\'{o}gico (CNPq), Financiadora de Estudos e Projetos (FINEP),
Funda\c{c}\~{a}o de Amparo \`{a} Pesquisa do Estado de S\~{a}o Paulo (FAPESP);
National Natural Science Foundation of China (NSFC), the Chinese Ministry of Education (CMOE)
and the Ministry of Science and Technology of China (MSTC);
Ministry of Education and Youth of the Czech Republic;
Danish Natural Science Research Council, the Carlsberg Foundation and the Danish National Research Foundation;
The European Research Council under the European Community's Seventh Framework Programme;
Helsinki Institute of Physics and the Academy of Finland;
French CNRS-IN2P3, the `Region Pays de Loire', `Region Alsace', `Region Auvergne' and CEA, France;
German BMBF and the Helmholtz Association;
General Secretariat for Research and Technology, Ministry of
Development, Greece;
Hungarian OTKA and National Office for Research and Technology (NKTH);
Department of Atomic Energy and Department of Science and Technology of the Government of India;
Istituto Nazionale di Fisica Nucleare (INFN) and Centro Fermi -
Museo Storico della Fisica e Centro Studi e Ricerche "Enrico
Fermi", Italy;
MEXT Grant-in-Aid for Specially Promoted Research, Ja\-pan;
Joint Institute for Nuclear Research, Dubna;
National Research Foundation of Korea (NRF);
CONACYT, DGAPA, M\'{e}xico, ALFA-EC and the EPLANET Program
(European Particle Physics Latin American Network)
Stichting voor Fundamenteel Onderzoek der Materie (FOM) and the Nederlandse Organisatie voor Wetenschappelijk Onderzoek (NWO), Netherlands;
Research Council of Norway (NFR);
Polish Ministry of Science and Higher Education;
National Science Centre, Poland;
 Ministry of National Education/Institute for Atomic Physics and CNCS-UEFISCDI - Romania;
Ministry of Education and Science of Russian Federation, Russian
Academy of Sciences, Russian Federal Agency of Atomic Energy,
Russian Federal Agency for Science and Innovations and The Russian
Foundation for Basic Research;
Ministry of Education of Slovakia;
Department of Science and Technology, South Africa;
CIEMAT, EELA, Ministerio de Econom\'{i}a y Competitividad (MINECO) of Spain, Xunta de Galicia (Conseller\'{\i}a de Educaci\'{o}n),
CEA\-DEN, Cubaenerg\'{\i}a, Cuba, and IAEA (International Atomic Energy Agency);
Swedish Research Council (VR) and Knut $\&$ Alice Wallenberg
Foundation (KAW);
Ukraine Ministry of Education and Science;
United Kingdom Science and Technology Facilities Council (STFC);
The United States Department of Energy, the United States National
Science Foundation, the State of Texas, and the State of Ohio.

%% file: section/references.tex

%% file: section/appendixB.tex
\section{Additional figures}
\label{Sec:AppendixB}

\subsection{Plots from Fig.~\ref{fig:pTDifferentialv2AllSpecies}}
\begin{center}
\includegraphics[width=\textwidth]{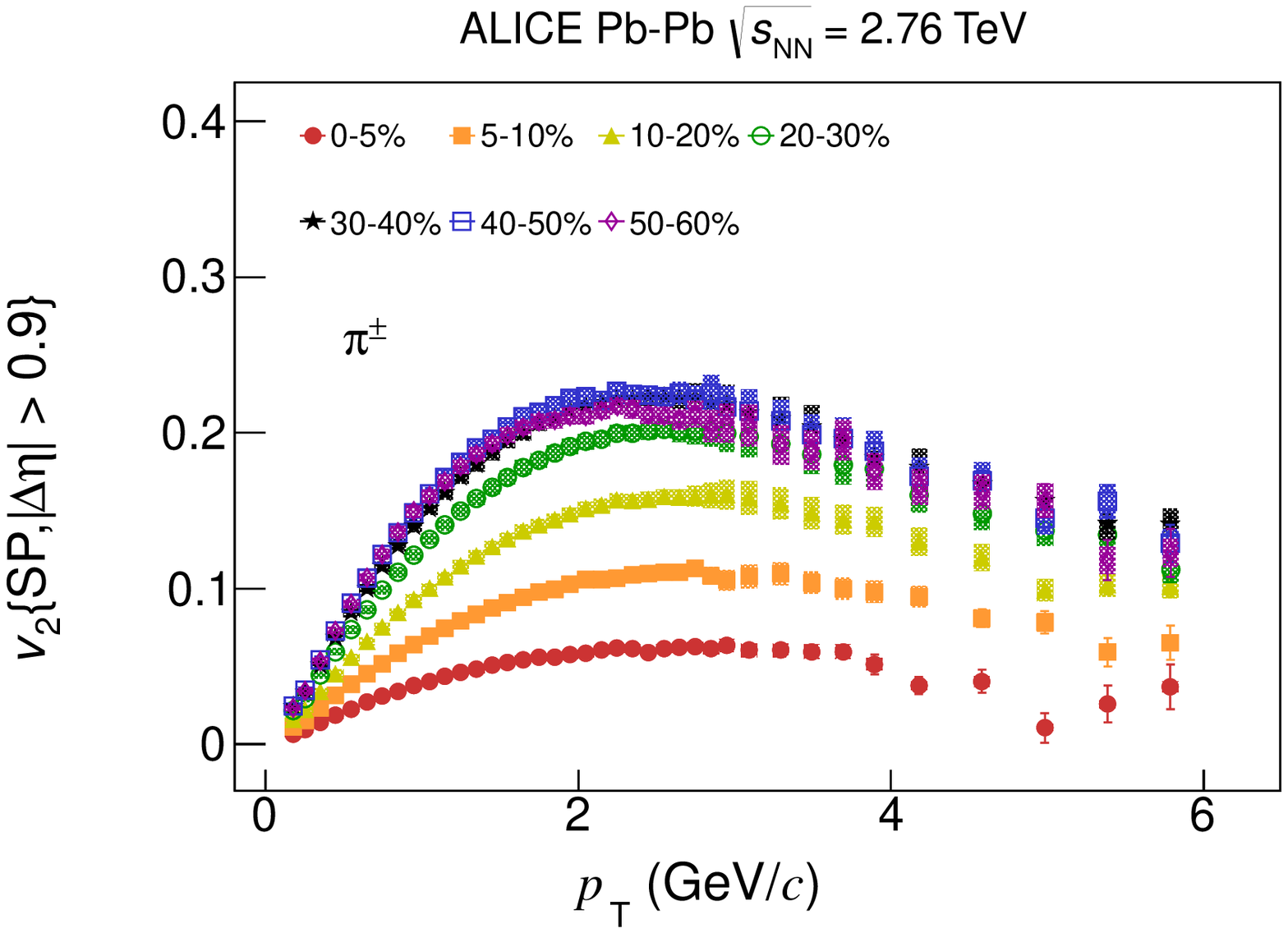}
\captionof{figure}{The $p_{\rm{T}}$-differential \vtwo~for different centralities of Pb--Pb collisions at $\sqrt{s_{\mathrm{NN}}} = 2.76$~TeV, 
represented by the different symbols and colors for $\pi^{\pm}$.}
\label{fig:pTDifferentialv2Pion} 
\end{center}

\begin{center}
\includegraphics[width=\textwidth]{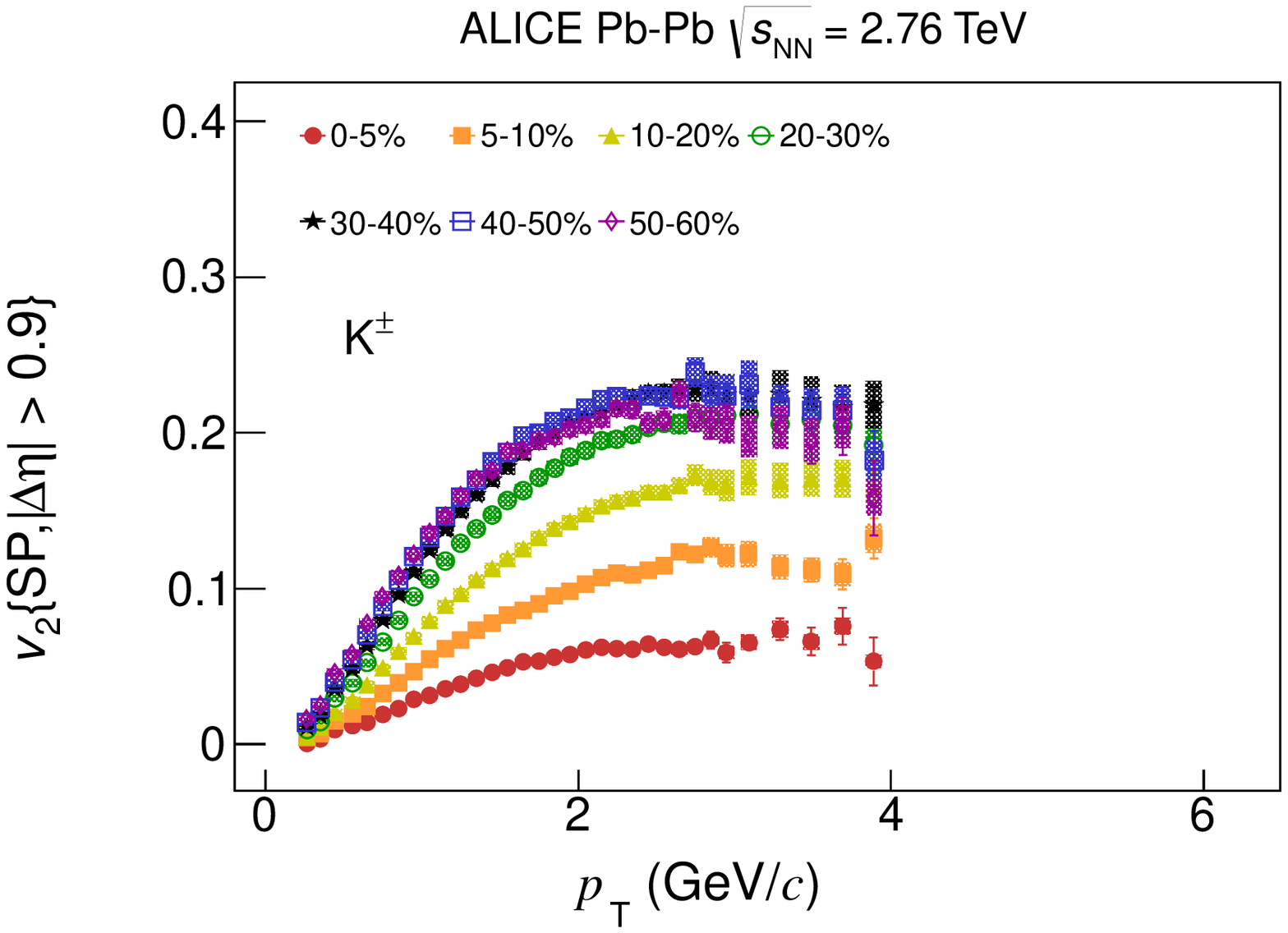}
\captionof{figure}{The $p_{\rm{T}}$-differential \vtwo~for different centralities of Pb--Pb collisions at $\sqrt{s_{\mathrm{NN}}} = 2.76$~TeV, 
represented by the different symbols and colors for $\mathrm{K}^{\pm}$.}
\label{fig:pTDifferentialv2Kaon} 
\end{center}

\begin{center}
\includegraphics[width=\textwidth]{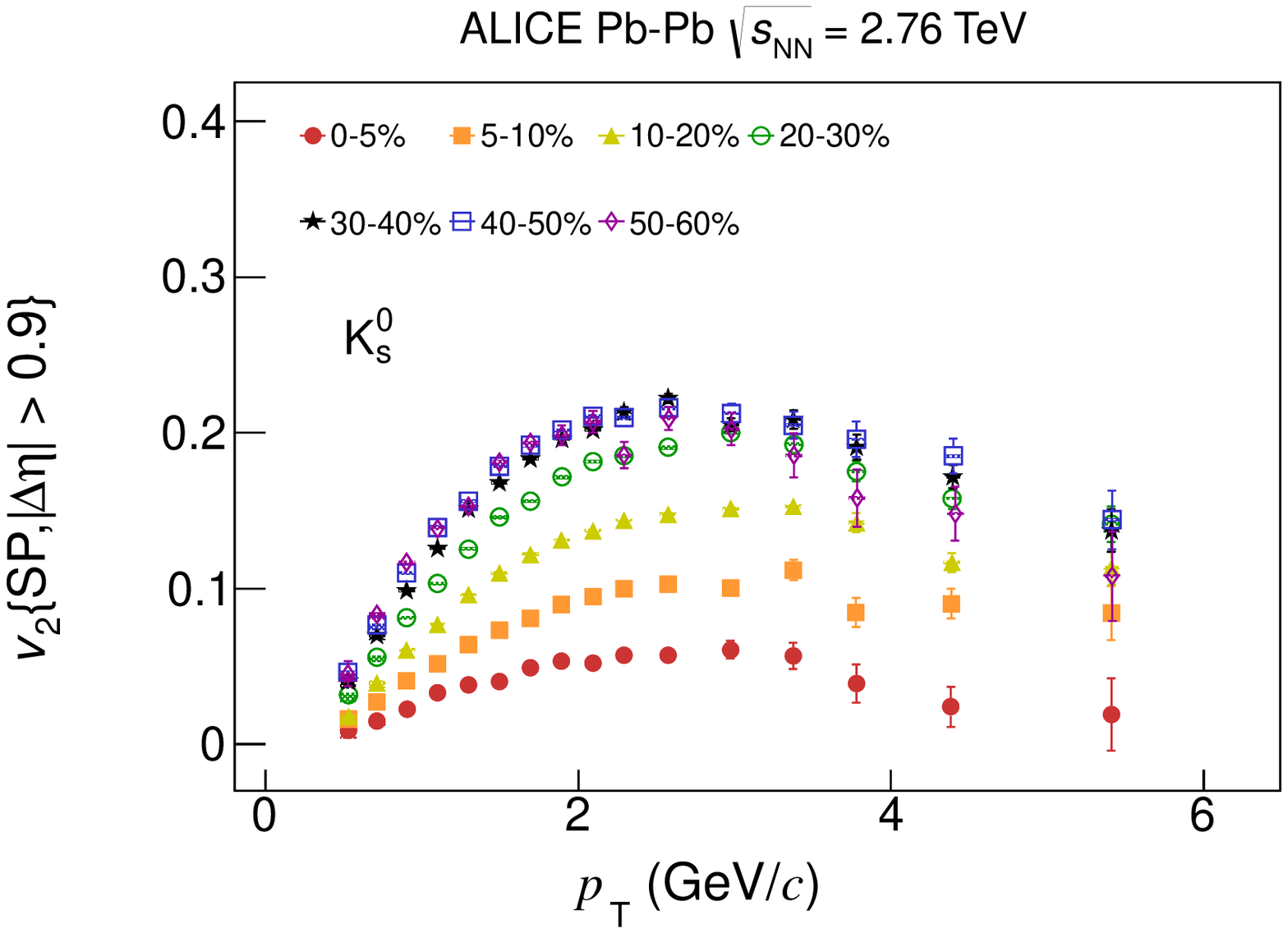}
\captionof{figure}{The $p_{\rm{T}}$-differential \vtwo~for different centralities of Pb--Pb collisions at $\sqrt{s_{\mathrm{NN}}} = 2.76$~TeV, 
represented by the different symbols and colors for $\mathrm{K}^{0}_{s}$.}
\label{fig:pTDifferentialv2Kzero} 
\end{center}

\begin{center}
\includegraphics[width=\textwidth]{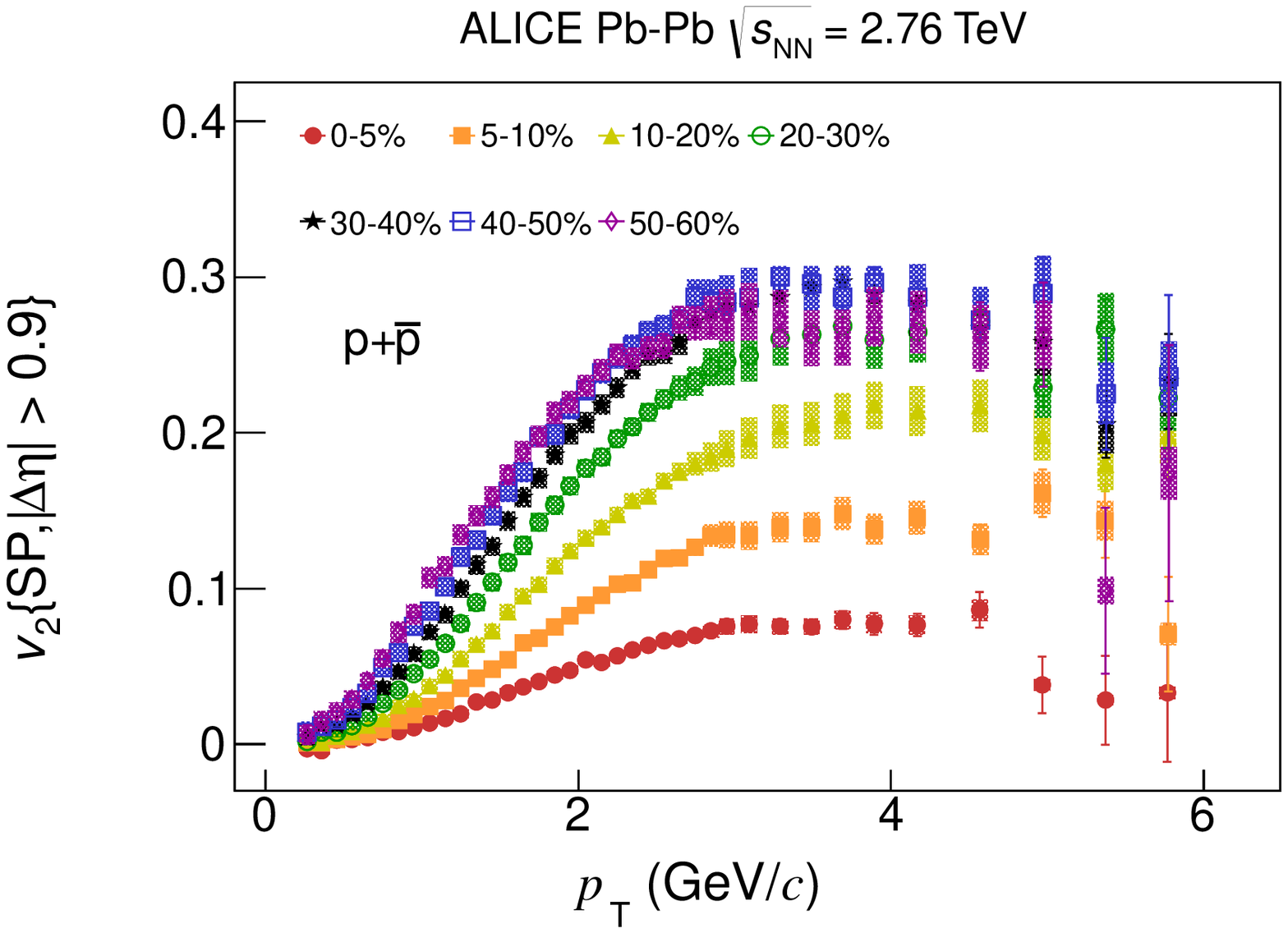}
\captionof{figure}{The $p_{\rm{T}}$-differential \vtwo~for different centralities of Pb--Pb collisions at $\sqrt{s_{\mathrm{NN}}} = 2.76$~TeV, represented by the different symbols and colors for p+$\overline{\mathrm{p}}$.}
\label{fig:pTDifferentialv2Proton} 
\end{center}

\begin{center}
\includegraphics[width=\textwidth]{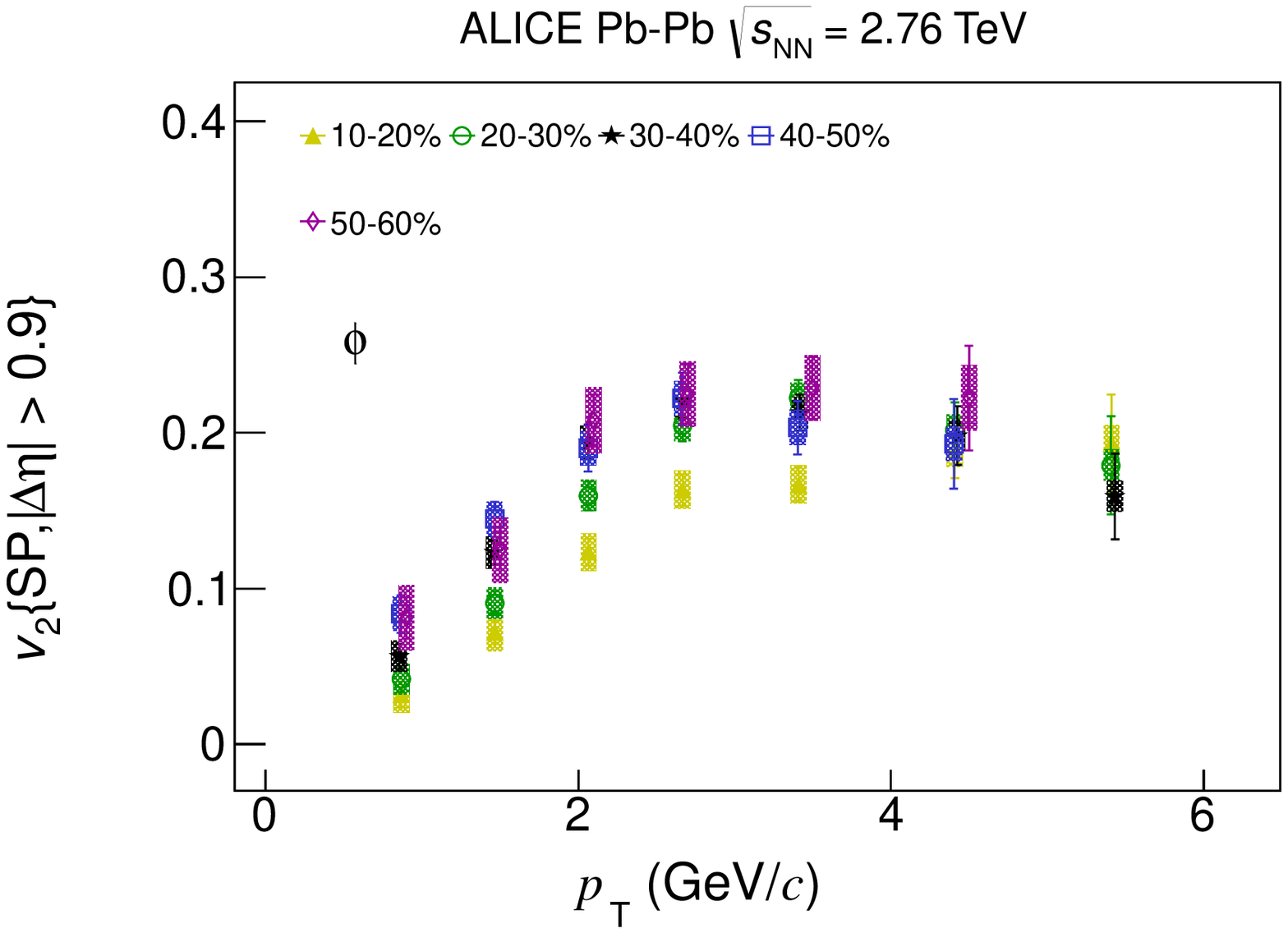}
\captionof{figure}{The $p_{\rm{T}}$-differential \vtwo~for different centralities of Pb--Pb collisions at $\sqrt{s_{\mathrm{NN}}} = 2.76$~TeV, 
represented by the different symbols and colors for $\phi$.}
\label{fig:pTDifferentialv2Phi} 
\end{center}

\begin{center}
\includegraphics[width=\textwidth]{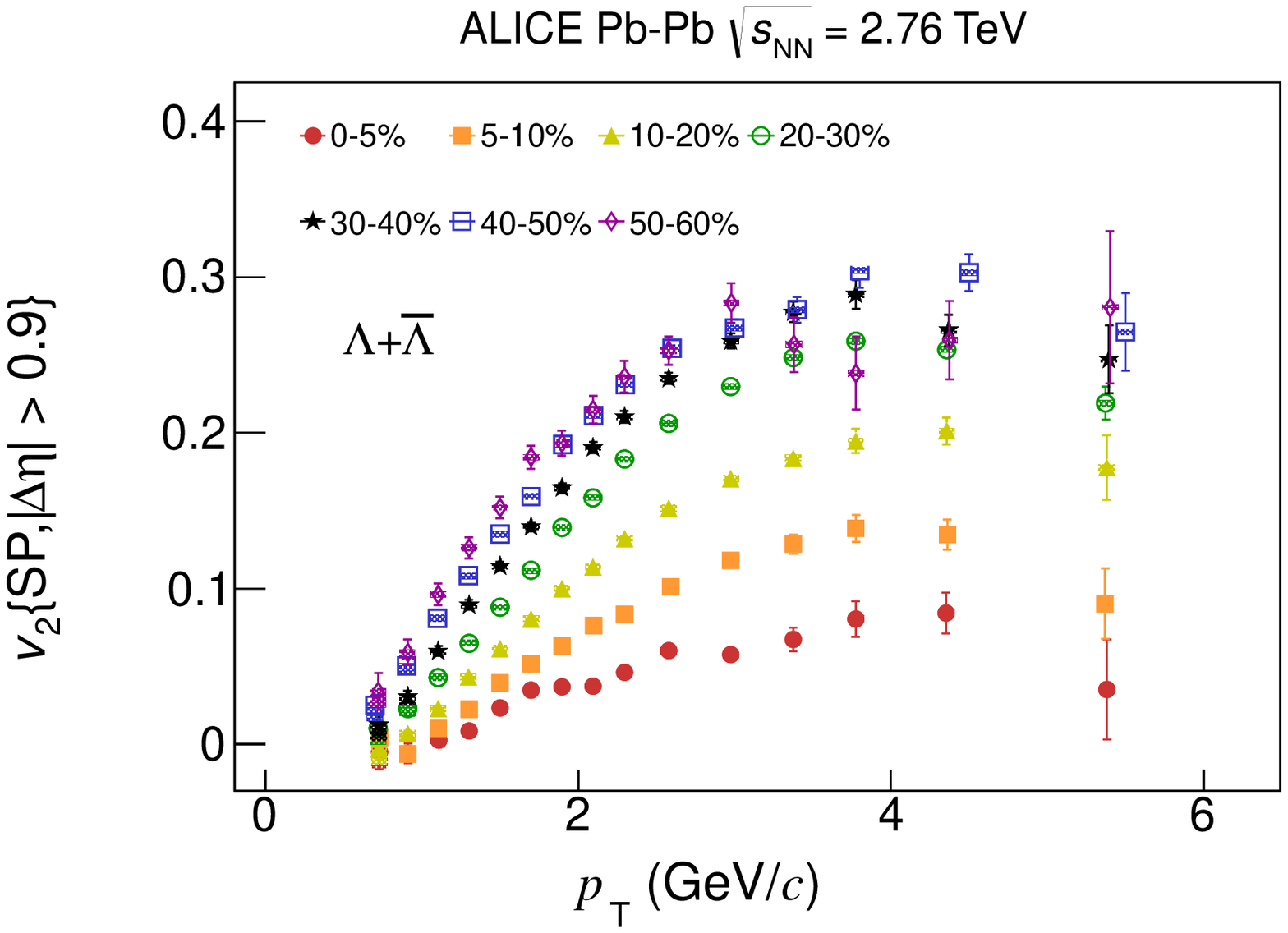}
\captionof{figure}{The $p_{\rm{T}}$-differential \vtwo~for different centralities of Pb--Pb collisions at $\sqrt{s_{\mathrm{NN}}} = 2.76$~TeV, 
represented by the different symbols and colors for $\mathrm{\Lambda}$+$\overline{\mathrm{\Lambda}}$.}
\label{fig:pTDifferentialv2Lambda} 
\end{center}

\begin{center}
\includegraphics[width=\textwidth]{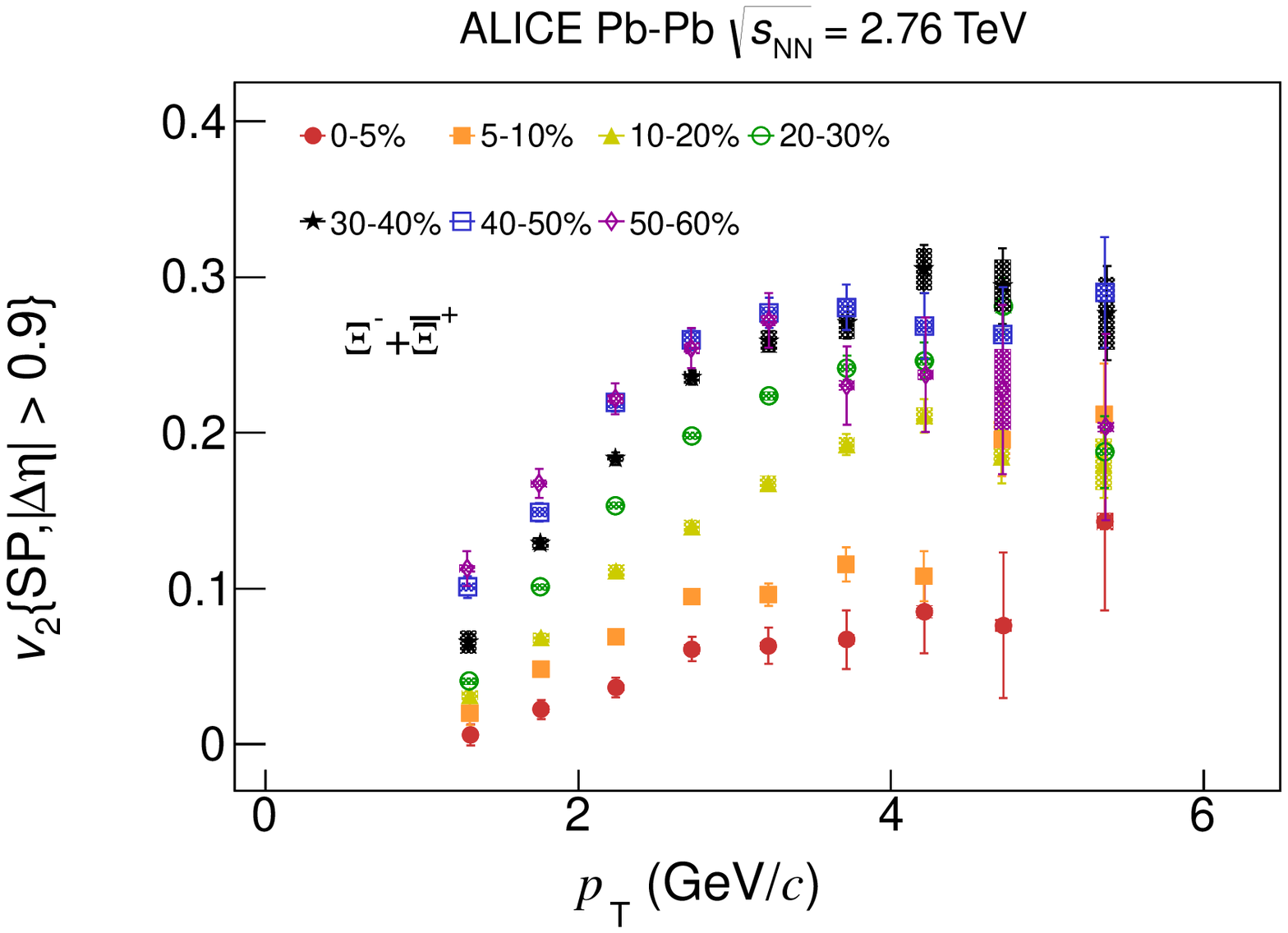}
\captionof{figure}{The $p_{\rm{T}}$-differential \vtwo~for different centralities of Pb--Pb collisions at $\sqrt{s_{\mathrm{NN}}} = 2.76$~TeV, 
represented by the different symbols and colors for $\mathrm{\Xi^-}$+$\overline{\mathrm{\Xi}}^+$.}
\label{fig:pTDifferentialv2Xi} 
\end{center}

\begin{center}
\includegraphics[width=\textwidth]{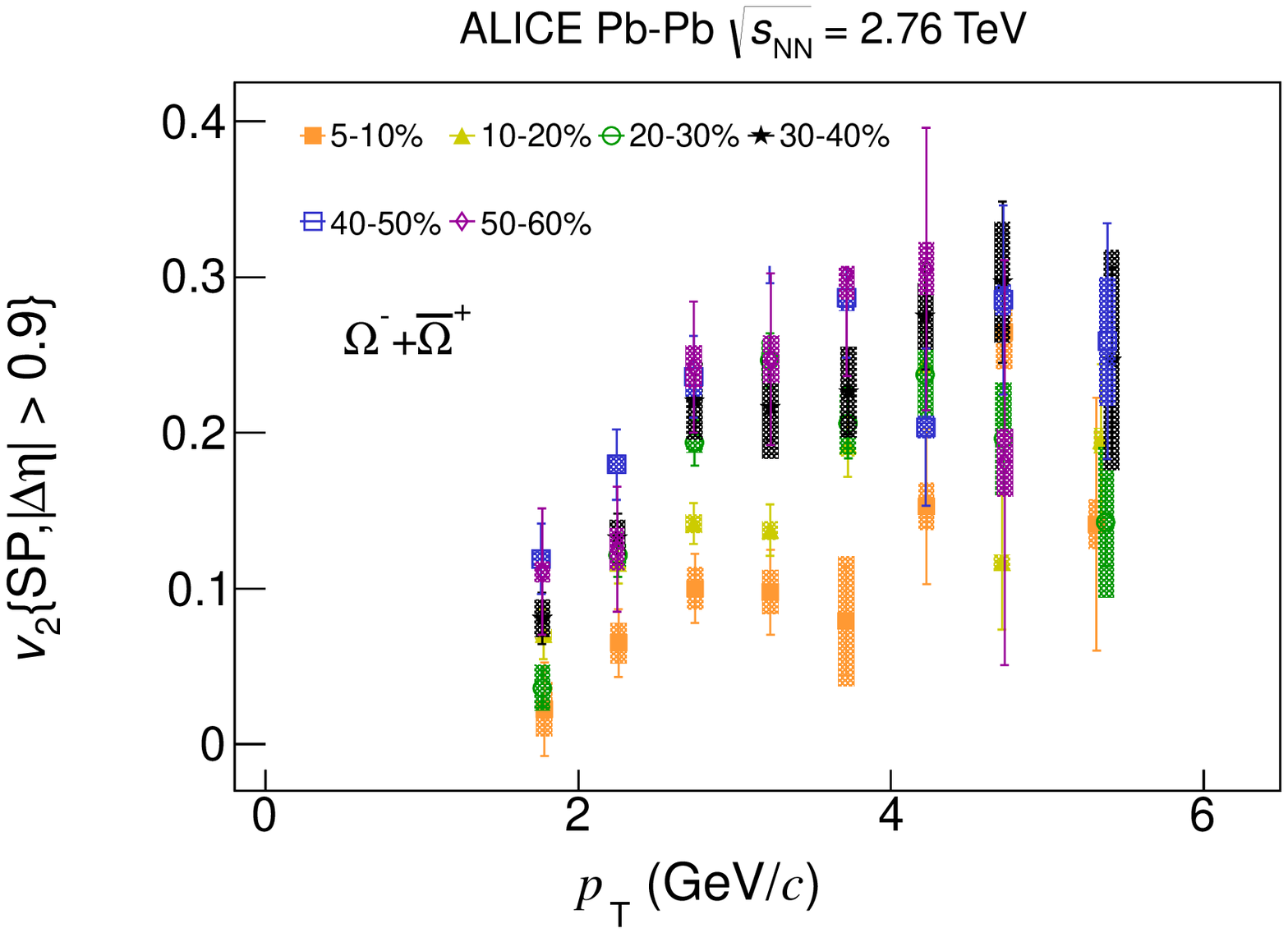}
\captionof{figure}{The $p_{\rm{T}}$-differential \vtwo~for different centralities of Pb--Pb collisions at $\sqrt{s_{\mathrm{NN}}} = 2.76$~TeV, 
represented by the different symbols and colors for $\mathrm{\Omega}^-$+$\overline{\mathrm{\Omega}}^+$.}
\label{fig:pTDifferentialv2Omega} 
\end{center}

\subsection{Plots from Fig.~\ref{fig:pTDifferentialv2AllSpecies2}}
\begin{center}
\includegraphics[width=\textwidth]{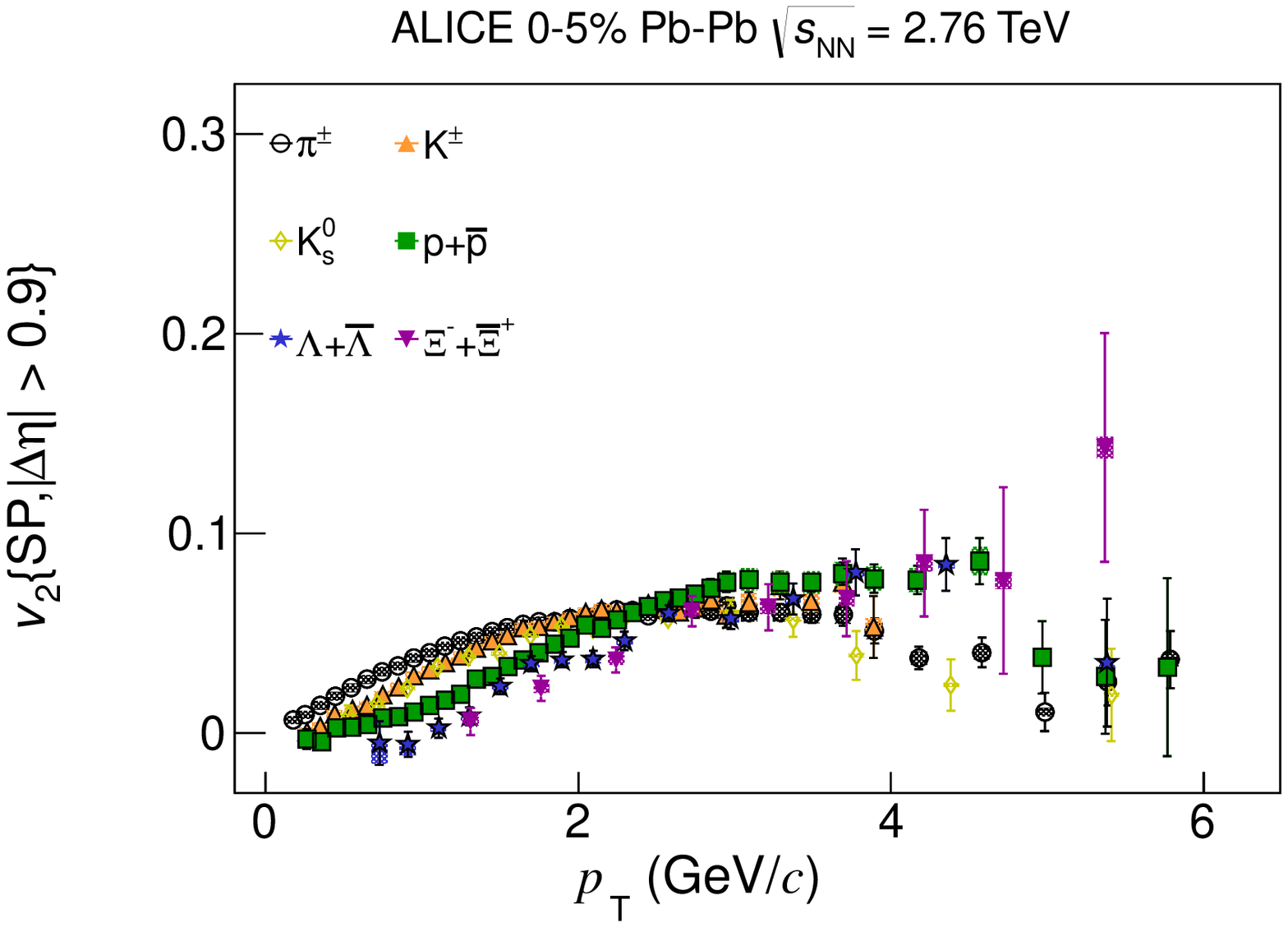}
\captionof{figure}{The $p_{\rm{T}}$-differential \vtwo~for different particle species, represented by the different symbols 
and colors, for the 0--5$\%$ centrality interval of Pb--Pb collisions at $\sqrt{s_{\mathrm{NN}}} = 2.76$~TeV.}
\label{fig:v2Centrality0To5}
\end{center}

\begin{center}
\includegraphics[width=\textwidth]{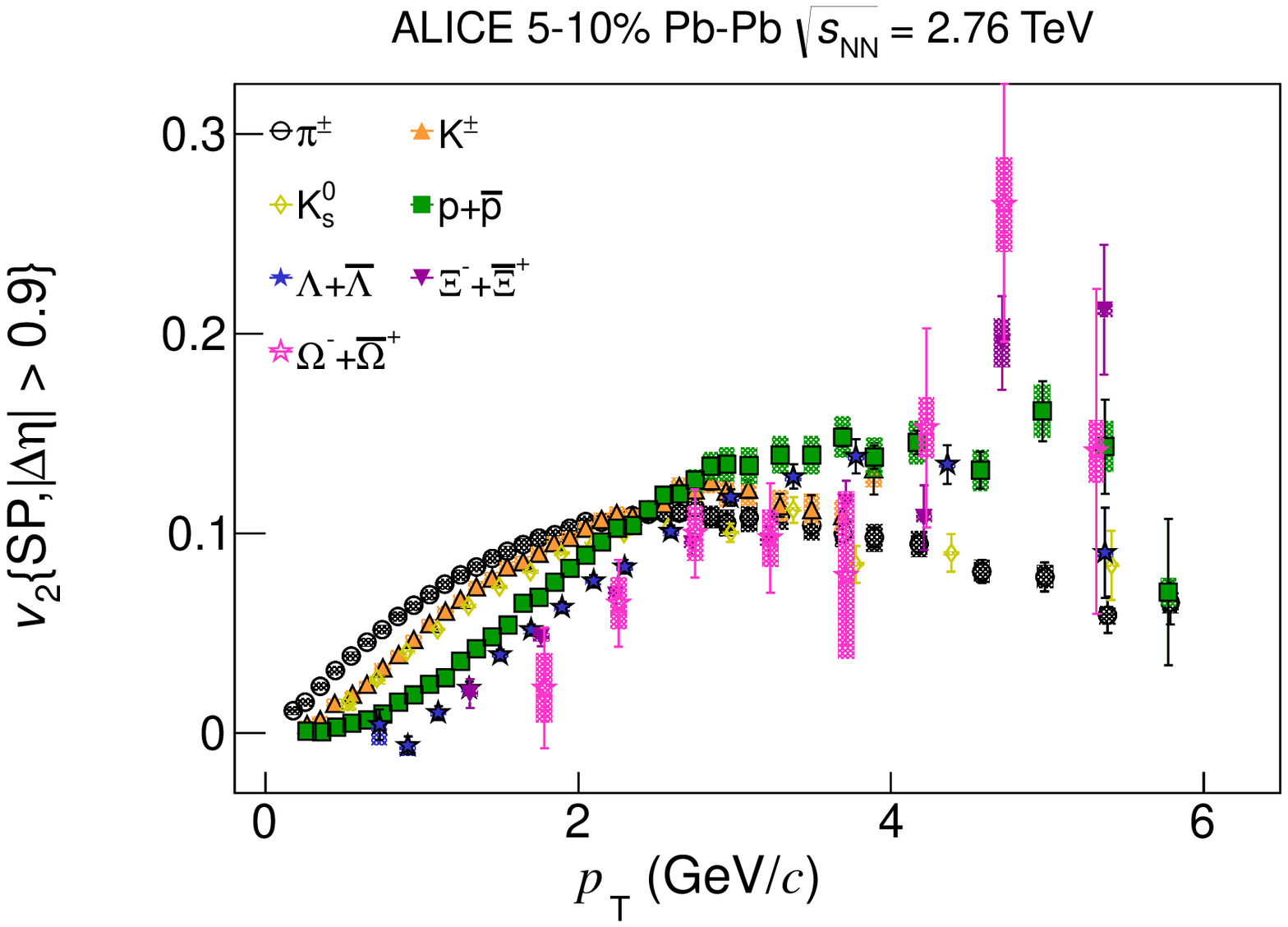}
\captionof{figure}{The $p_{\rm{T}}$-differential \vtwo~for different particle species, represented by the different symbols 
and colors, for the 5--10$\%$ centrality interval of Pb--Pb collisions at $\sqrt{s_{\mathrm{NN}}} = 2.76$~TeV.}
\label{fig:v2Centrality5To10}
\end{center}

\begin{center}
\includegraphics[width=\textwidth]{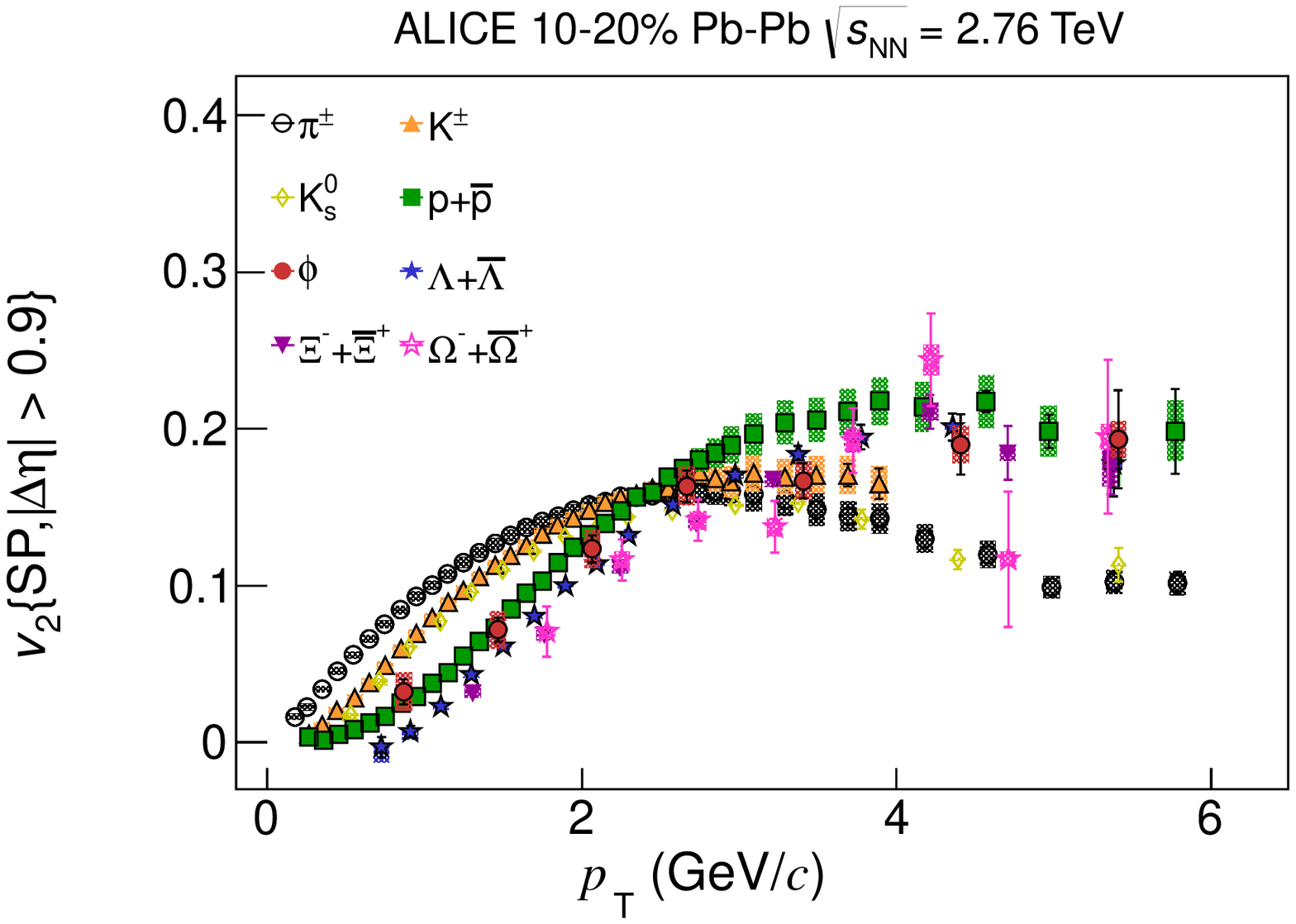}
\captionof{figure}{The $p_{\rm{T}}$-differential \vtwo~for different particle species, represented by the different symbols 
and colors, for the 10--20$\%$ centrality interval of Pb--Pb collisions at $\sqrt{s_{\mathrm{NN}}} = 2.76$~TeV.}
\label{fig:v2Centrality10To20}
\end{center}

\begin{center}
\includegraphics[width=\textwidth]{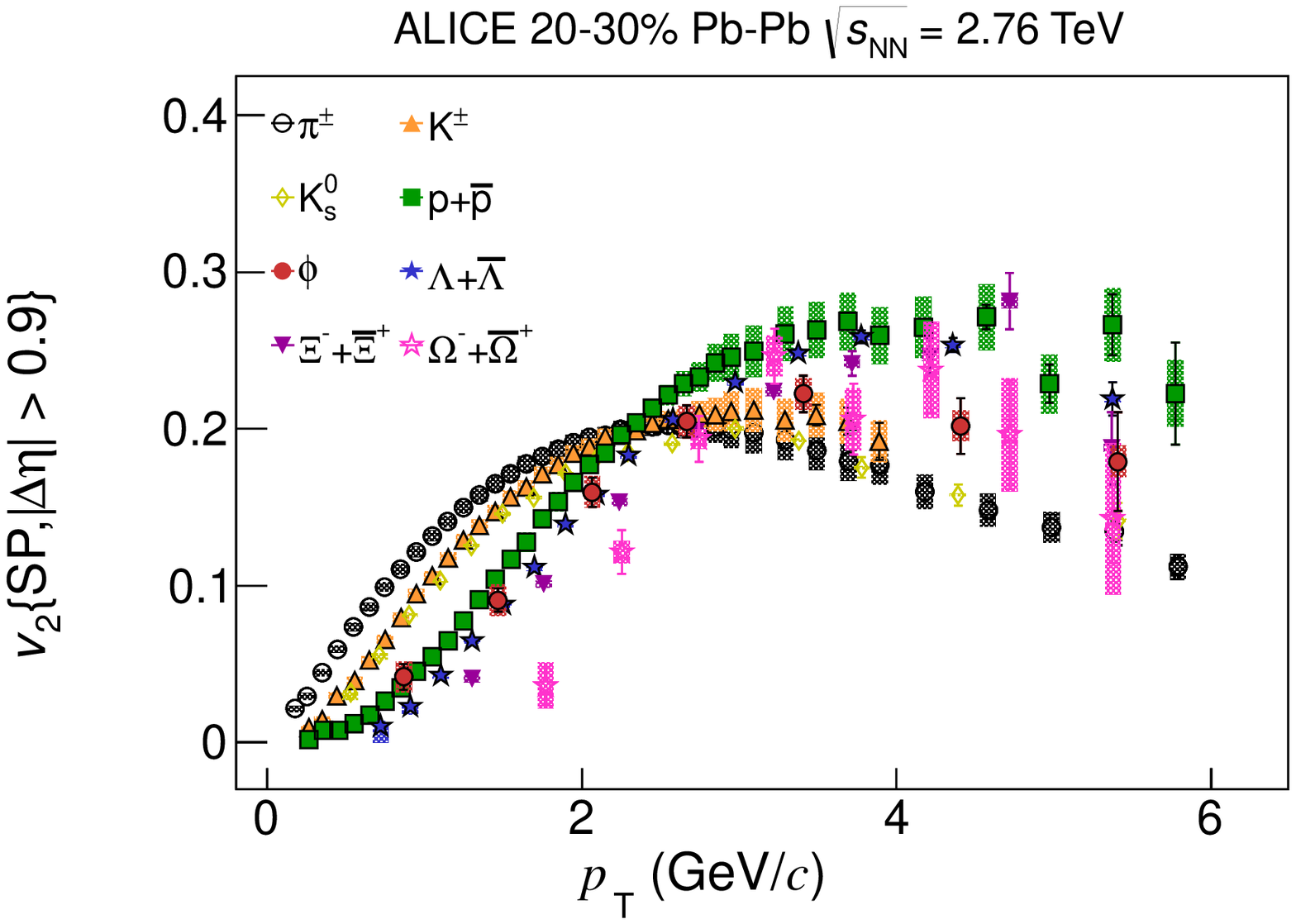}
\captionof{figure}{The $p_{\rm{T}}$-differential \vtwo~for different particle species, represented by the different symbols 
and colors, for the 20--30$\%$ centrality interval of Pb--Pb collisions at $\sqrt{s_{\mathrm{NN}}} = 2.76$~TeV.}
\label{fig:v2Centrality20To30}
\end{center}

\begin{center}
\includegraphics[width=\textwidth]{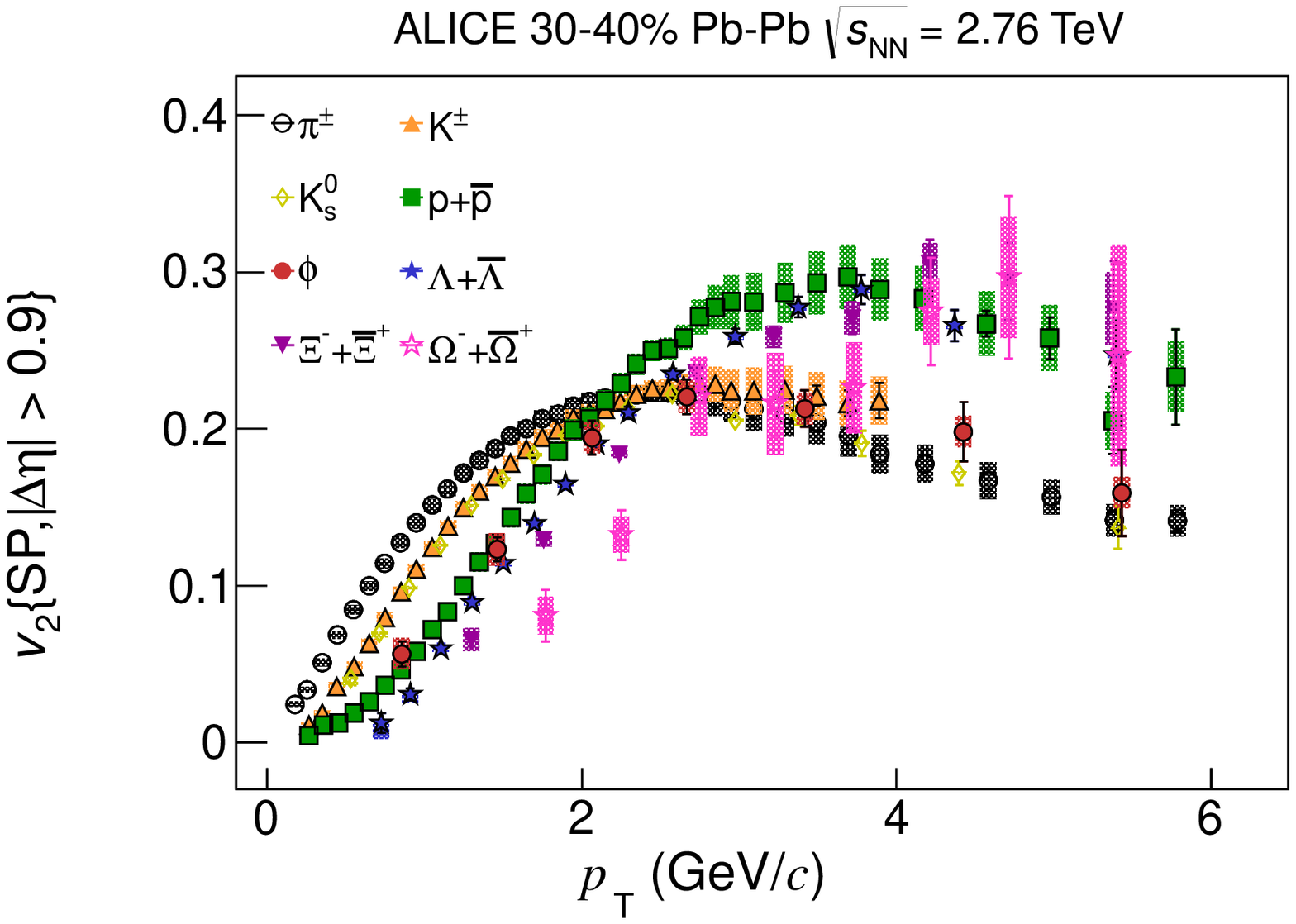}
\captionof{figure}{The $p_{\rm{T}}$-differential \vtwo~for different particle species, represented by the different symbols 
and colors, for the 30--40$\%$ centrality interval of Pb--Pb collisions at $\sqrt{s_{\mathrm{NN}}} = 2.76$~TeV.}
\label{fig:v2Centrality30To40}
\end{center}

\begin{center}
\includegraphics[width=\textwidth]{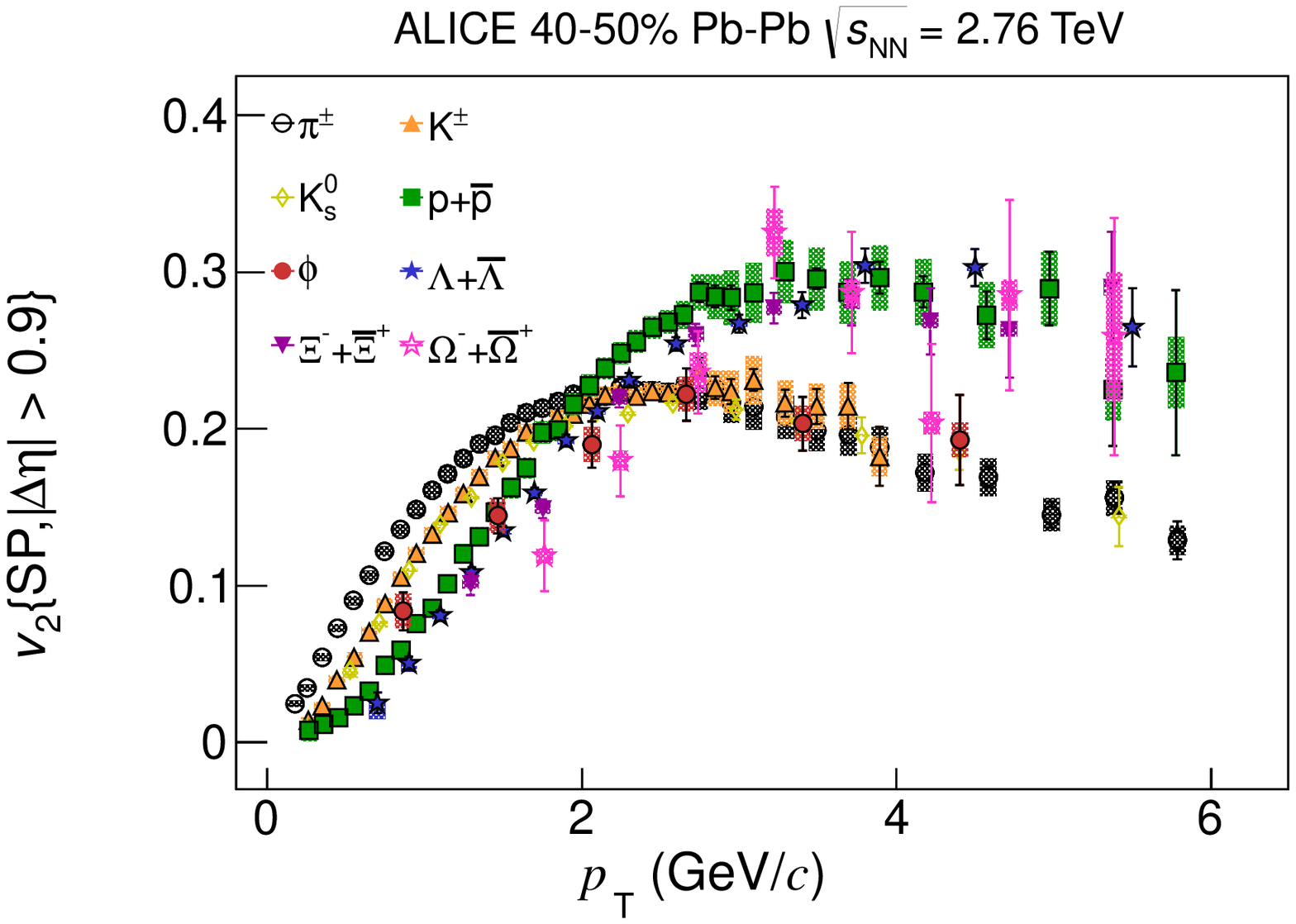}
\captionof{figure}{The $p_{\rm{T}}$-differential \vtwo~for different particle species, represented by the different symbols 
and colors, for the 40--50$\%$ centrality interval of Pb--Pb collisions at $\sqrt{s_{\mathrm{NN}}} = 2.76$~TeV.}
\label{fig:v2Centrality40To50}
\end{center}

\begin{center}
\includegraphics[width=\textwidth]{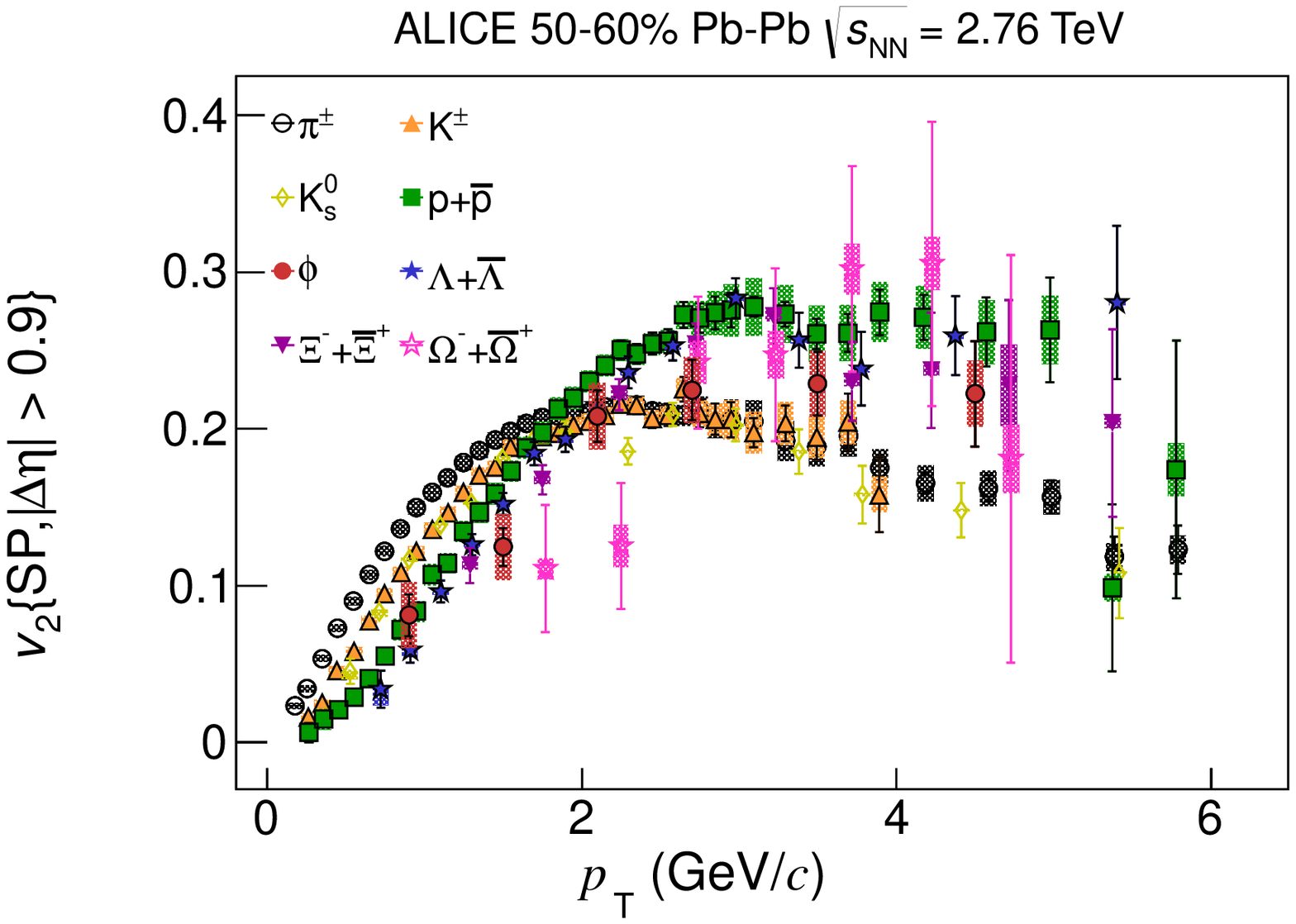}
\captionof{figure}{The $p_{\rm{T}}$-differential \vtwo~for different particle species, represented by the different symbols 
and colors, for the 50--60$\%$ centrality interval of Pb--Pb collisions at $\sqrt{s_{\mathrm{NN}}} = 2.76$~TeV.}
\label{fig:v2Centrality50To60}
\end{center}

\subsection{Plots from Fig.~\ref{fig:v2DataVsHydro}}

\begin{center}
  \includegraphics[width=\textwidth]{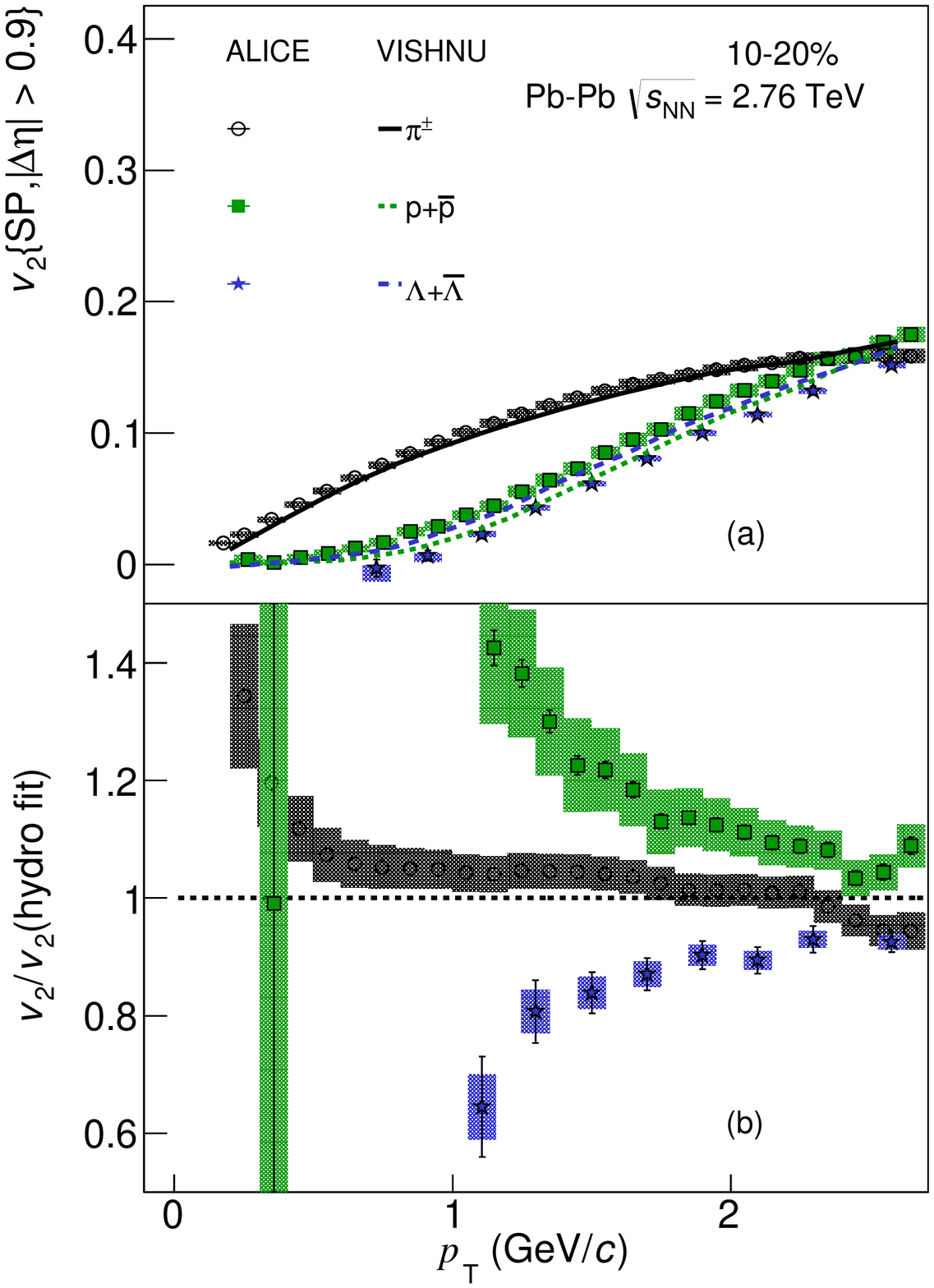}
  \captionof{figure}{The $p_{\rm{T}}$-differential \vtwo~for different particle species in (a) measured with the scalar product 
  method with a minimum pseudorapidity gap $|\Delta\eta| > 0.9$ in Pb--Pb collisions at $\sqrt{s_{\mathrm{NN}}} = 2.76$~TeV, 
  compared to theoretical, hydrodynamical calculations coupled to a hadronic cascade model \cite{Song:2007fn,Song:2007ux,Song:2008si}. Panel (b) shows the dependence of the ratio of the experimental points to a fit over the theoretical calculations as a function of $p_{\rm{T}}$. }
  \label{fig:v2DataVsHydroGroup1}
\end{center}

\begin{center}
  \includegraphics[width=\textwidth]{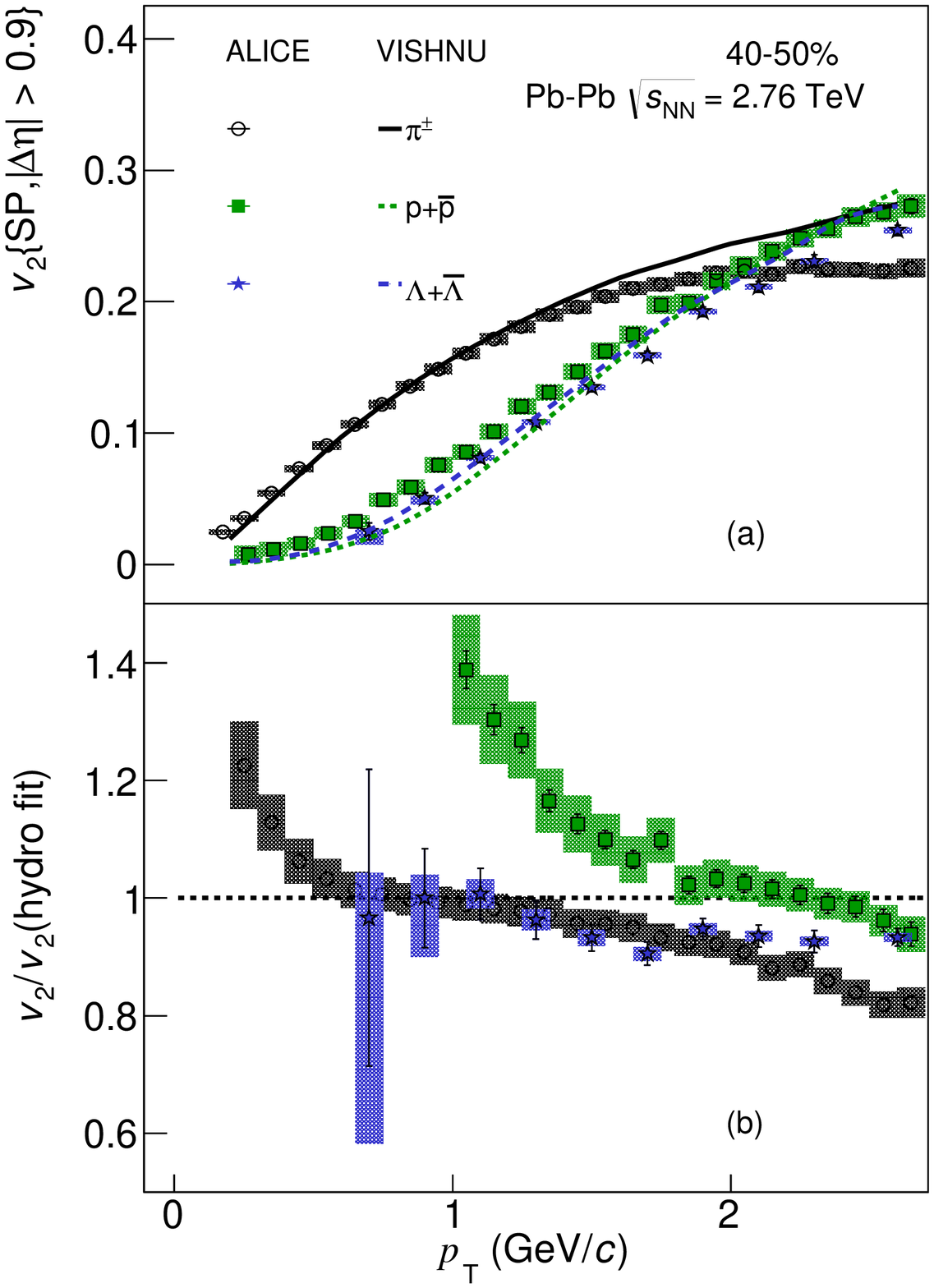}
  \captionof{figure}{The $p_{\rm{T}}$-differential \vtwo~for different particle species in (a) measured with the scalar product 
  method with a minimum pseudorapidity gap $|\Delta\eta| > 0.9$ in Pb--Pb collisions at $\sqrt{s_{\mathrm{NN}}} = 2.76$~TeV, 
  compared to theoretical, hydrodynamical calculations coupled to a hadronic cascade model \cite{Song:2007fn,Song:2007ux,Song:2008si}. Panel (b) shows the dependence of the ratio of the experimental points to a fit over the theoretical calculations as a function of $p_{\rm{T}}$. }
  \label{fig:v2DataVsHydroGroup2}
\end{center}

\begin{center}
  \includegraphics[width=\textwidth]{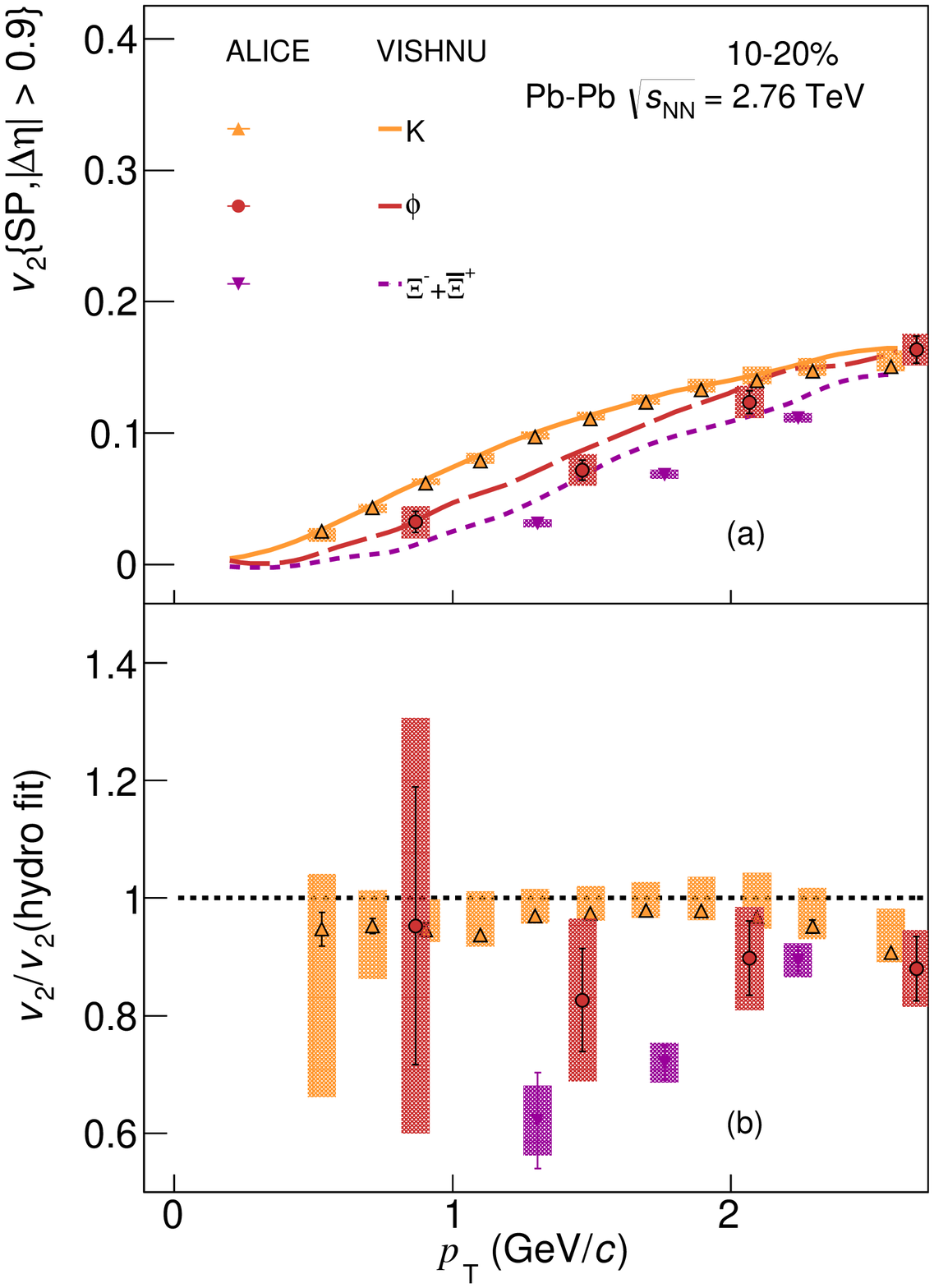}
  \captionof{figure}{The $p_{\rm{T}}$-differential \vtwo~for different particle species in (a) measured with the scalar product 
  method with a minimum pseudorapidity gap $|\Delta\eta| > 0.9$ in Pb--Pb collisions at $\sqrt{s_{\mathrm{NN}}} = 2.76$~TeV, 
  compared to theoretical, hydrodynamical calculations coupled to a hadronic cascade model \cite{Song:2007fn,Song:2007ux,Song:2008si}. Panel (b) shows the dependence of the ratio of the experimental points to a fit over the theoretical calculations as a function of $p_{\rm{T}}$. }
  \label{fig:v2DataVsHydroGroup3}
\end{center}

\begin{center}
  \includegraphics[width=\textwidth]{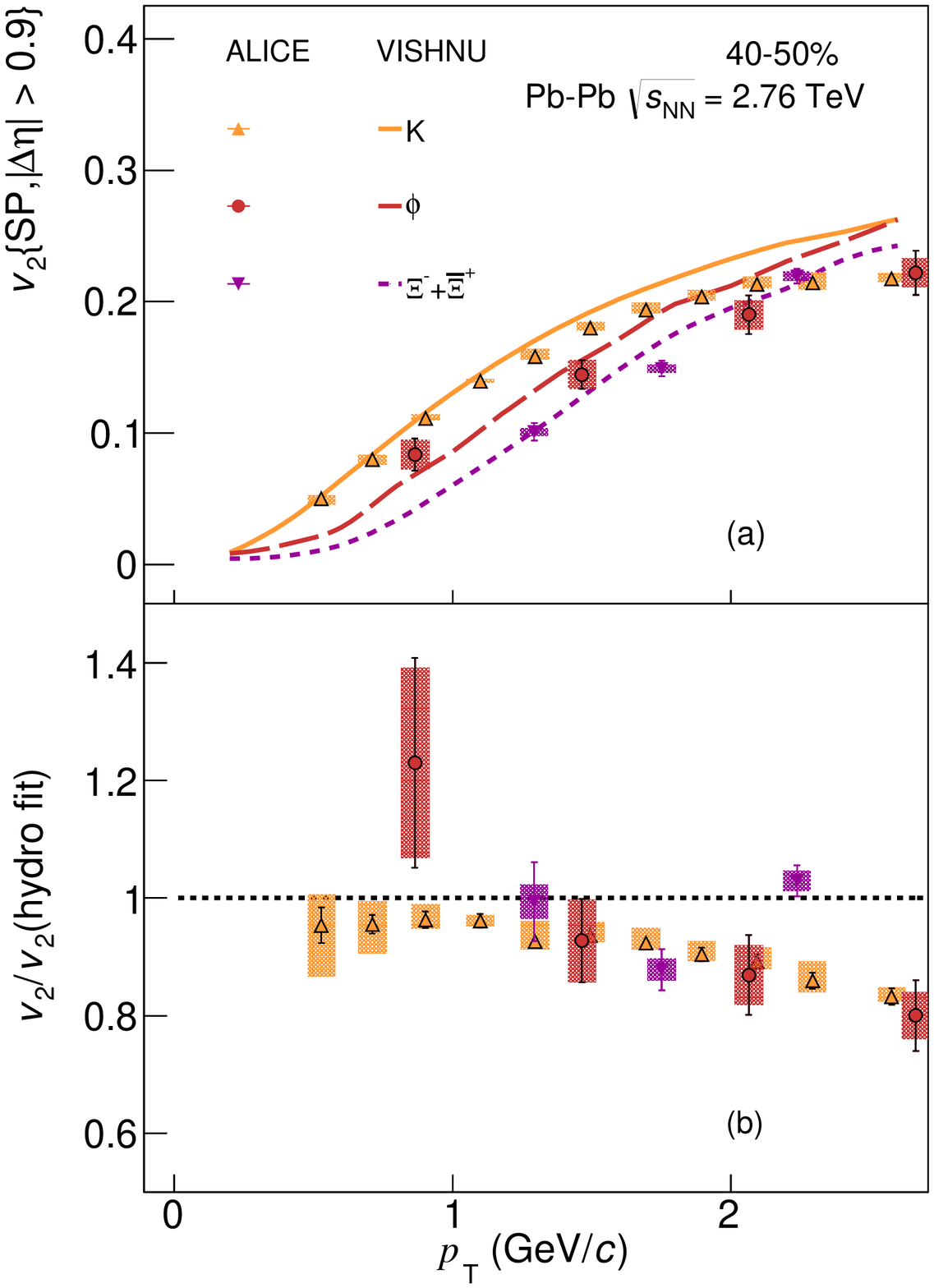}
  \captionof{figure}{The $p_{\rm{T}}$-differential \vtwo~for different particle species in (a) measured with the scalar product 
  method with a minimum pseudorapidity gap $|\Delta\eta| > 0.9$ in Pb--Pb collisions at $\sqrt{s_{\mathrm{NN}}} = 2.76$~TeV, 
  compared to theoretical, hydrodynamical calculations coupled to a hadronic cascade model \cite{Song:2007fn,Song:2007ux,Song:2008si}. Panel (b) shows the dependence of the ratio of the experimental points to a fit over the theoretical calculations as a function of $p_{\rm{T}}$. }
  \label{fig:v2DataVsHydroGroup4}
\end{center}

\subsection{Plots from Fig.~\ref{fig:v2DataVsRHIC}}

\begin{center}
  \includegraphics[width=\textwidth]{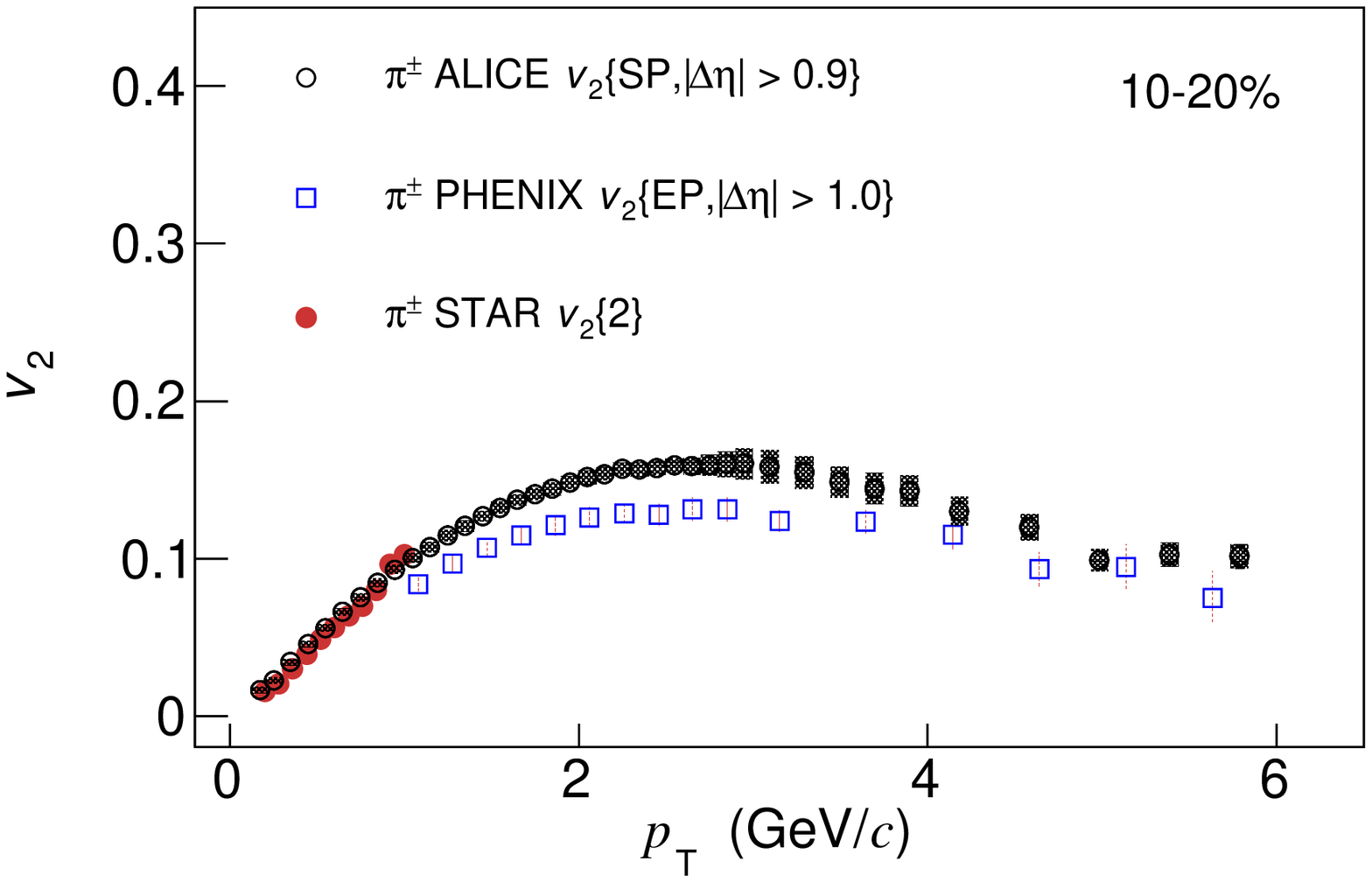}
  \captionof{figure}{The comparison of the $p_{\rm{T}}$-differential \vtwo~for pions for the 10--20$\%$ centrality interval of Pb--Pb and Au--Au collisions at the LHC and RHIC, respectively. The RHIC 
  points are extracted from \cite{Adams:2004bi} (STAR) and \cite{Adare:2012vq} (PHENIX).}
\label{fig:v2DataVsRHICPions}
\end{center}

\begin{center}
  \includegraphics[width=\textwidth]{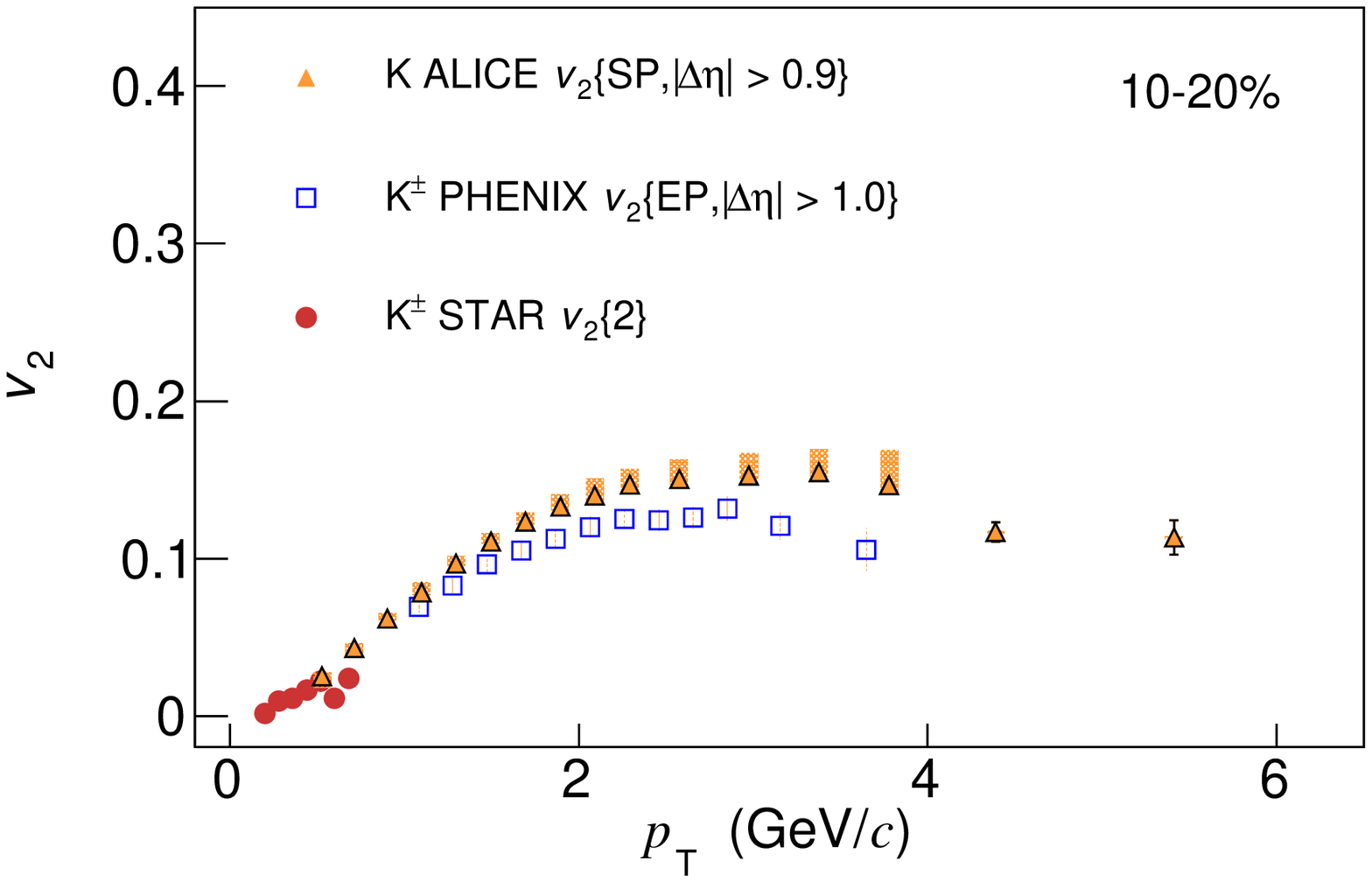}
  \captionof{figure}{The comparison of the $p_{\rm{T}}$-differential \vtwo~for kaons for the 10--20$\%$ centrality interval of Pb--Pb and Au--Au collisions at the LHC and RHIC, respectively. The RHIC 
  points are extracted from \cite{Adams:2004bi} (STAR) and \cite{Adare:2012vq} (PHENIX).}
\label{fig:v2DataVsRHICKaons}
\end{center}

\begin{center}
  \includegraphics[width=\textwidth]{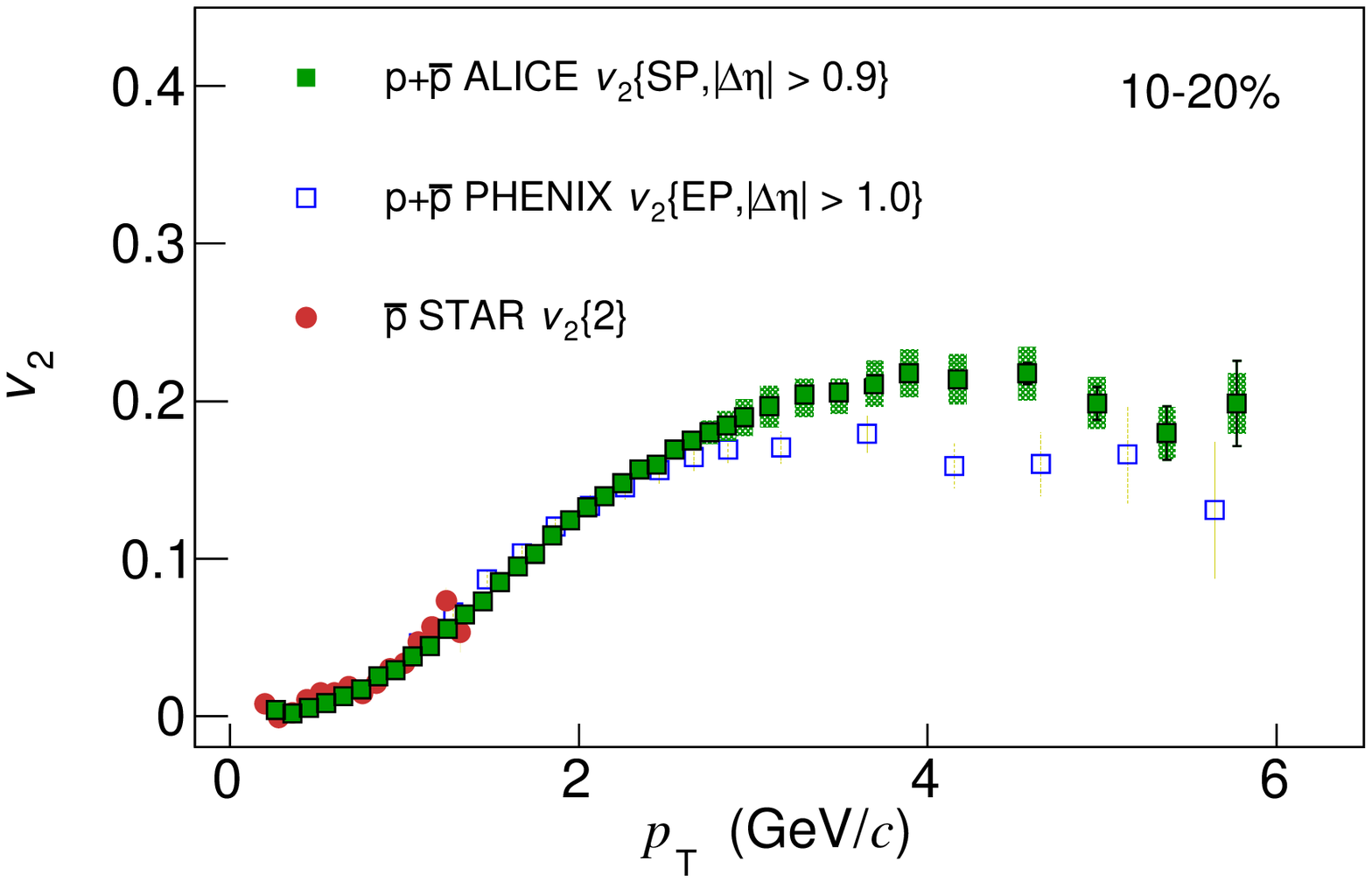}
  \captionof{figure}{The comparison of the $p_{\rm{T}}$-differential \vtwo~for p+$\overline{\mathrm{p}}$ 
  for the 10--20$\%$ centrality interval of Pb--Pb and Au--Au collisions at the LHC and RHIC, respectively. The RHIC 
  points are extracted from \cite{Adams:2004bi} (STAR) and \cite{Adare:2012vq} (PHENIX).}
\label{fig:v2DataVsRHICProtons}
\end{center}

\subsection{Plots from Fig.~\ref{fig:pTScaling}}

\begin{center}
  \includegraphics[width=\textwidth]{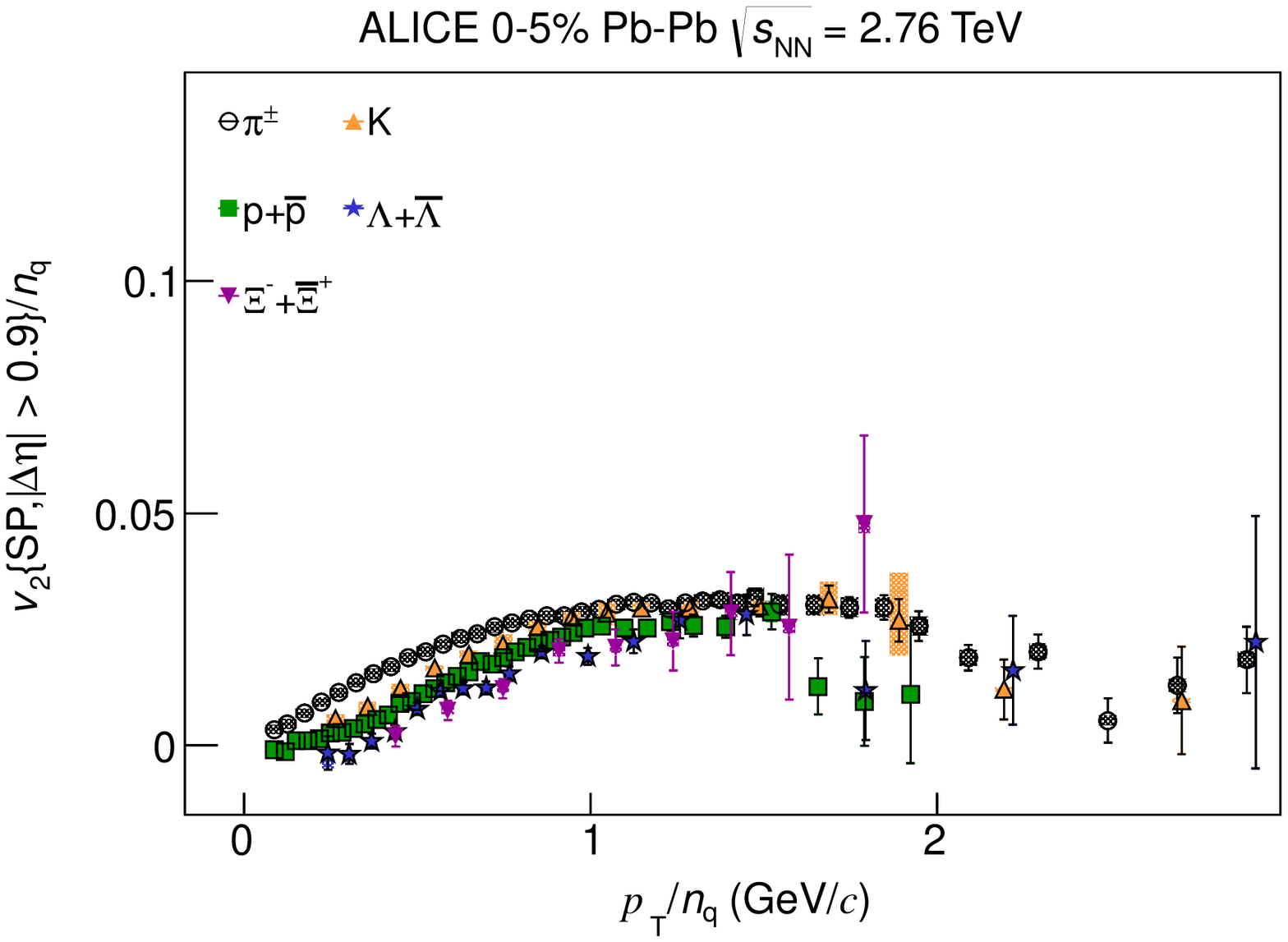}
  \captionof{figure}{The $p_{\rm{T}}/n_q$ dependence of $v_2/n_q$ for $\pi^{\pm}$, $\mathrm{K}$, p+$\overline{\mathrm{p}}$, $\phi$, $\Lambda$+$\overline{\mathrm{\Lambda}}$, $\mathrm{\Xi^-}$+$\overline{\mathrm{\Xi}}^+$ and $\mathrm{\Omega}^-$+$\overline{\mathrm{\Omega}}^+$ for the 0--5$\%$ centrality interval in Pb--Pb collisions at $\sqrt{s_{\mathrm{NN}}} = 2.76$~TeV.}
\label{fig:pTScalingCent0To5}
\end{center}

\begin{center}
  \includegraphics[width=\textwidth]{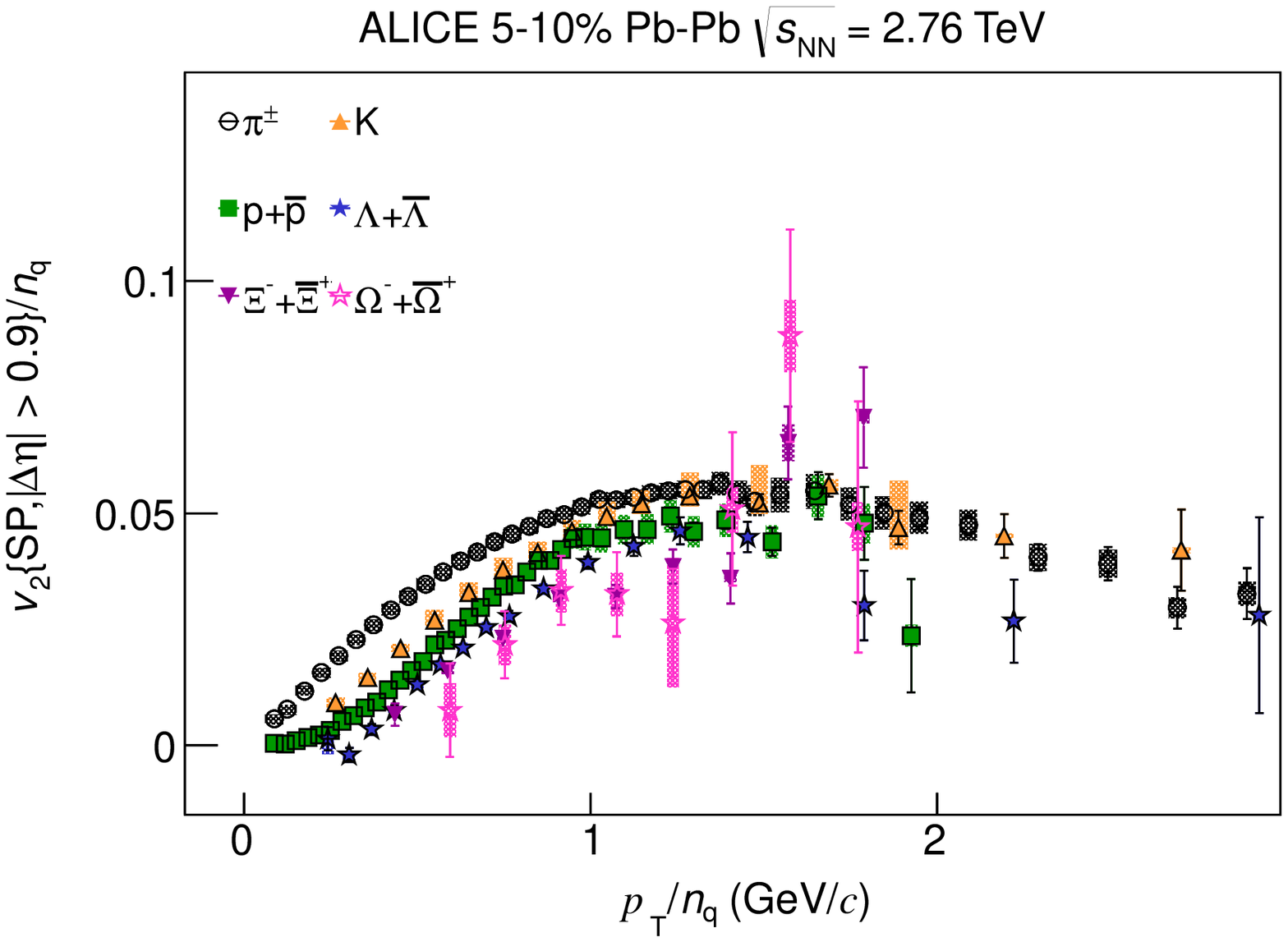}
  \captionof{figure}{The $p_{\rm{T}}/n_q$ dependence of $v_2/n_q$ for $\pi^{\pm}$, $\mathrm{K}$, p+$\overline{\mathrm{p}}$, $\phi$, $\Lambda$+$\overline{\mathrm{\Lambda}}$, $\mathrm{\Xi^-}$+$\overline{\mathrm{\Xi}}^+$ and $\mathrm{\Omega}^-$+$\overline{\mathrm{\Omega}}^+$ for the 5--10$\%$ centrality interval in Pb--Pb collisions at $\sqrt{s_{\mathrm{NN}}} = 2.76$~TeV.}
\label{fig:pTScalingCent5To10}
\end{center}

\begin{center}
  \includegraphics[width=\textwidth]{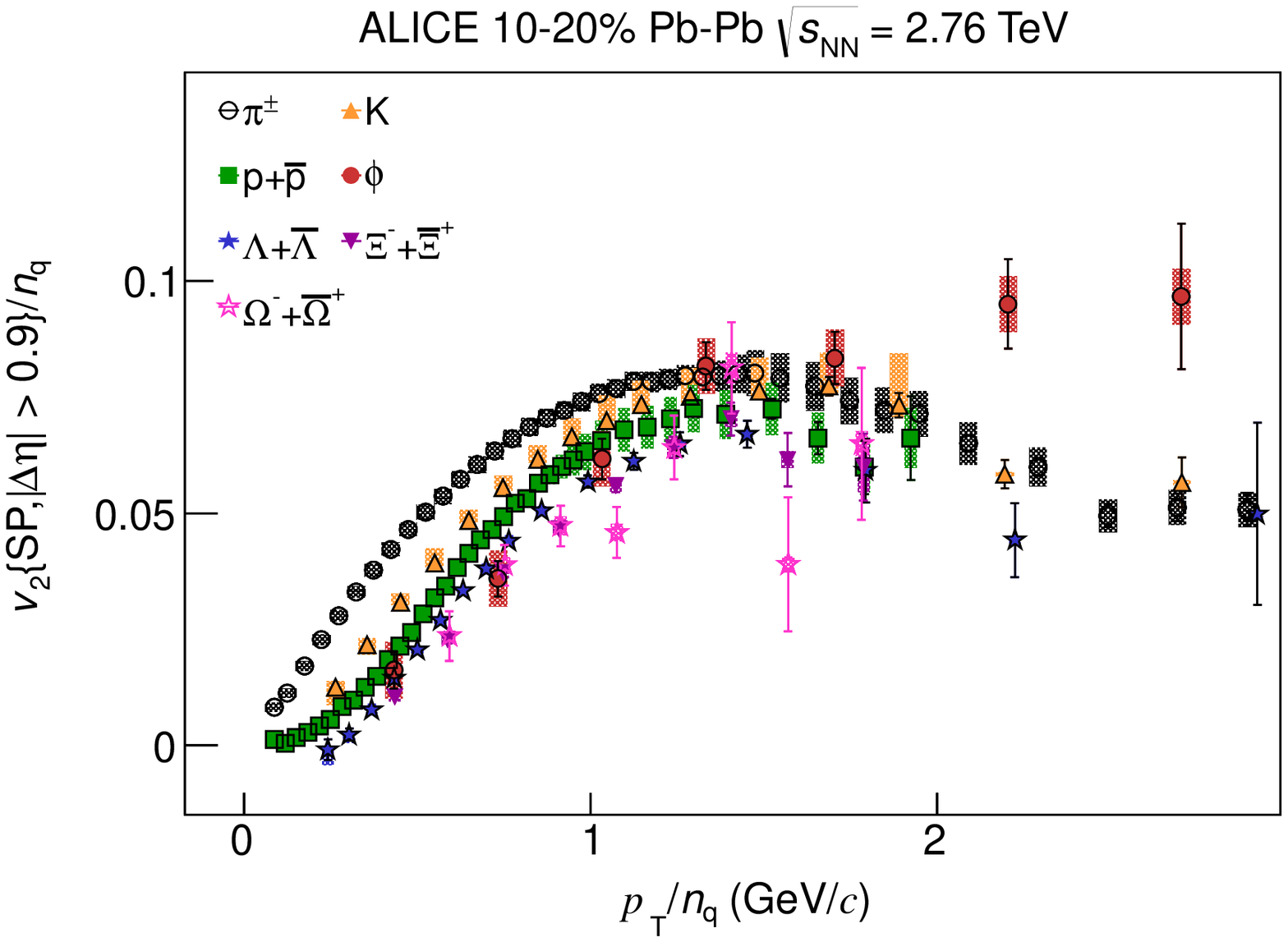}
  \captionof{figure}{The $p_{\rm{T}}/n_q$ dependence of $v_2/n_q$ for $\pi^{\pm}$, $\mathrm{K}$, p+$\overline{\mathrm{p}}$, $\phi$, $\Lambda$+$\overline{\mathrm{\Lambda}}$, $\mathrm{\Xi^-}$+$\overline{\mathrm{\Xi}}^+$ and $\mathrm{\Omega}^-$+$\overline{\mathrm{\Omega}}^+$ for the 10--20$\%$ centrality interval in Pb--Pb collisions at $\sqrt{s_{\mathrm{NN}}} = 2.76$~TeV.}
\label{fig:pTScalingCent10To20}
\end{center}

\begin{center}
  \includegraphics[width=\textwidth]{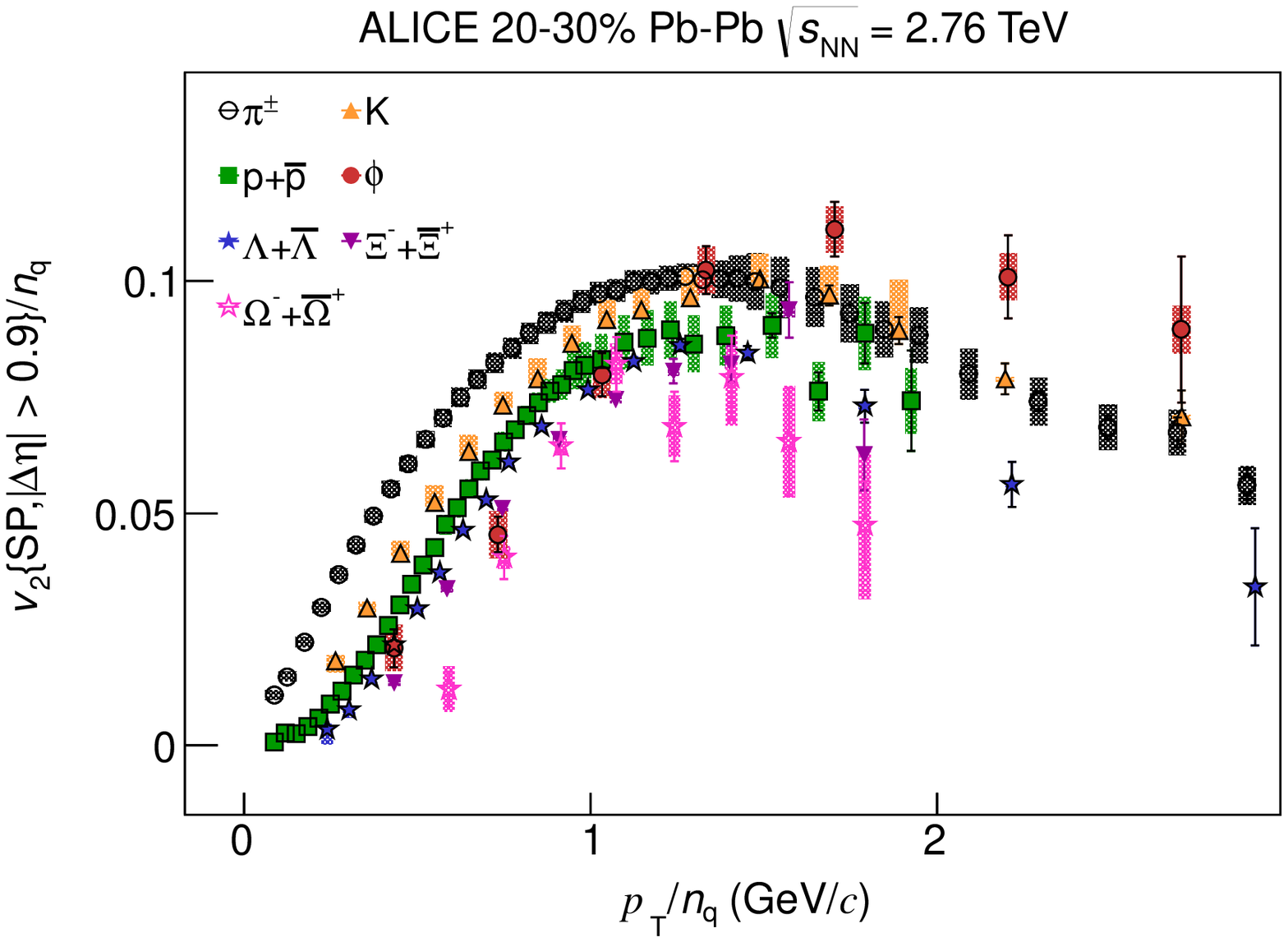}
  \captionof{figure}{The $p_{\rm{T}}/n_q$ dependence of $v_2/n_q$ for $\pi^{\pm}$, $\mathrm{K}$, p+$\overline{\mathrm{p}}$, $\phi$, $\Lambda$+$\overline{\mathrm{\Lambda}}$, $\mathrm{\Xi^-}$+$\overline{\mathrm{\Xi}}^+$ and $\mathrm{\Omega}^-$+$\overline{\mathrm{\Omega}}^+$ for the 20--30$\%$ centrality interval in Pb--Pb collisions at $\sqrt{s_{\mathrm{NN}}} = 2.76$~TeV.}
\label{fig:pTScalingCent20To30}
\end{center}

\begin{center}
  \includegraphics[width=\textwidth]{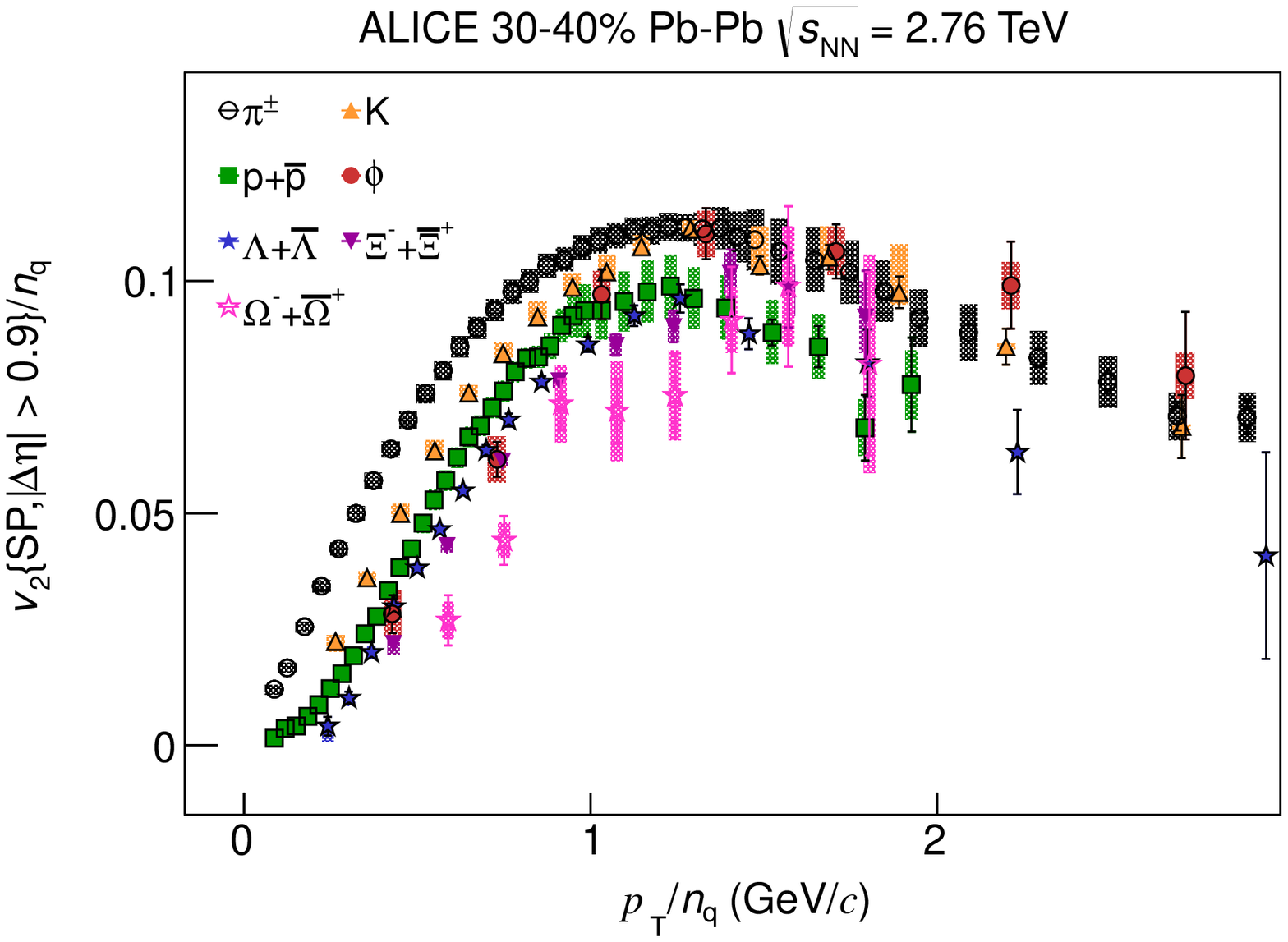}
  \captionof{figure}{The $p_{\rm{T}}/n_q$ dependence of $v_2/n_q$ for $\pi^{\pm}$, $\mathrm{K}$, p+$\overline{\mathrm{p}}$, $\phi$, $\Lambda$+$\overline{\mathrm{\Lambda}}$, $\mathrm{\Xi^-}$+$\overline{\mathrm{\Xi}}^+$ and $\mathrm{\Omega}^-$+$\overline{\mathrm{\Omega}}^+$ for the 30--40$\%$ centrality interval in Pb--Pb collisions at $\sqrt{s_{\mathrm{NN}}} = 2.76$~TeV.}
\label{fig:pTScalingCent30To40}
\end{center}

\begin{center}
  \includegraphics[width=\textwidth]{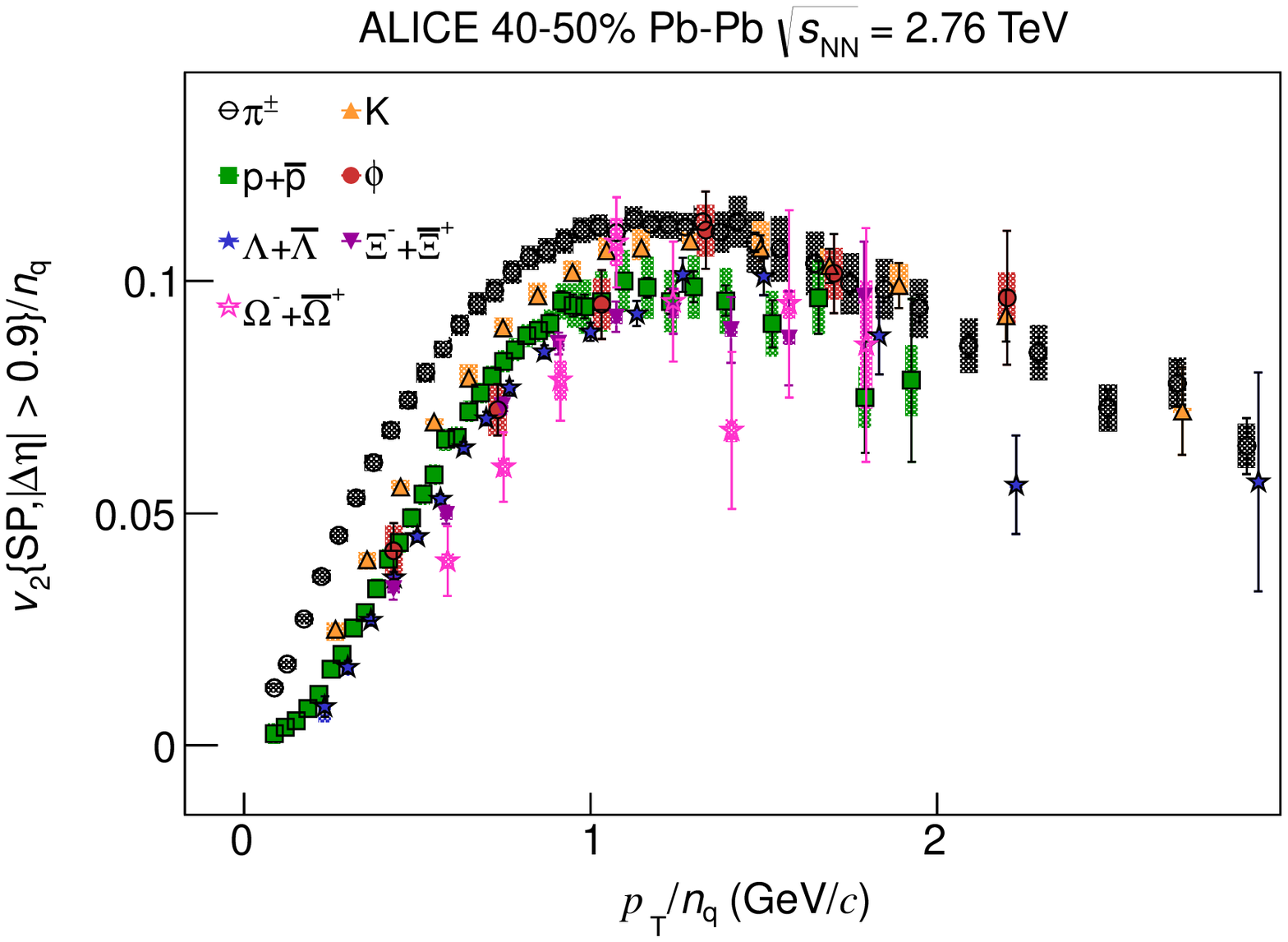}
  \captionof{figure}{The $p_{\rm{T}}/n_q$ dependence of $v_2/n_q$ for $\pi^{\pm}$, $\mathrm{K}$, p+$\overline{\mathrm{p}}$, $\phi$, $\Lambda$+$\overline{\mathrm{\Lambda}}$, $\mathrm{\Xi^-}$+$\overline{\mathrm{\Xi}}^+$ and $\mathrm{\Omega}^-$+$\overline{\mathrm{\Omega}}^+$ for the 40--50$\%$ centrality interval in Pb--Pb collisions at $\sqrt{s_{\mathrm{NN}}} = 2.76$~TeV.}
\label{fig:pTScalingCent40To50}
\end{center}

\begin{center}
  \includegraphics[width=\textwidth]{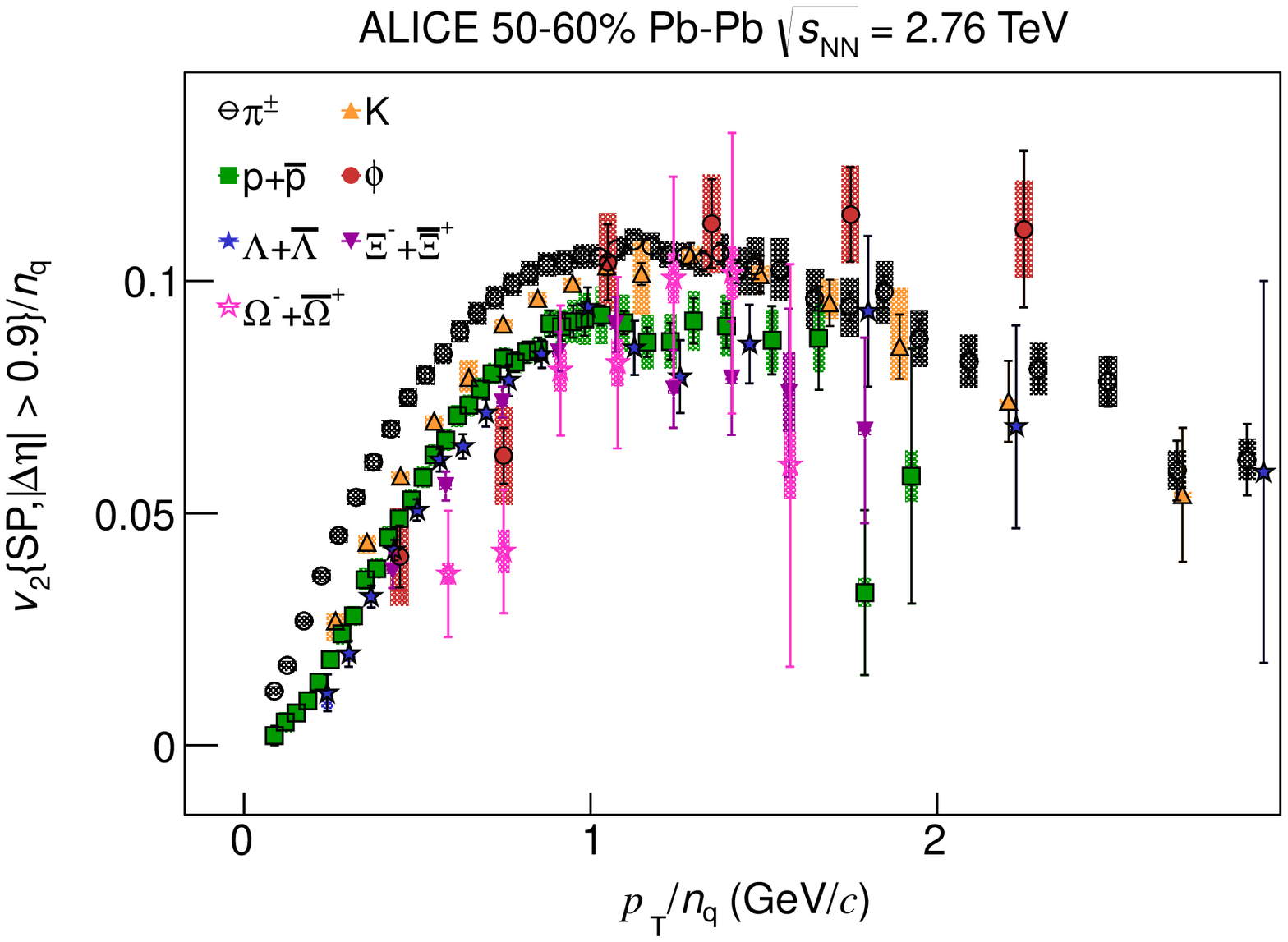}
  \captionof{figure}{The $p_{\rm{T}}/n_q$ dependence of $v_2/n_q$ for $\pi^{\pm}$, $\mathrm{K}$, p+$\overline{\mathrm{p}}$, $\phi$, $\Lambda$+$\overline{\mathrm{\Lambda}}$, $\mathrm{\Xi^-}$+$\overline{\mathrm{\Xi}}^+$ and $\mathrm{\Omega}^-$+$\overline{\mathrm{\Omega}}^+$ for the 50--60$\%$ centrality interval in Pb--Pb collisions at $\sqrt{s_{\mathrm{NN}}} = 2.76$~TeV.}
\label{fig:pTScalingCent50To60}
\end{center}

\subsection{Plots from Fig.~\ref{fig:pTScaling2}}

\begin{center}
  \includegraphics[width=\textwidth]{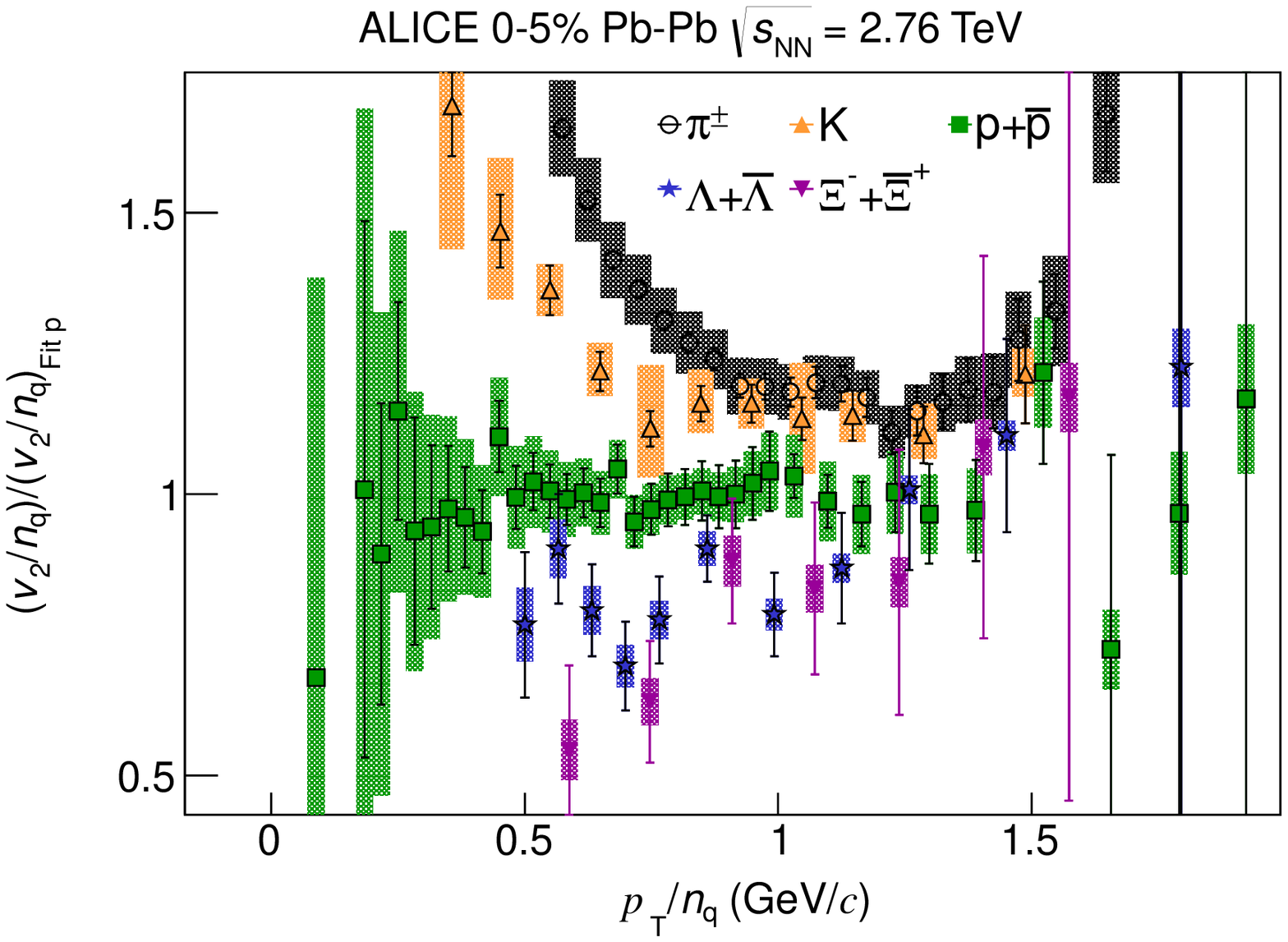}
  \captionof{figure}{The $p_{\rm{T}}/n_q$ dependence of the double ratio of $v_2/n_q$ for every particle 
  species relative to a fit to $v_2/n_q$ of p and $\overline{\mathrm{p}}$ (see text for details) for $\pi^{\pm}$, $\mathrm{K}$, p+$\overline{\mathrm{p}}$, $\phi$, $\Lambda$+$\overline{\mathrm{\Lambda}}$, $\mathrm{\Xi^-}$+$\overline{\mathrm{\Xi}}^+$ and $\mathrm{\Omega}^-$+$\overline{\mathrm{\Omega}}^+$ for the 0--5$\%$ centrality interval in Pb--Pb collisions at $\sqrt{s_{\mathrm{NN}}} = 2.76$~TeV. }
\label{fig:pTScaling2Cent0To5}
\end{center}

\begin{center}
  \includegraphics[width=\textwidth]{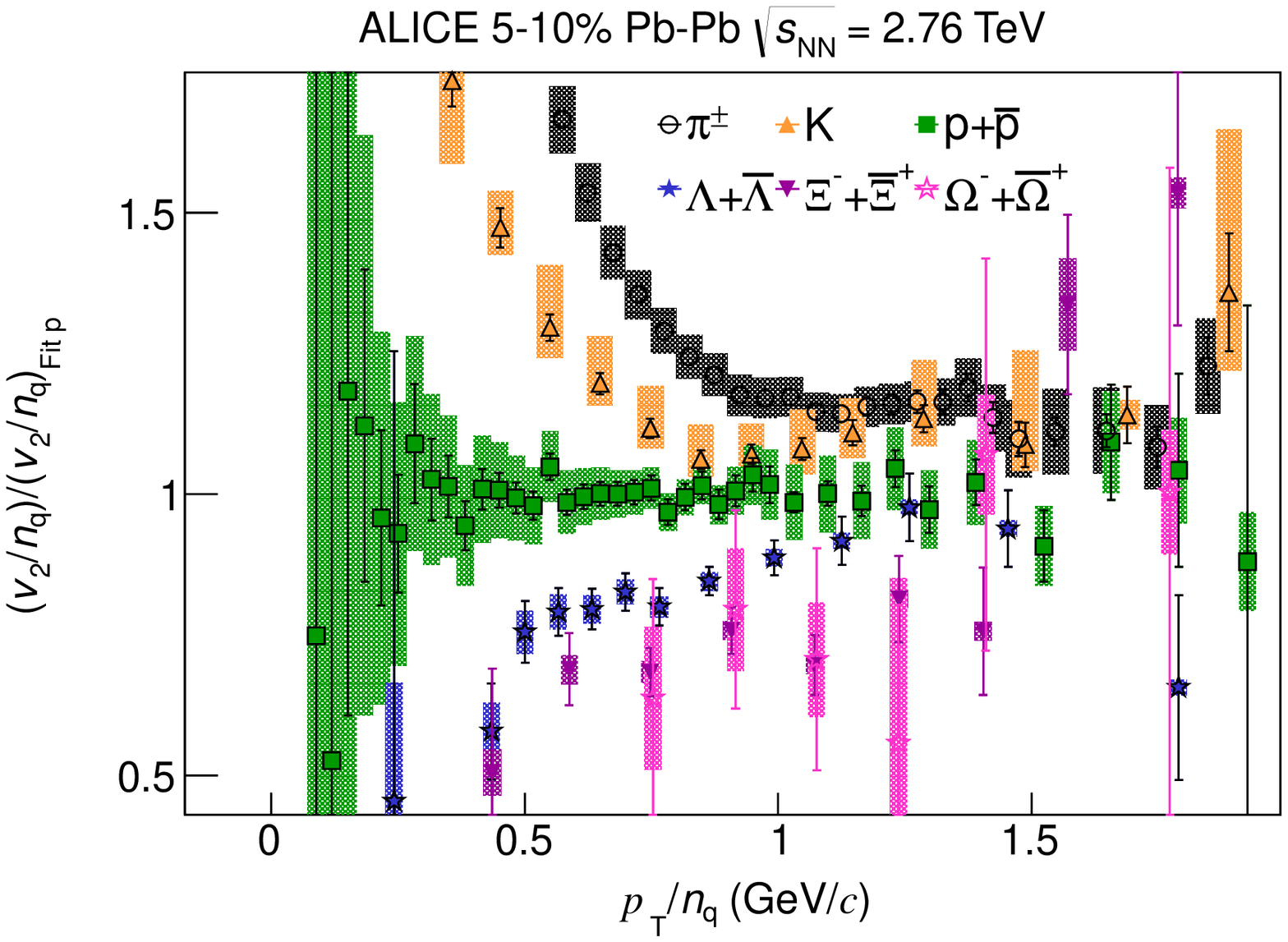}
  \captionof{figure}{The $p_{\rm{T}}/n_q$ dependence of the double ratio of $v_2/n_q$ for every particle 
  species relative to a fit to $v_2/n_q$ of p and $\overline{\mathrm{p}}$ (see text for details) for $\pi^{\pm}$, $\mathrm{K}$, p+$\overline{\mathrm{p}}$, $\phi$, $\Lambda$+$\overline{\mathrm{\Lambda}}$, $\mathrm{\Xi^-}$+$\overline{\mathrm{\Xi}}^+$ and $\mathrm{\Omega}^-$+$\overline{\mathrm{\Omega}}^+$ for the 5--10$\%$ centrality interval in Pb--Pb collisions at $\sqrt{s_{\mathrm{NN}}} = 2.76$~TeV. }
\label{fig:pTScaling2Cent5To10}
\end{center}

\begin{center}
  \includegraphics[width=\textwidth]{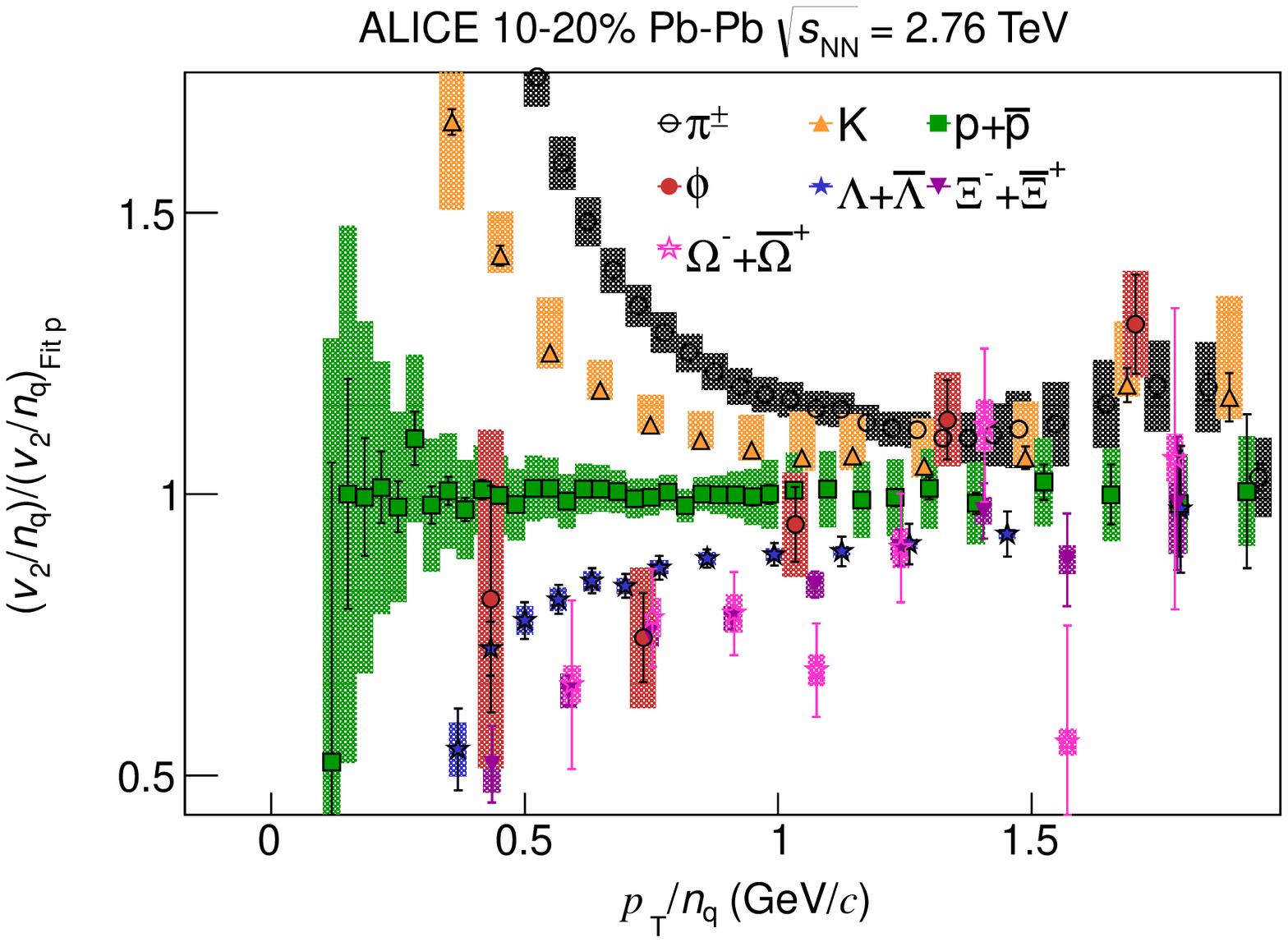}
  \captionof{figure}{The $p_{\rm{T}}/n_q$ dependence of the double ratio of $v_2/n_q$ for every particle 
  species relative to a fit to $v_2/n_q$ of p and $\overline{\mathrm{p}}$ (see text for details) for $\pi^{\pm}$, $\mathrm{K}$, p+$\overline{\mathrm{p}}$, $\phi$, $\Lambda$+$\overline{\mathrm{\Lambda}}$, $\mathrm{\Xi^-}$+$\overline{\mathrm{\Xi}}^+$ and $\mathrm{\Omega}^-$+$\overline{\mathrm{\Omega}}^+$ for the 10--20$\%$ centrality interval in Pb--Pb collisions at $\sqrt{s_{\mathrm{NN}}} = 2.76$~TeV. }
\label{fig:pTScaling2Cent10To20}
\end{center}

\begin{center}
  \includegraphics[width=\textwidth]{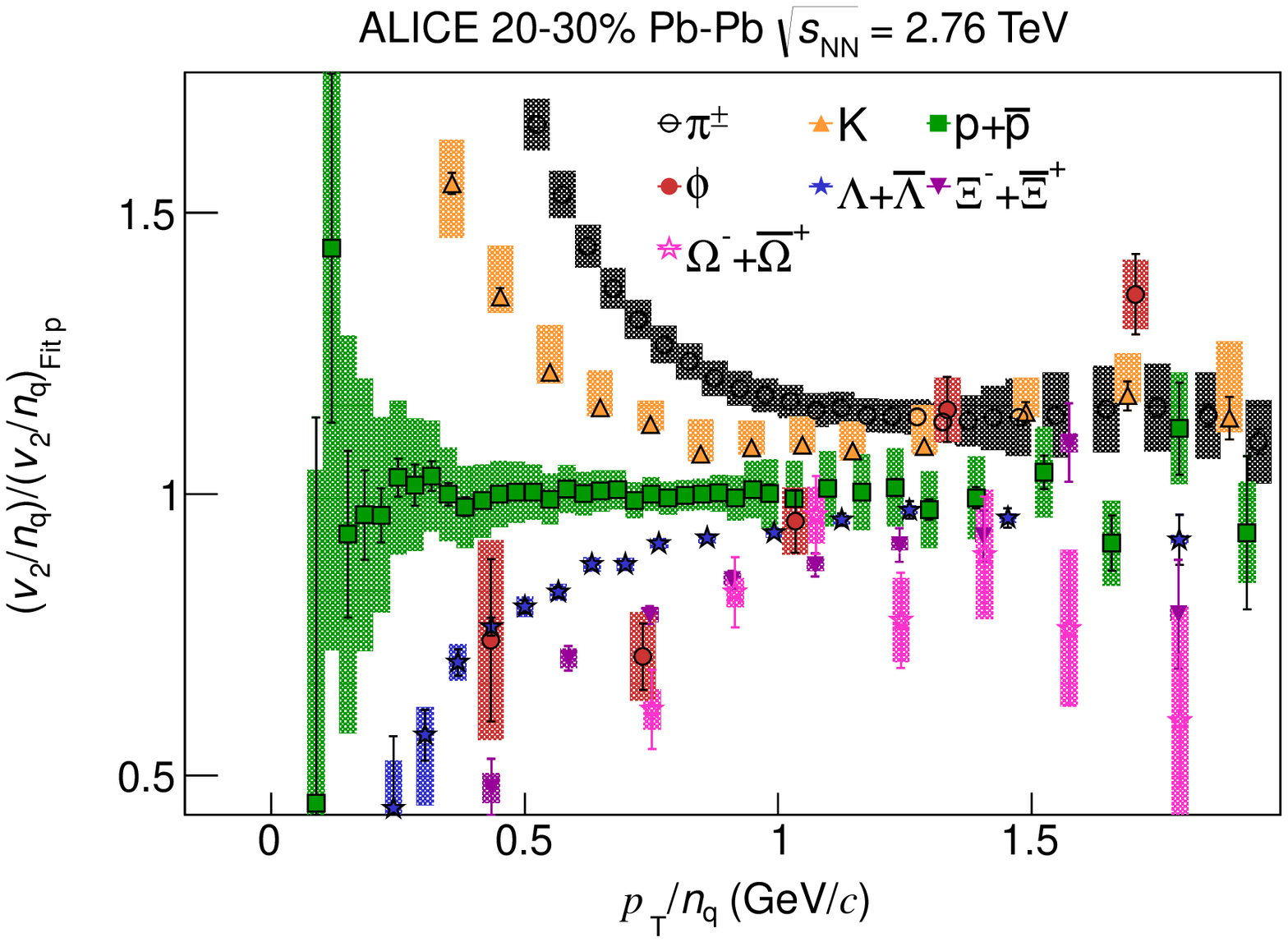}
  \captionof{figure}{The $p_{\rm{T}}/n_q$ dependence of the double ratio of $v_2/n_q$ for every particle 
  species relative to a fit to $v_2/n_q$ of p and $\overline{\mathrm{p}}$ (see text for details) for $\pi^{\pm}$, $\mathrm{K}$, p+$\overline{\mathrm{p}}$, $\phi$, $\Lambda$+$\overline{\mathrm{\Lambda}}$, $\mathrm{\Xi^-}$+$\overline{\mathrm{\Xi}}^+$ and $\mathrm{\Omega}^-$+$\overline{\mathrm{\Omega}}^+$ for the 20--30$\%$ centrality interval in Pb--Pb collisions at $\sqrt{s_{\mathrm{NN}}} = 2.76$~TeV. }
\label{fig:pTScaling2Cent20To30}
\end{center}

\begin{center}
  \includegraphics[width=\textwidth]{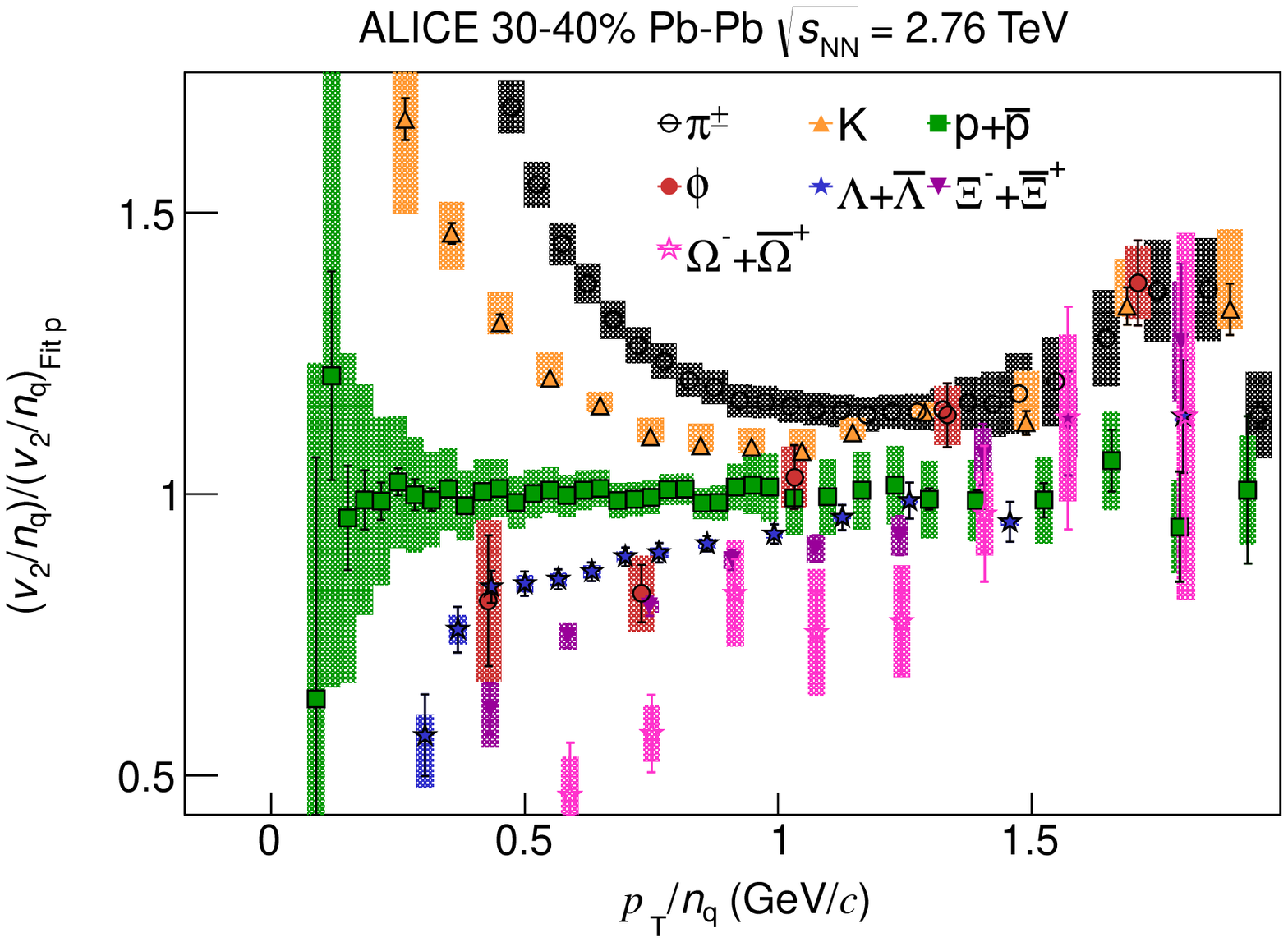}
  \captionof{figure}{The $p_{\rm{T}}/n_q$ dependence of the double ratio of $v_2/n_q$ for every particle 
  species relative to a fit to $v_2/n_q$ of p and $\overline{\mathrm{p}}$ (see text for details) for $\pi^{\pm}$, $\mathrm{K}$, p+$\overline{\mathrm{p}}$, $\phi$, $\Lambda$+$\overline{\mathrm{\Lambda}}$, $\mathrm{\Xi^-}$+$\overline{\mathrm{\Xi}}^+$ and $\mathrm{\Omega}^-$+$\overline{\mathrm{\Omega}}^+$ for the 30--40$\%$ centrality interval in Pb--Pb collisions at $\sqrt{s_{\mathrm{NN}}} = 2.76$~TeV. }
\label{fig:pTScaling2Cent30To40}
\end{center}

\begin{center}
  \includegraphics[width=\textwidth]{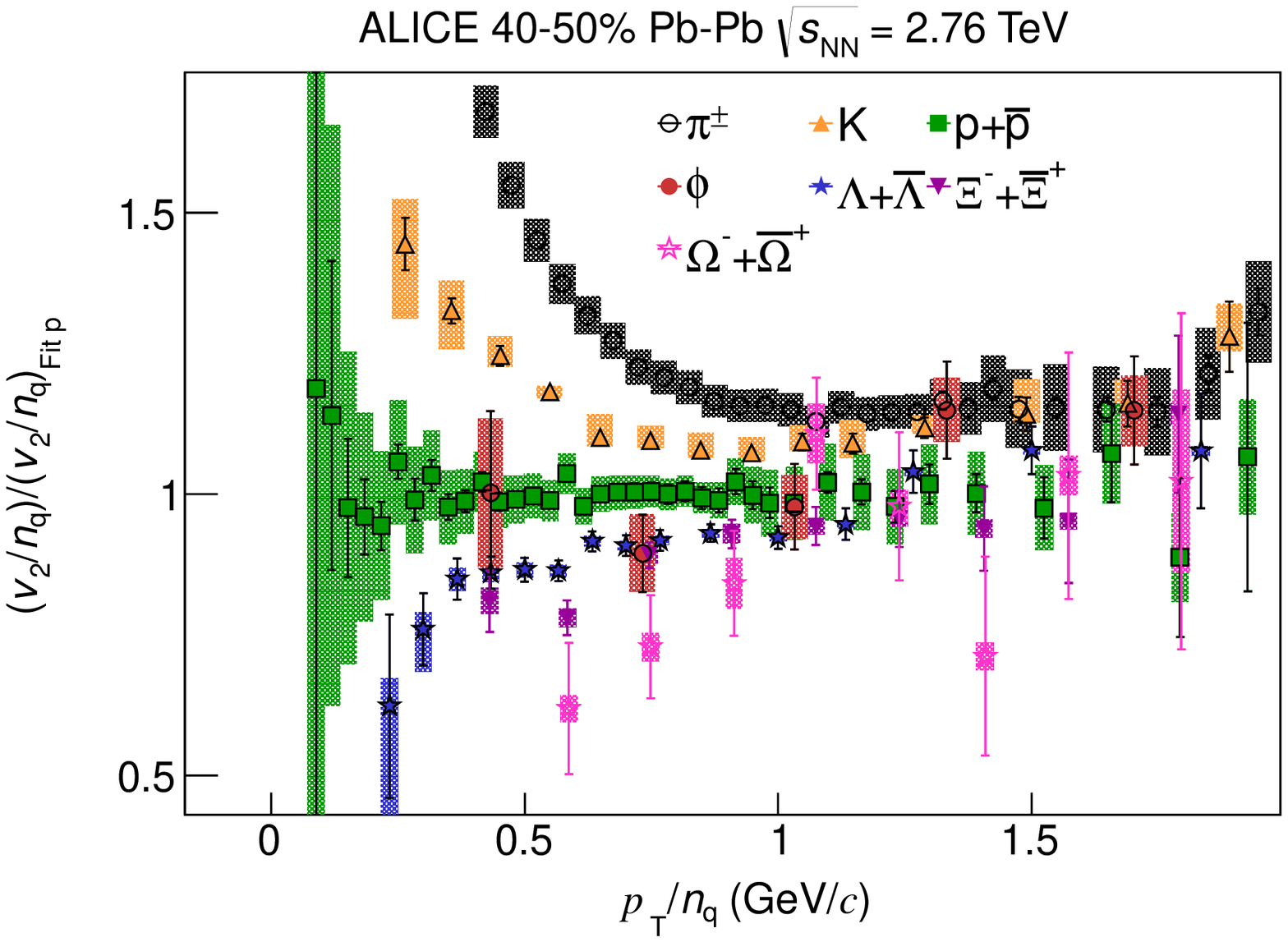}
  \captionof{figure}{The $p_{\rm{T}}/n_q$ dependence of the double ratio of $v_2/n_q$ for every particle 
  species relative to a fit to $v_2/n_q$ of p and $\overline{\mathrm{p}}$ (see text for details) for $\pi^{\pm}$, $\mathrm{K}$, p+$\overline{\mathrm{p}}$, $\phi$, $\Lambda$+$\overline{\mathrm{\Lambda}}$, $\mathrm{\Xi^-}$+$\overline{\mathrm{\Xi}}^+$ and $\mathrm{\Omega}^-$+$\overline{\mathrm{\Omega}}^+$ for the 40--50$\%$ centrality interval in Pb--Pb collisions at $\sqrt{s_{\mathrm{NN}}} = 2.76$~TeV. }
\label{fig:pTScaling2Cent40To50}
\end{center}

\begin{center}
  \includegraphics[width=\textwidth]{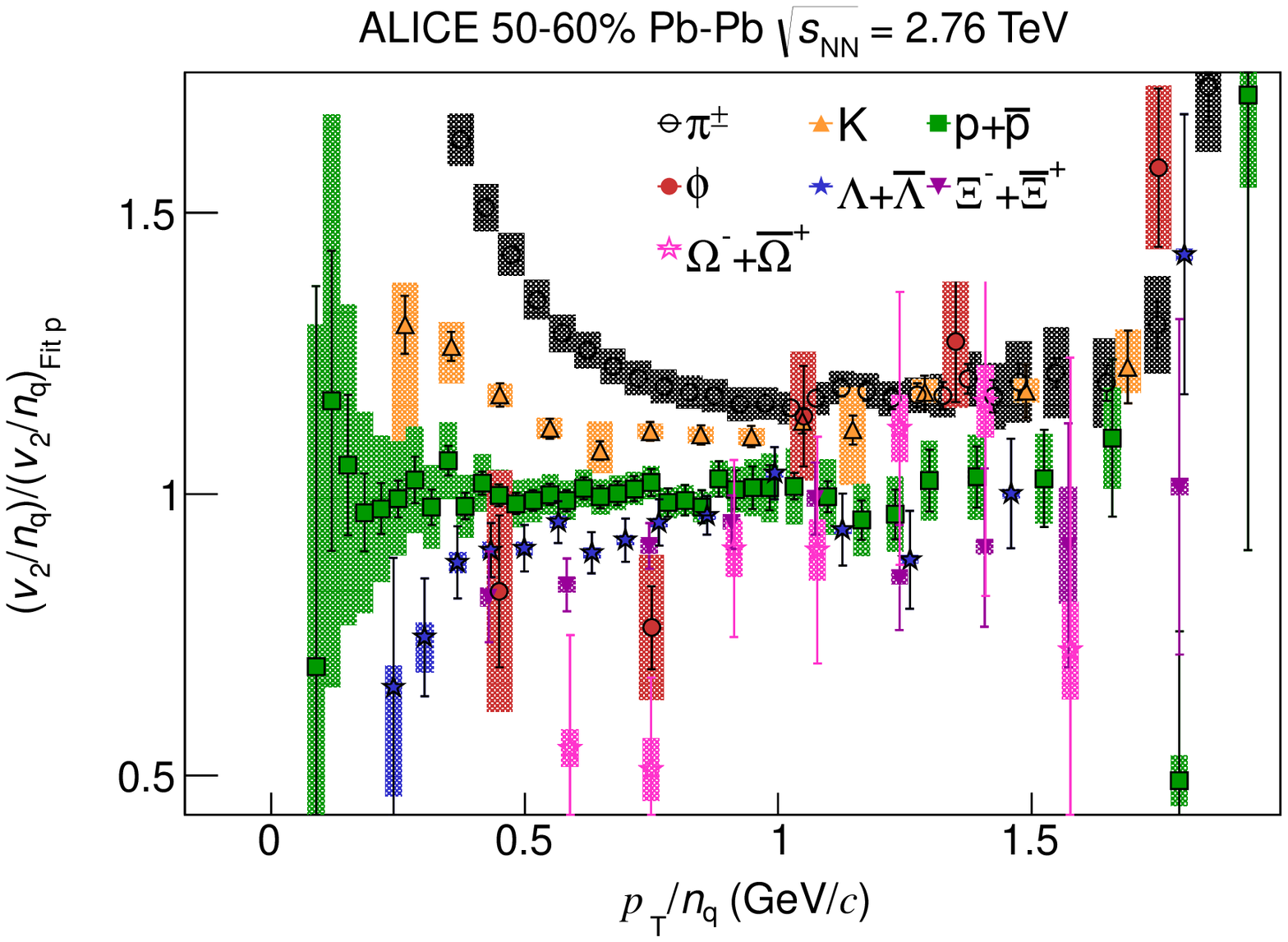}
  \captionof{figure}{The $p_{\rm{T}}/n_q$ dependence of the double ratio of $v_2/n_q$ for every particle 
  species relative to a fit to $v_2/n_q$ of p and $\overline{\mathrm{p}}$ (see text for details) for $\pi^{\pm}$, $\mathrm{K}$, p+$\overline{\mathrm{p}}$, $\phi$, $\Lambda$+$\overline{\mathrm{\Lambda}}$, $\mathrm{\Xi^-}$+$\overline{\mathrm{\Xi}}^+$ and $\mathrm{\Omega}^-$+$\overline{\mathrm{\Omega}}^+$ for the 50--60$\%$ centrality interval in Pb--Pb collisions at $\sqrt{s_{\mathrm{NN}}} = 2.76$~TeV. }
\label{fig:pTScaling2Cent50To60}
\end{center}

\subsection{Plots from Fig.~\ref{fig:mTScaling}}

\begin{center}
  \includegraphics[width=\textwidth]{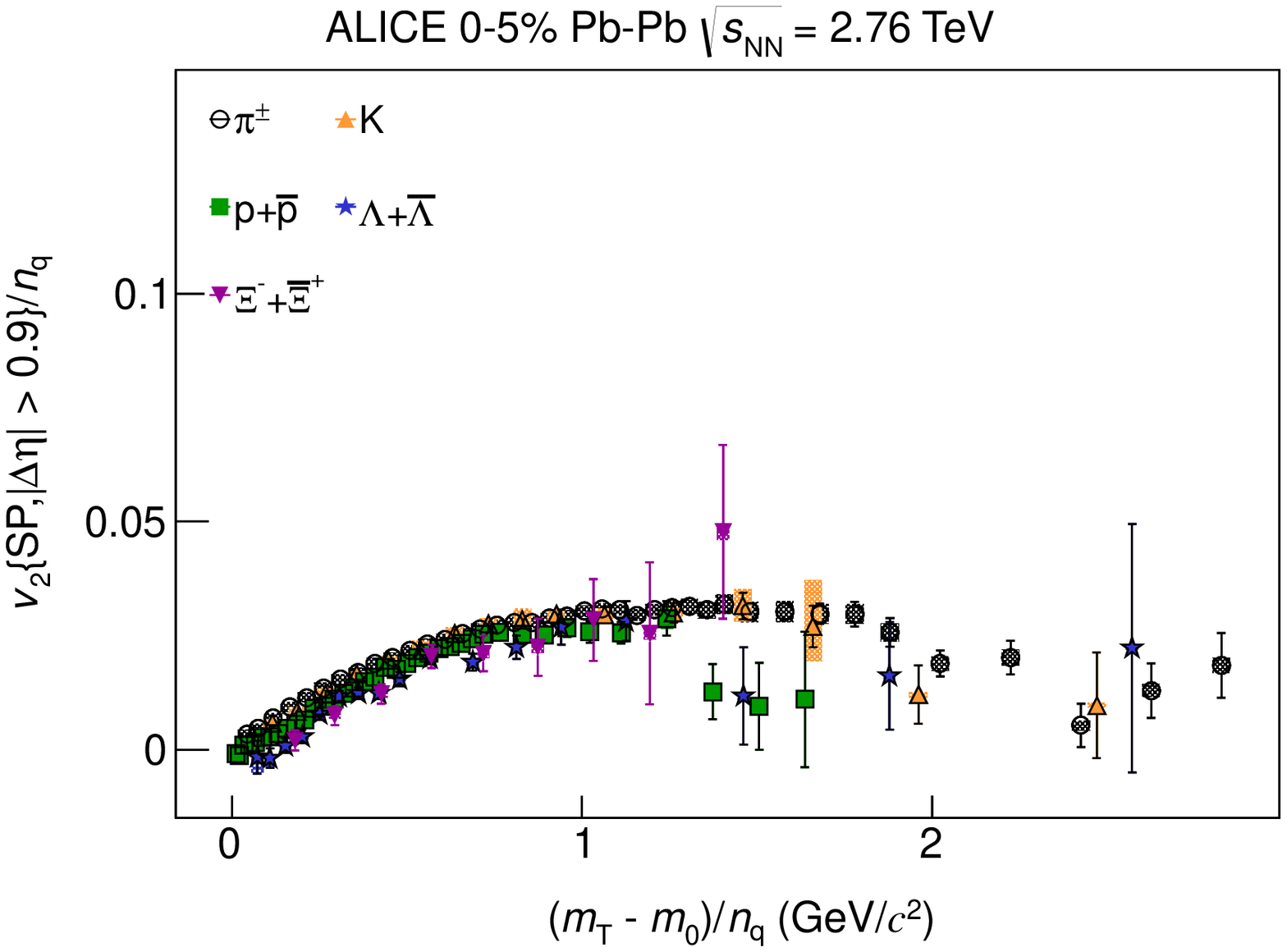}
  \captionof{figure}{The $(m_{\rm{T}} - m_0)/n_q$ dependence of $v_2/n_q$ for
    $\pi^{\pm}$, $\mathrm{K}$, p+$\overline{\mathrm{p}}$, 
  $\phi$, $\Lambda$+$\overline{\mathrm{\Lambda}}$, $\mathrm{\Xi^-}$+$\overline{\mathrm{\Xi}}^+$ and $\mathrm{\Omega}^-$+$\overline{\mathrm{\Omega}}^+$ for the 0--5$\%$ centrality interval in 
     Pb--Pb collisions at $\sqrt{s_{\mathrm{NN}}} = 2.76$~TeV.}
\label{fig:mTScalingCent0To5}
\end{center}

\begin{center}
  \includegraphics[width=\textwidth]{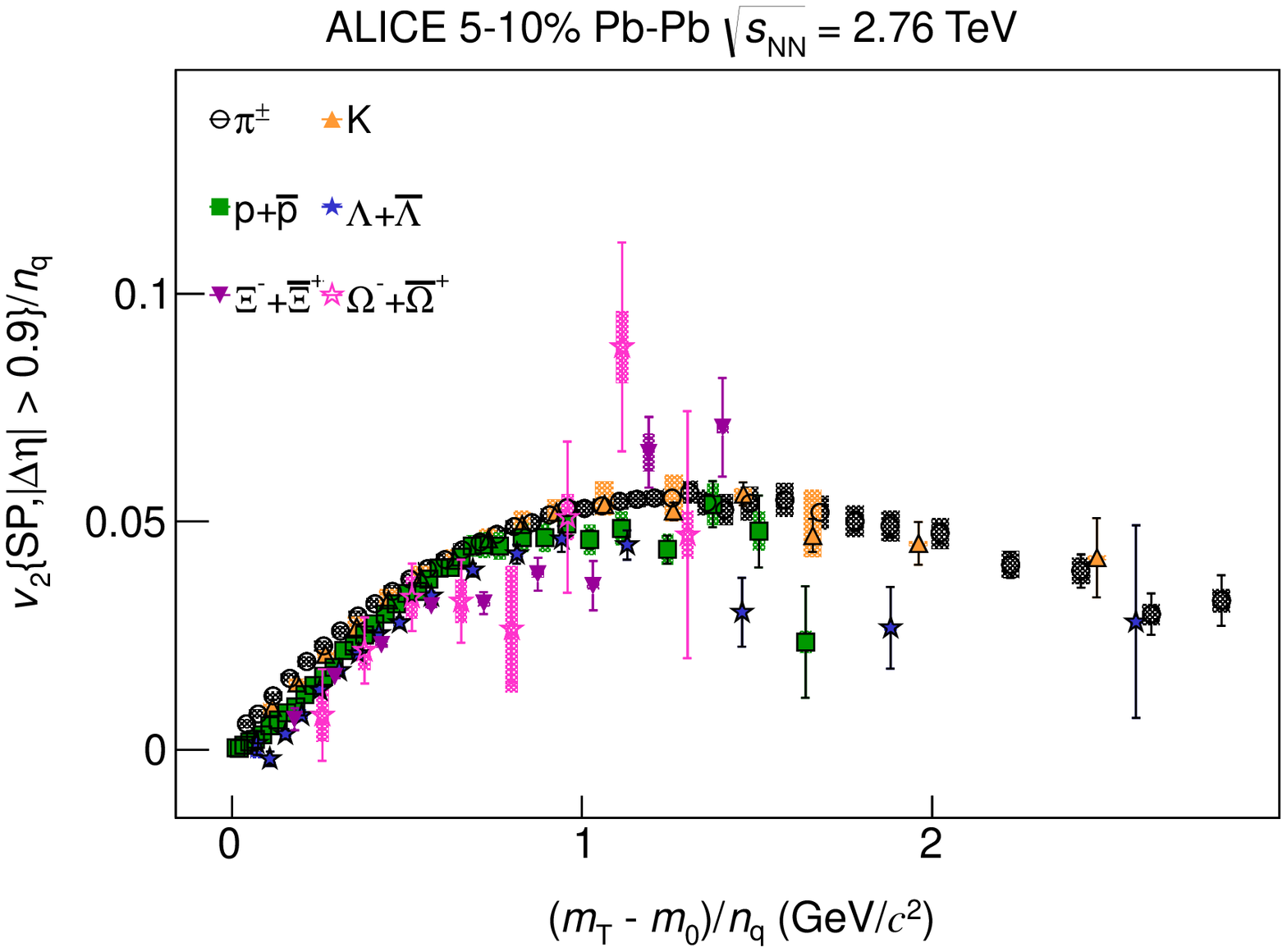}
  \captionof{figure}{The $(m_{\rm{T}} - m_0)/n_q$ dependence of $v_2/n_q$ for
    $\pi^{\pm}$, $\mathrm{K}$, p+$\overline{\mathrm{p}}$, 
  $\phi$, $\Lambda$+$\overline{\mathrm{\Lambda}}$, $\mathrm{\Xi^-}$+$\overline{\mathrm{\Xi}}^+$ and $\mathrm{\Omega}^-$+$\overline{\mathrm{\Omega}}^+$ for the 5--10$\%$ centrality interval in 
     Pb--Pb collisions at $\sqrt{s_{\mathrm{NN}}} = 2.76$~TeV.}
\label{fig:mTScalingCent5To10}
\end{center}

\begin{center}
  \includegraphics[width=\textwidth]{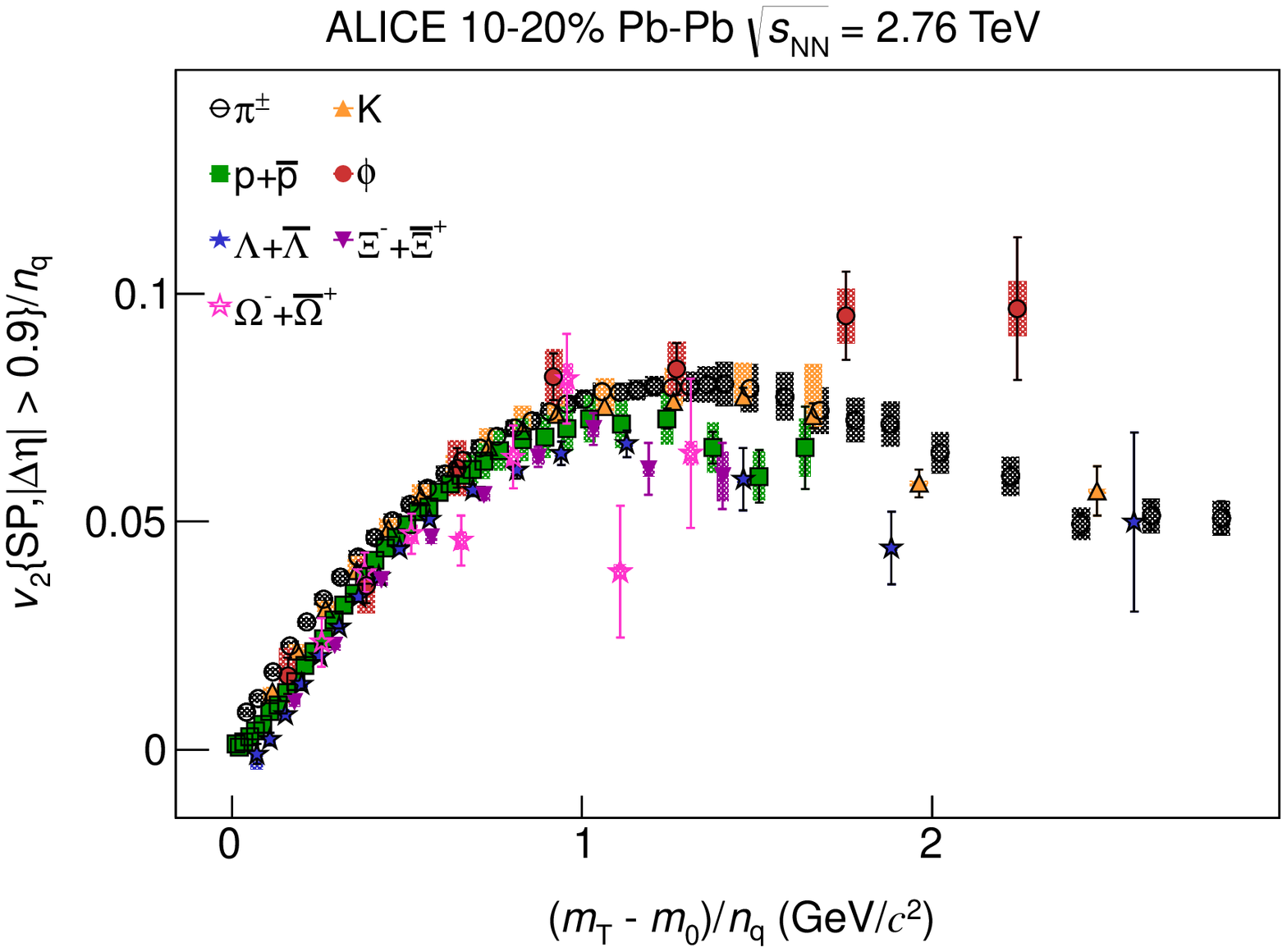}
  \captionof{figure}{The $(m_{\rm{T}} - m_0)/n_q$ dependence of $v_2/n_q$ for
    $\pi^{\pm}$, $\mathrm{K}$, p+$\overline{\mathrm{p}}$, 
  $\phi$, $\Lambda$+$\overline{\mathrm{\Lambda}}$, $\mathrm{\Xi^-}$+$\overline{\mathrm{\Xi}}^+$ and $\mathrm{\Omega}^-$+$\overline{\mathrm{\Omega}}^+$ for the 10--20$\%$ centrality interval in 
     Pb--Pb collisions at $\sqrt{s_{\mathrm{NN}}} = 2.76$~TeV.}
\label{fig:mTScalingCent10To20}
\end{center}

\begin{center}
  \includegraphics[width=\textwidth]{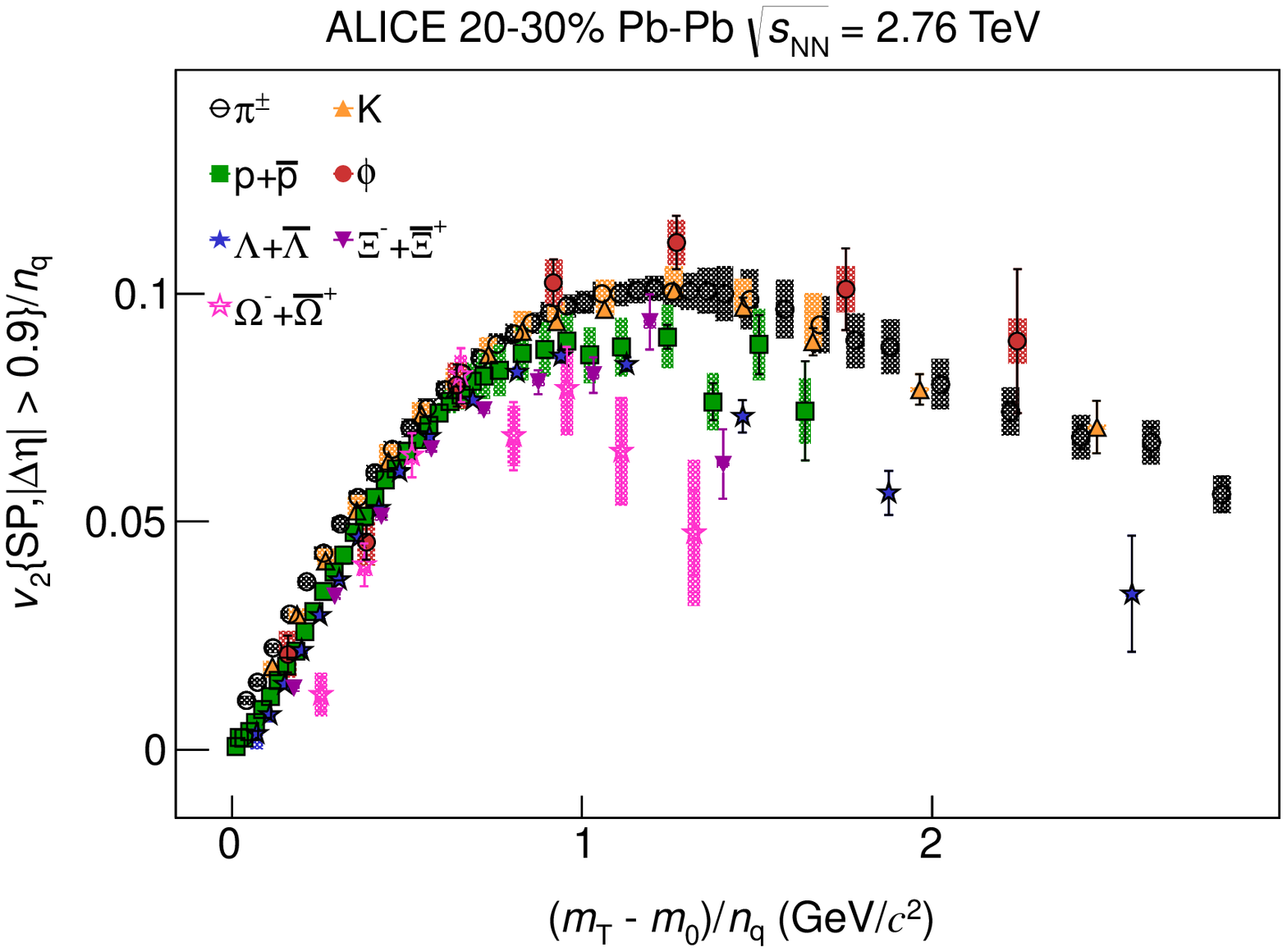}
  \captionof{figure}{The $(m_{\rm{T}} - m_0)/n_q$ dependence of $v_2/n_q$ for
    $\pi^{\pm}$, $\mathrm{K}$, p+$\overline{\mathrm{p}}$, 
  $\phi$, $\Lambda$+$\overline{\mathrm{\Lambda}}$, $\mathrm{\Xi^-}$+$\overline{\mathrm{\Xi}}^+$ and $\mathrm{\Omega}^-$+$\overline{\mathrm{\Omega}}^+$ for the 20--30$\%$ centrality interval in 
     Pb--Pb collisions at $\sqrt{s_{\mathrm{NN}}} = 2.76$~TeV.}
\label{fig:mTScalingCent20To30}
\end{center}

\begin{center}
  \includegraphics[width=\textwidth]{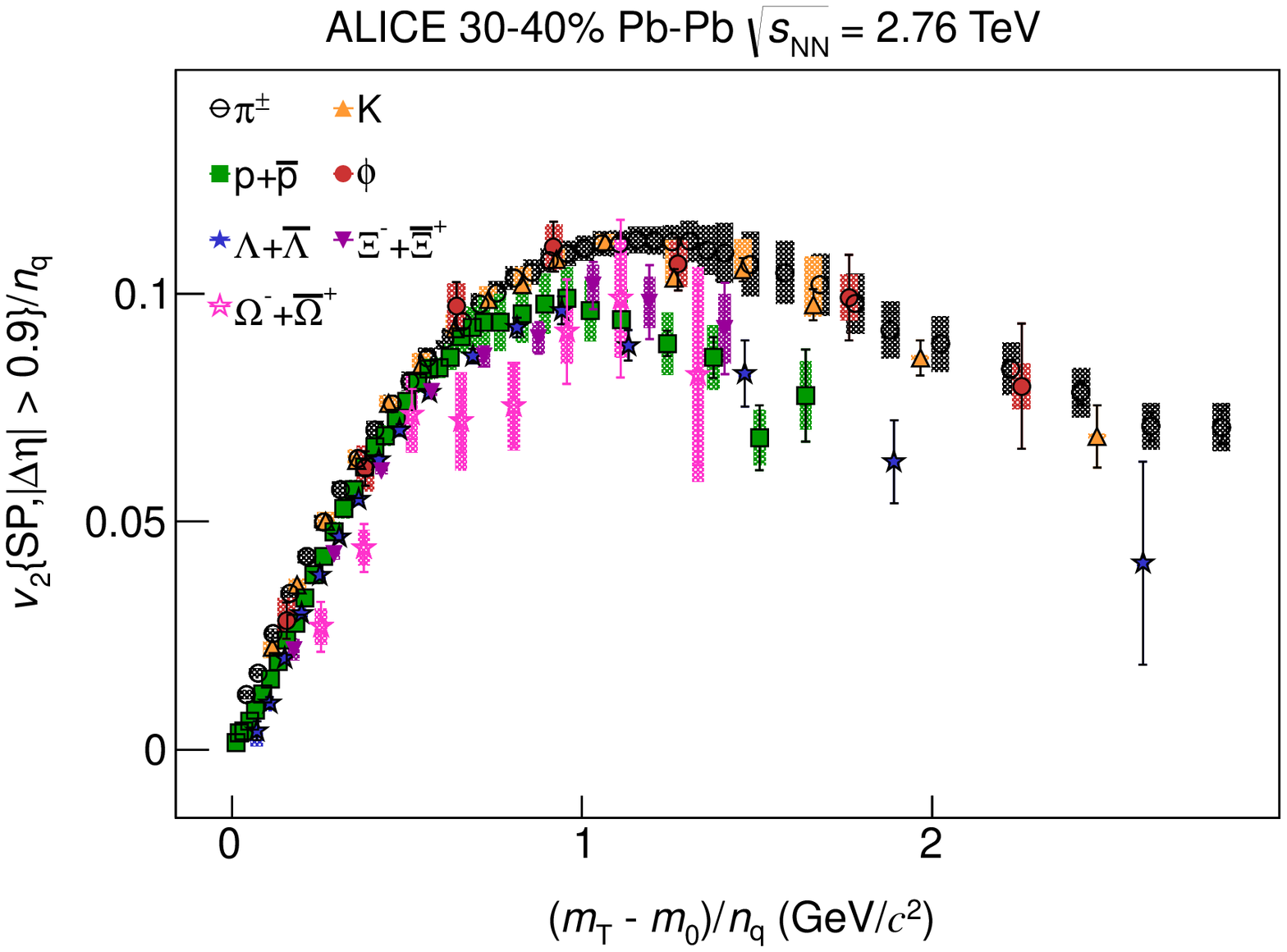}
  \captionof{figure}{The $(m_{\rm{T}} - m_0)/n_q$ dependence of $v_2/n_q$ for
    $\pi^{\pm}$, $\mathrm{K}$, p+$\overline{\mathrm{p}}$, 
  $\phi$, $\Lambda$+$\overline{\mathrm{\Lambda}}$, $\mathrm{\Xi^-}$+$\overline{\mathrm{\Xi}}^+$ and $\mathrm{\Omega}^-$+$\overline{\mathrm{\Omega}}^+$ for the 30--40$\%$ centrality interval in 
     Pb--Pb collisions at $\sqrt{s_{\mathrm{NN}}} = 2.76$~TeV.}
\label{fig:mTScalingCent30To40}
\end{center}

\begin{center}
  \includegraphics[width=\textwidth]{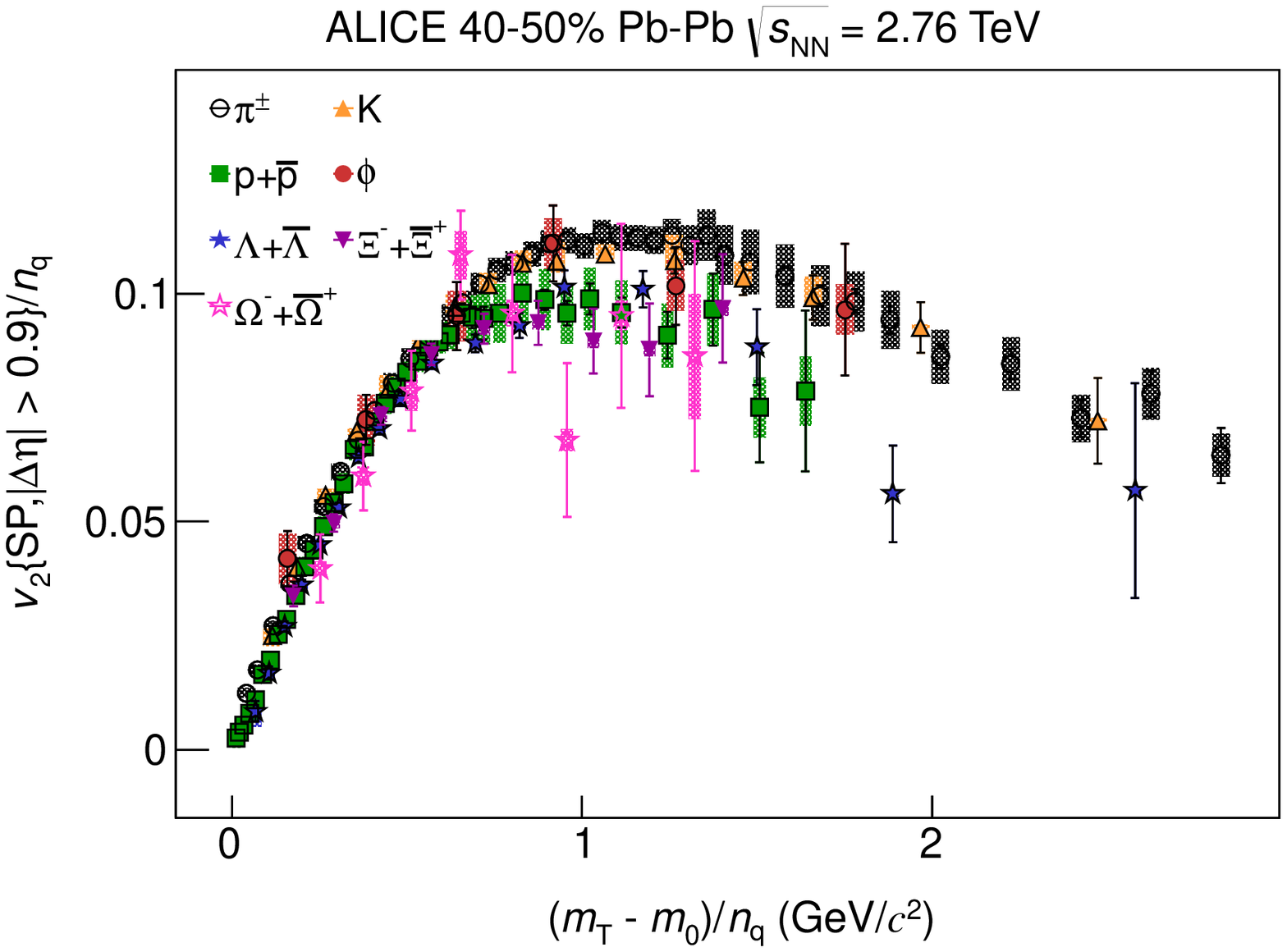}
  \captionof{figure}{The $(m_{\rm{T}} - m_0)/n_q$ dependence of $v_2/n_q$ for
    $\pi^{\pm}$, $\mathrm{K}$, p+$\overline{\mathrm{p}}$, 
  $\phi$, $\Lambda$+$\overline{\mathrm{\Lambda}}$, $\mathrm{\Xi^-}$+$\overline{\mathrm{\Xi}}^+$ and $\mathrm{\Omega}^-$+$\overline{\mathrm{\Omega}}^+$ for the 40--50$\%$ centrality interval in 
     Pb--Pb collisions at $\sqrt{s_{\mathrm{NN}}} = 2.76$~TeV.}
\label{fig:mTScalingCent40To50}
\end{center}

\begin{center}
  \includegraphics[width=\textwidth]{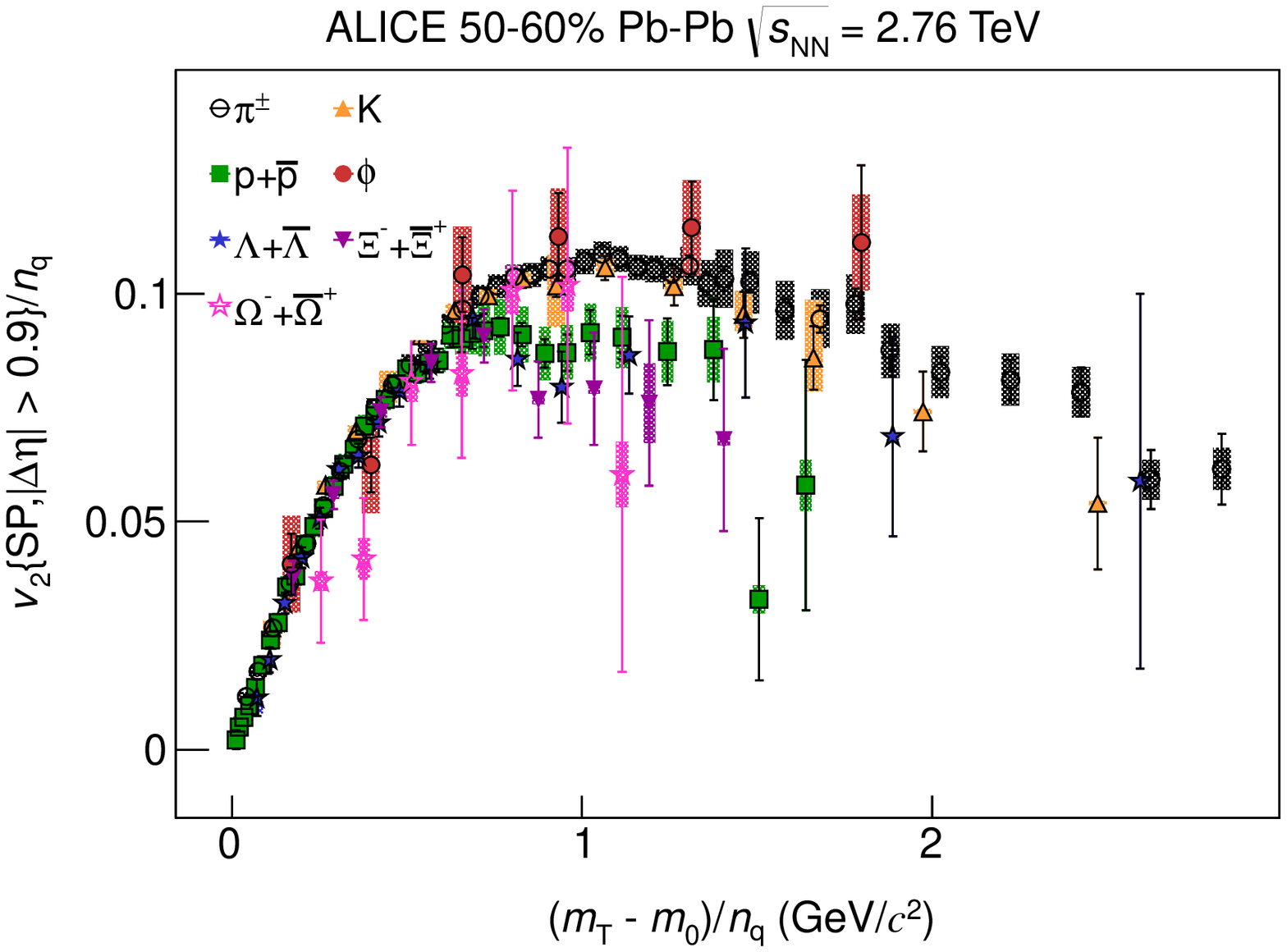}
  \captionof{figure}{The $(m_{\rm{T}} - m_0)/n_q$ dependence of $v_2/n_q$ for
    $\pi^{\pm}$, $\mathrm{K}$, p+$\overline{\mathrm{p}}$, 
  $\phi$, $\Lambda$+$\overline{\mathrm{\Lambda}}$, $\mathrm{\Xi^-}$+$\overline{\mathrm{\Xi}}^+$ and $\mathrm{\Omega}^-$+$\overline{\mathrm{\Omega}}^+$ for the 50--60$\%$ centrality interval in 
     Pb--Pb collisions at $\sqrt{s_{\mathrm{NN}}} = 2.76$~TeV.}
\label{fig:mTScalingCent50To60}
\end{center}

\subsection{Plots from Fig.~\ref{fig:mTScaling2}}

\begin{center}
  \includegraphics[width=\textwidth]{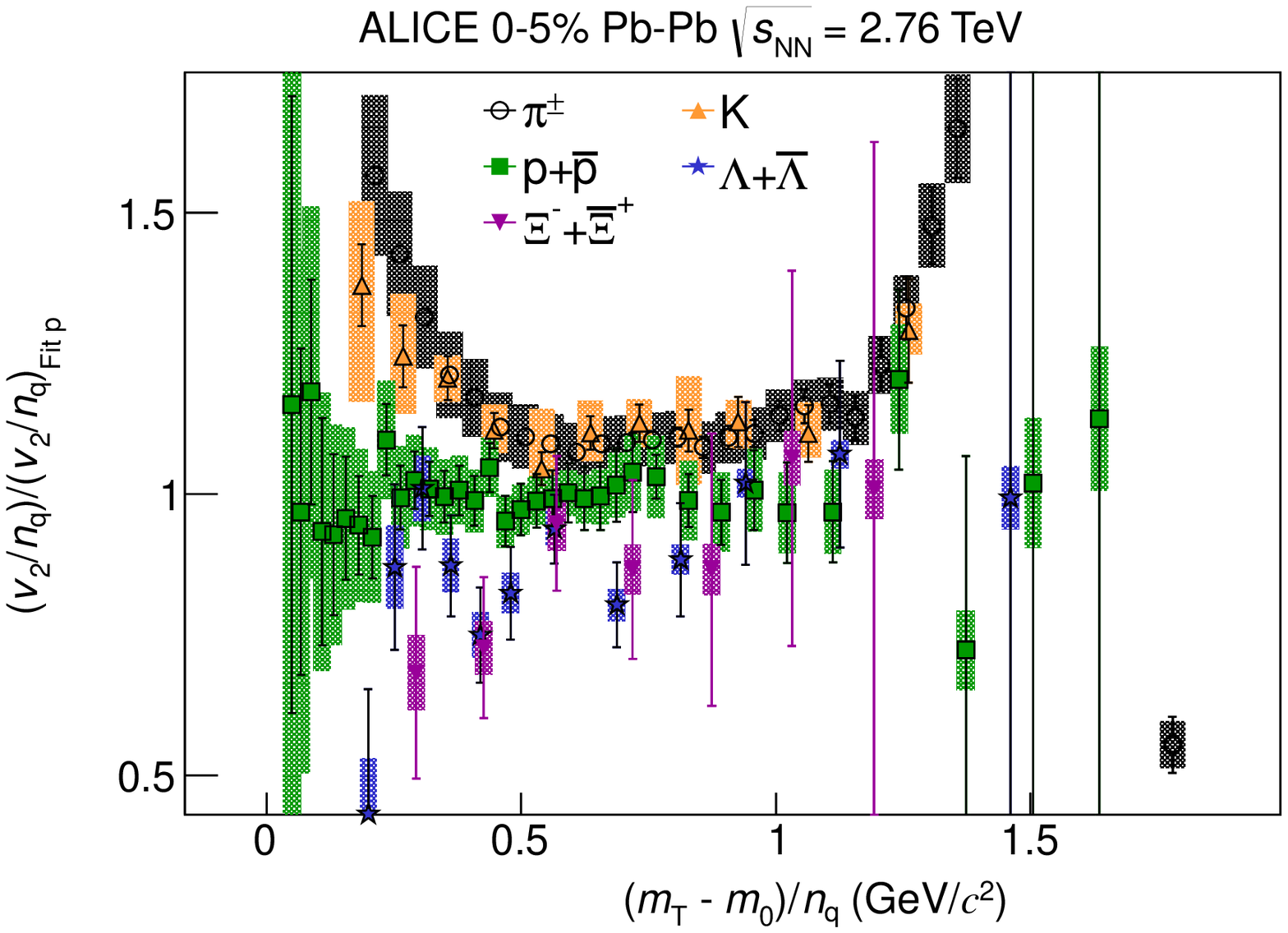}
  \captionof{figure}{The $(m_{\rm{T}} - m_0)/n_q$ dependence of the double ratio of $v_2/n_q$ for every particle species
    relative to a fit to $v_2/n_q$ of p and $\overline{\mathrm{p}}$ (see text for
    details) for
    $\pi^{\pm}$, $\mathrm{K}$, p+$\overline{\mathrm{p}}$, 
  $\phi$, $\Lambda$+$\overline{\mathrm{\Lambda}}$, $\mathrm{\Xi^-}$+$\overline{\mathrm{\Xi}}^+$ and $\mathrm{\Omega}^-$+$\overline{\mathrm{\Omega}}^+$ for the 0--5$\%$ centrality interval in 
    Pb--Pb collisions at $\sqrt{s_{\mathrm{NN}}} = 2.76$~TeV.}
\label{fig:mTScaling2Cent0To5}
\end{center}

\begin{center}
  \includegraphics[width=\textwidth]{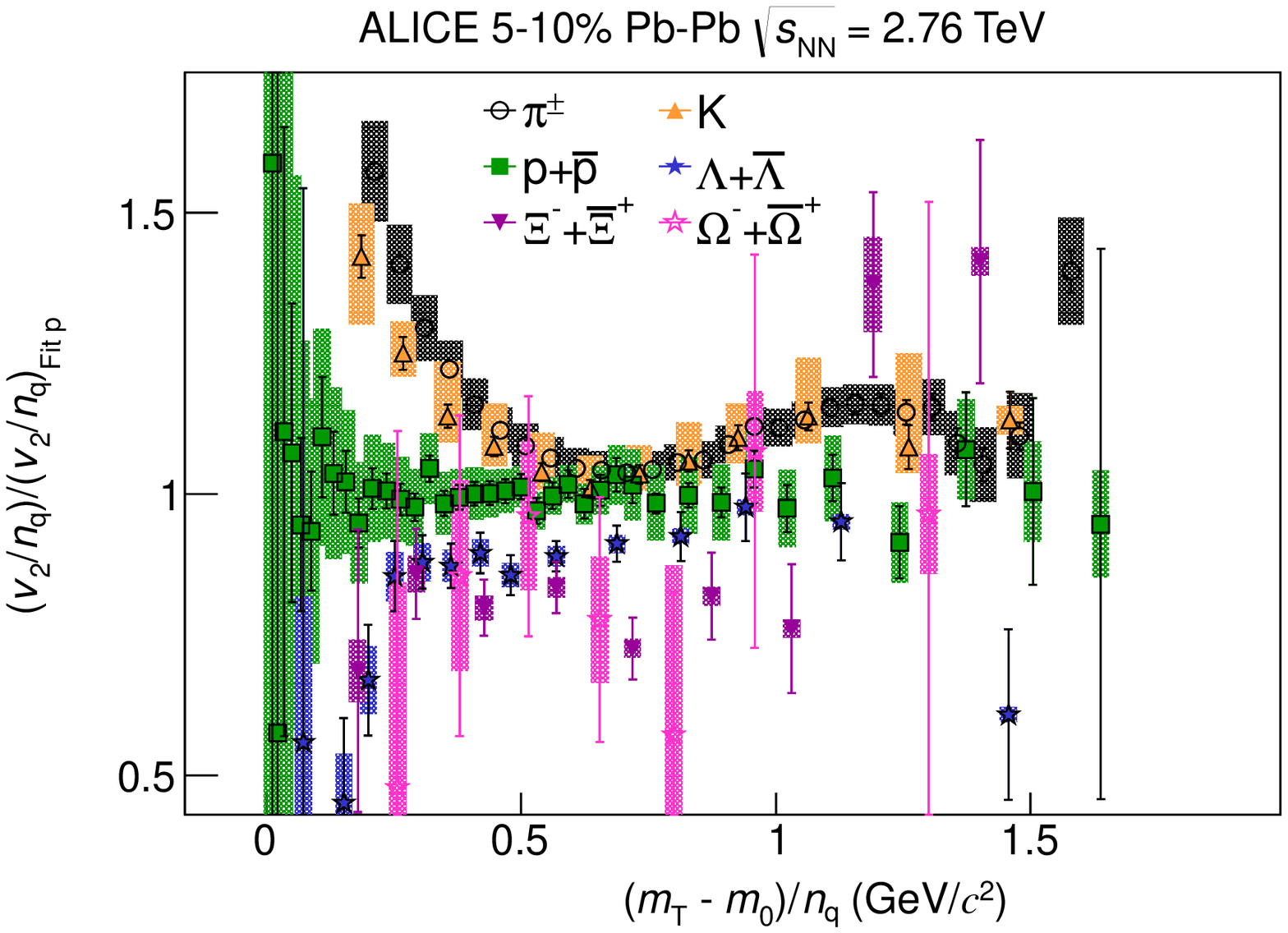}
  \captionof{figure}{The $(m_{\rm{T}} - m_0)/n_q$ dependence of the double ratio of $v_2/n_q$ for every particle species
    relative to a fit to $v_2/n_q$ of p and $\overline{\mathrm{p}}$ (see text for
    details) for
    $\pi^{\pm}$, $\mathrm{K}$, p+$\overline{\mathrm{p}}$, 
  $\phi$, $\Lambda$+$\overline{\mathrm{\Lambda}}$, $\mathrm{\Xi^-}$+$\overline{\mathrm{\Xi}}^+$ and $\mathrm{\Omega}^-$+$\overline{\mathrm{\Omega}}^+$ for the 5--10$\%$ centrality interval in 
    Pb--Pb collisions at $\sqrt{s_{\mathrm{NN}}} = 2.76$~TeV.}
\label{fig:mTScaling2Cent5To10}
\end{center}

\begin{center}
  \includegraphics[width=\textwidth]{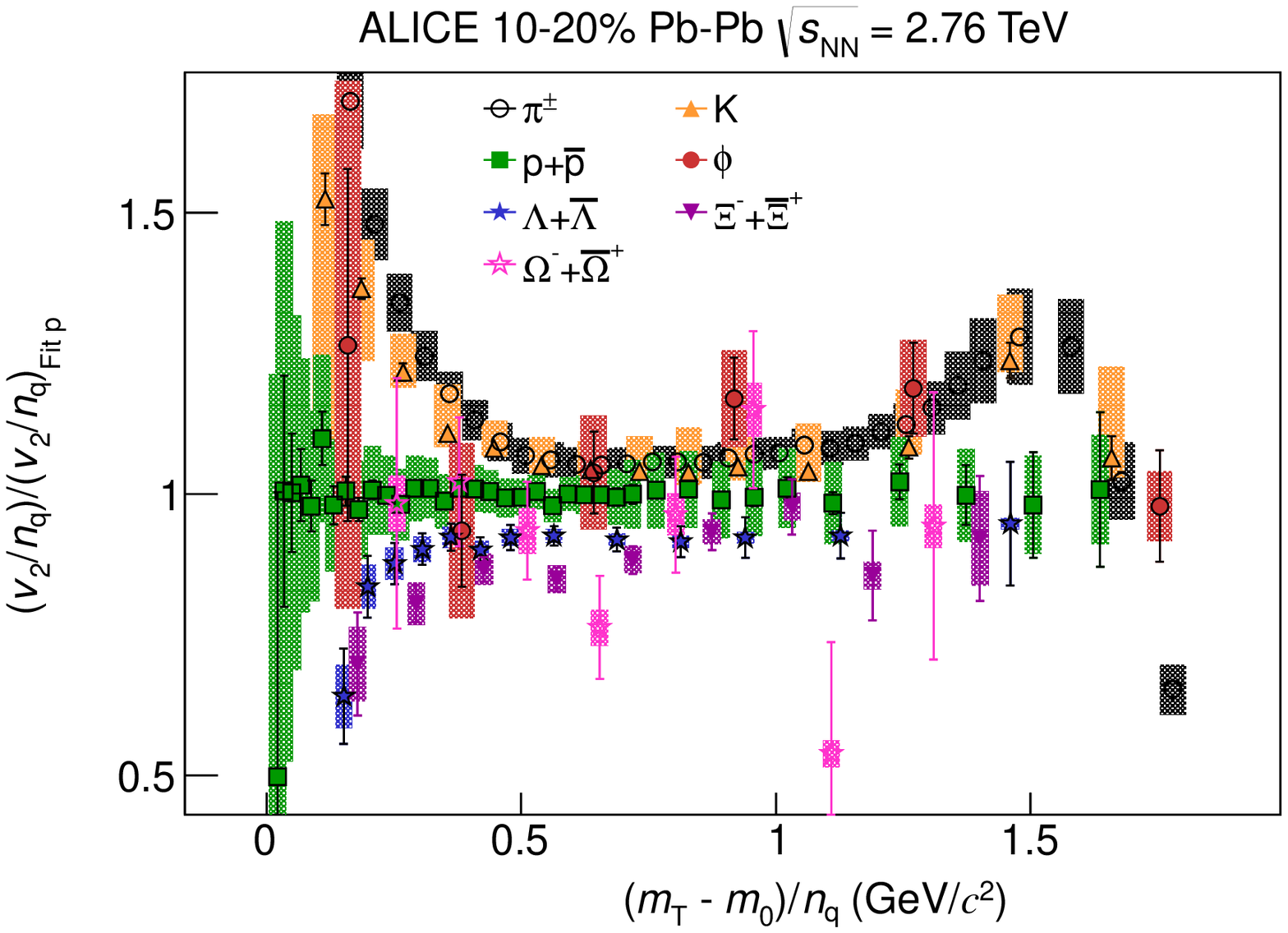}
  \captionof{figure}{The $(m_{\rm{T}} - m_0)/n_q$ dependence of the double ratio of $v_2/n_q$ for every particle species
    relative to a fit to $v_2/n_q$ of p and $\overline{\mathrm{p}}$ (see text for
    details) for
    $\pi^{\pm}$, $\mathrm{K}$, p+$\overline{\mathrm{p}}$, 
  $\phi$, $\Lambda$+$\overline{\mathrm{\Lambda}}$, $\mathrm{\Xi^-}$+$\overline{\mathrm{\Xi}}^+$ and $\mathrm{\Omega}^-$+$\overline{\mathrm{\Omega}}^+$ for the 10--20$\%$ centrality interval in 
    Pb--Pb collisions at $\sqrt{s_{\mathrm{NN}}} = 2.76$~TeV.}
\label{fig:mTScaling2Cent10To20}
\end{center}

\begin{center}
  \includegraphics[width=\textwidth]{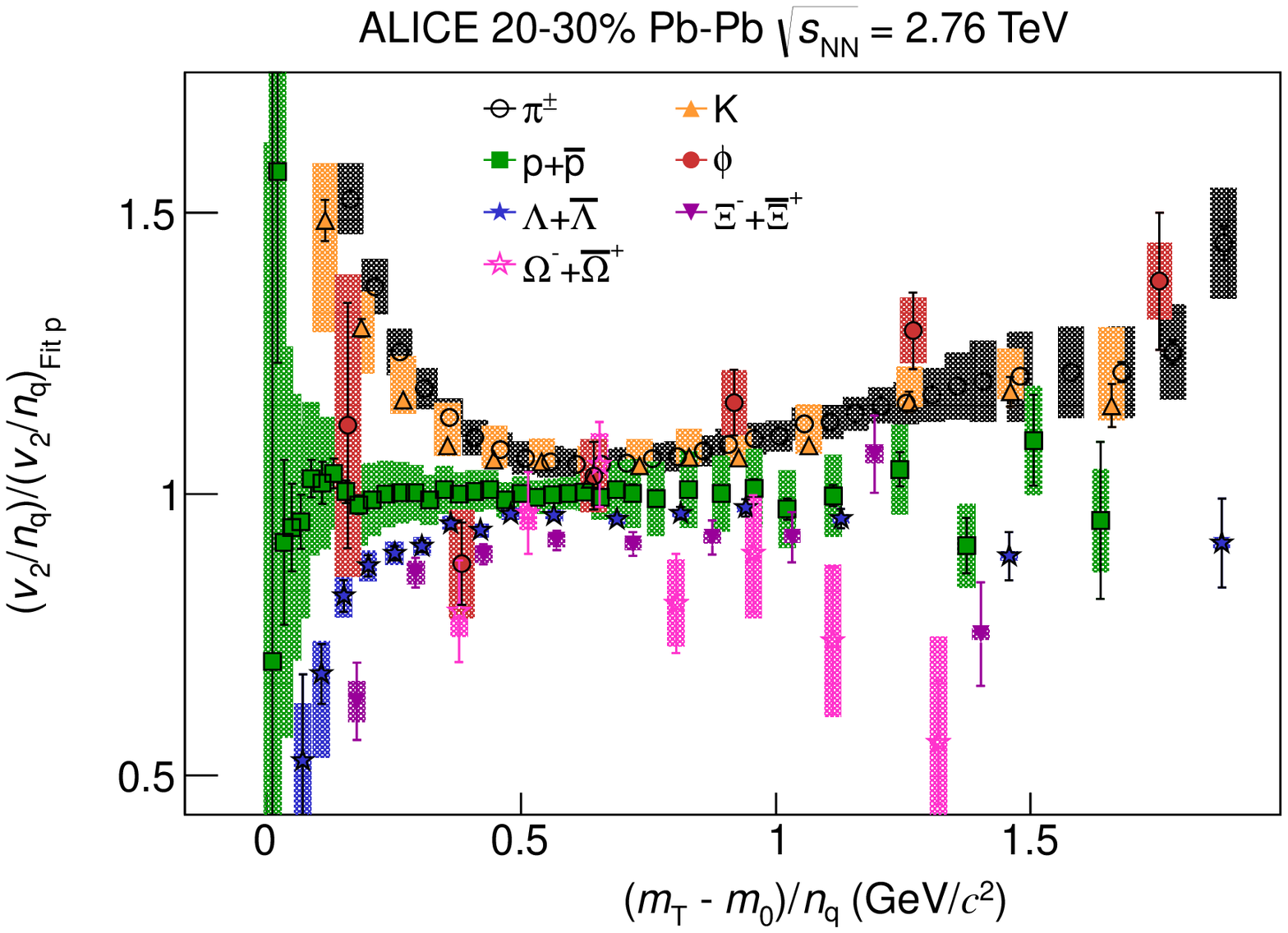}
  \captionof{figure}{The $(m_{\rm{T}} - m_0)/n_q$ dependence of the double ratio of $v_2/n_q$ for every particle species
    relative to a fit to $v_2/n_q$ of p and $\overline{\mathrm{p}}$ (see text for
    details) for
    $\pi^{\pm}$, $\mathrm{K}$, p+$\overline{\mathrm{p}}$, 
  $\phi$, $\Lambda$+$\overline{\mathrm{\Lambda}}$, $\mathrm{\Xi^-}$+$\overline{\mathrm{\Xi}}^+$ and $\mathrm{\Omega}^-$+$\overline{\mathrm{\Omega}}^+$ for the 20--30$\%$ centrality interval in 
    Pb--Pb collisions at $\sqrt{s_{\mathrm{NN}}} = 2.76$~TeV.}
\label{fig:mTScaling2Cent20To30}
\end{center}

\begin{center}
  \includegraphics[width=\textwidth]{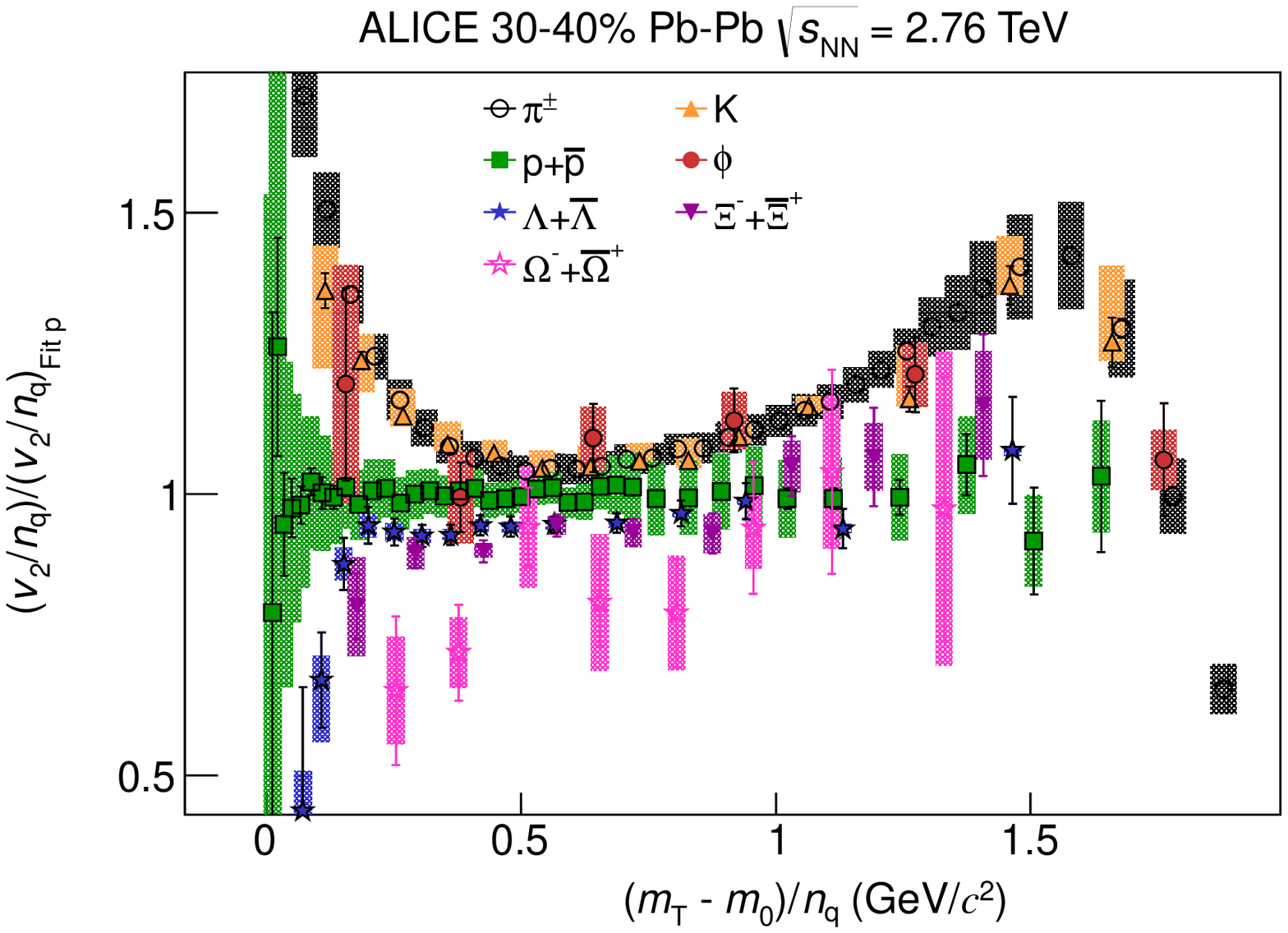}
  \captionof{figure}{The $(m_{\rm{T}} - m_0)/n_q$ dependence of the double ratio of $v_2/n_q$ for every particle species
    relative to a fit to $v_2/n_q$ of p and $\overline{\mathrm{p}}$ (see text for
    details) for
    $\pi^{\pm}$, $\mathrm{K}$, p+$\overline{\mathrm{p}}$, 
  $\phi$, $\Lambda$+$\overline{\mathrm{\Lambda}}$, $\mathrm{\Xi^-}$+$\overline{\mathrm{\Xi}}^+$ and $\mathrm{\Omega}^-$+$\overline{\mathrm{\Omega}}^+$ for the 30--40$\%$ centrality interval in 
    Pb--Pb collisions at $\sqrt{s_{\mathrm{NN}}} = 2.76$~TeV.}
\label{fig:mTScaling2Cent30To40}
\end{center}

\begin{center}
  \includegraphics[width=\textwidth]{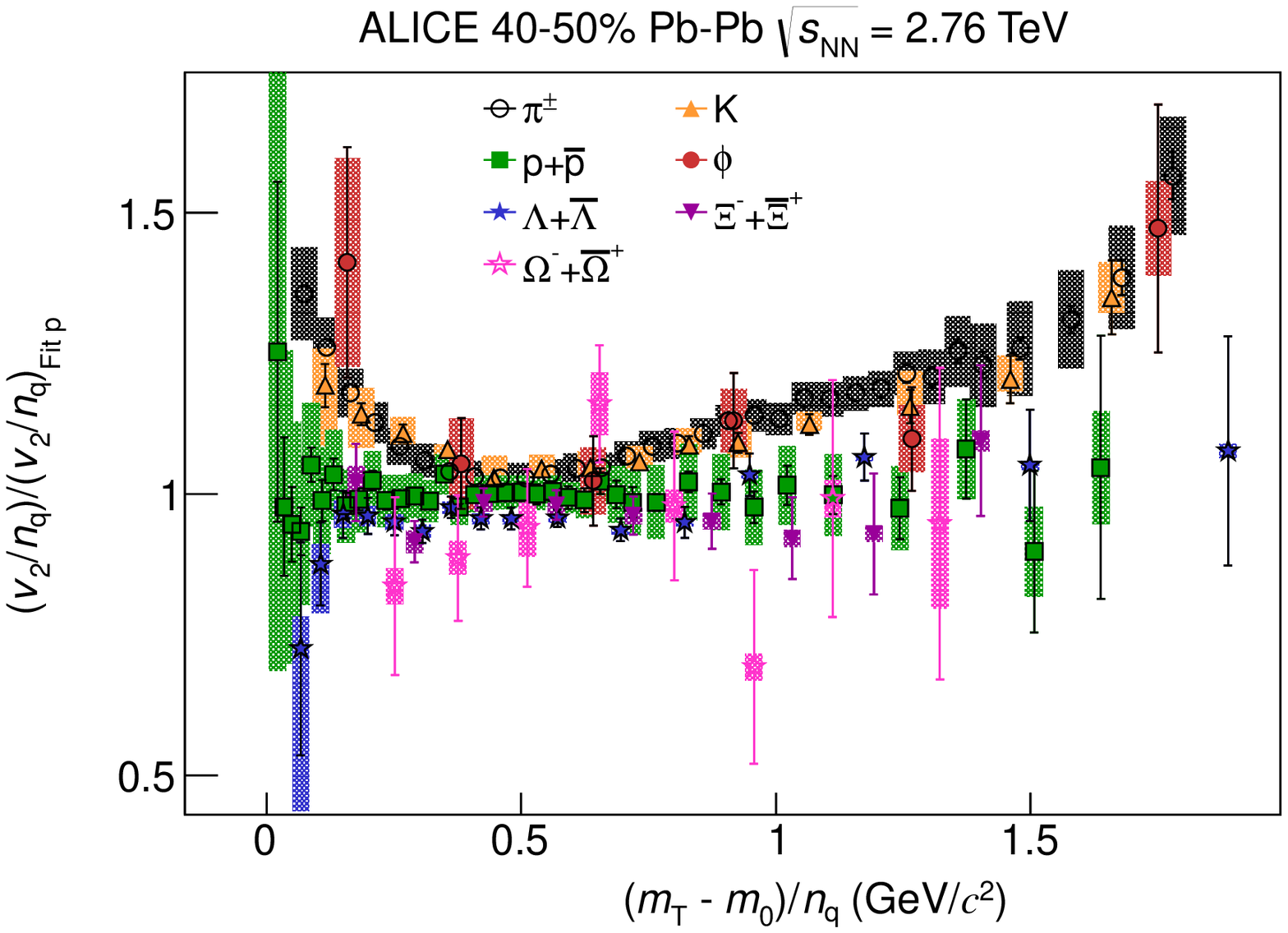}
  \captionof{figure}{The $(m_{\rm{T}} - m_0)/n_q$ dependence of the double ratio of $v_2/n_q$ for every particle species
    relative to a fit to $v_2/n_q$ of p and $\overline{\mathrm{p}}$ (see text for
    details) for
    $\pi^{\pm}$, $\mathrm{K}$, p+$\overline{\mathrm{p}}$, 
  $\phi$, $\Lambda$+$\overline{\mathrm{\Lambda}}$, $\mathrm{\Xi^-}$+$\overline{\mathrm{\Xi}}^+$ and $\mathrm{\Omega}^-$+$\overline{\mathrm{\Omega}}^+$ for the 40--50$\%$ centrality interval in 
    Pb--Pb collisions at $\sqrt{s_{\mathrm{NN}}} = 2.76$~TeV.}
\label{fig:mTScaling2Cent40To50}
\end{center}

\begin{center}
  \includegraphics[width=\textwidth]{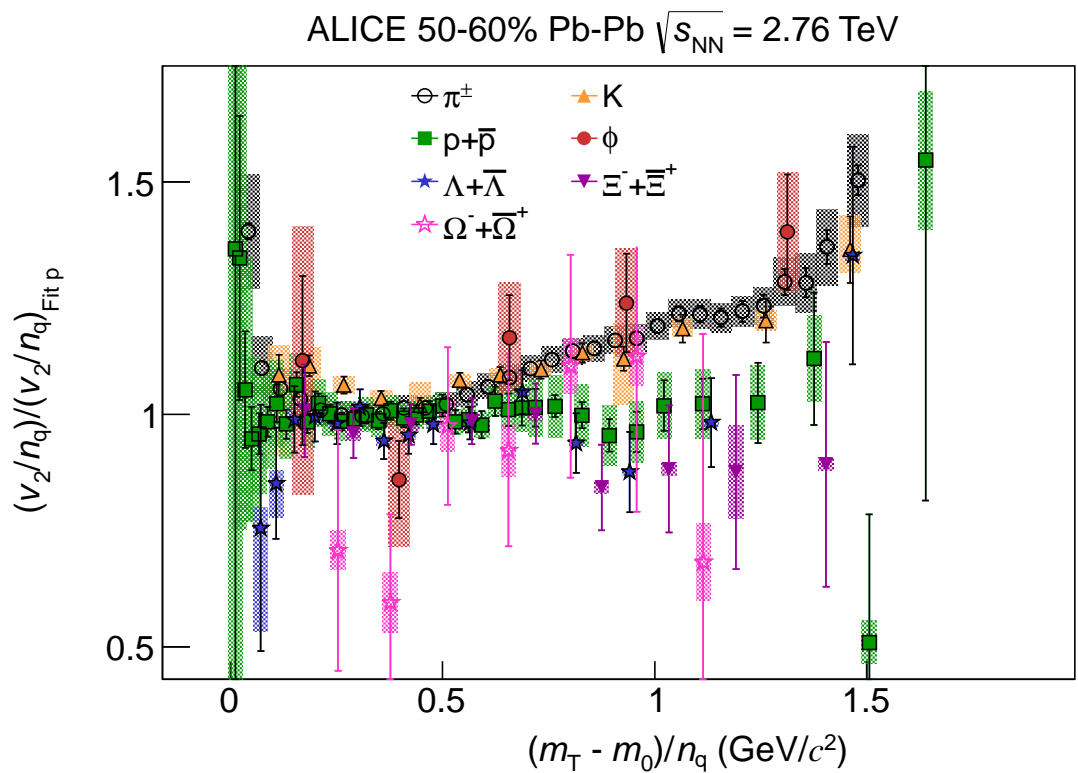}
  \captionof{figure}{The $(m_{\rm{T}} - m_0)/n_q$ dependence of the double ratio of $v_2/n_q$ for every particle species
    relative to a fit to $v_2/n_q$ of p and $\overline{\mathrm{p}}$ (see text for
    details) for
    $\pi^{\pm}$, $\mathrm{K}$, p+$\overline{\mathrm{p}}$, 
  $\phi$, $\Lambda$+$\overline{\mathrm{\Lambda}}$, $\mathrm{\Xi^-}$+$\overline{\mathrm{\Xi}}^+$ and $\mathrm{\Omega}^-$+$\overline{\mathrm{\Omega}}^+$ for the 50--60$\%$ centrality interval in 
    Pb--Pb collisions at $\sqrt{s_{\mathrm{NN}}} = 2.76$~TeV.}
\label{fig:mTScaling2Cent50To60}
\end{center}

%% file: AliceAuthorlistCERNPREP.tex


\begingroup
\small
\begin{flushleft}
B.~Abelev\Irefn{org69}\And
J.~Adam\Irefn{org37}\And
D.~Adamov\'{a}\Irefn{org77}\And
M.M.~Aggarwal\Irefn{org81}\And
M.~Agnello\Irefn{org105}\textsuperscript{,}\Irefn{org88}\And
A.~Agostinelli\Irefn{org26}\And
N.~Agrawal\Irefn{org44}\And
Z.~Ahammed\Irefn{org124}\And
N.~Ahmad\Irefn{org18}\And
I.~Ahmed\Irefn{org15}
% ...
\renewcommand\labelenumi{\textsuperscript{\theenumi}~}

\section*{Affiliation notes}
\renewcommand\theenumi{\roman{enumi}}
\begin{Authlist}
\item \Adef{0}Deceased
\item \Adef{idp1103456}{Also at: St. Petersburg State Polytechnical University}
\item \Adef{idp3006624}{Also at: Department of Applied Physics, Aligarh Muslim University, Aligarh, India}
\item \Adef{idp3689616}{Also at: M.V. Lomonosov Moscow State University, D.V. Skobeltsyn Institute of Nuclear Physics, Moscow, Russia}
\item \Adef{idp3933376}{Also at: University of Belgrade, Faculty of Physics and "Vin\v{c}a" Institute of Nuclear Sciences, Belgrade, Serbia}
\item \Adef{idp4238864}{Permanent Address: Permanent Address: Konkuk University, Seoul, Korea}
\item \Adef{idp4779200}{Also at: Institute of Theoretical Physics, University of Wroclaw, Wroclaw, Poland}
\item \Adef{idp5696608}{Also at: University of Kansas, Lawrence, KS, United States}
\end{Authlist}

\section*{Collaboration Institutes}
\renewcommand\theenumi{\arabic{enumi}~}
\begin{Authlist}

\item \Idef{org1}A.I. Alikhanyan National Science Laboratory (Yerevan Physics Institute) Foundation, Yerevan, Armenia
\item \Idef{org2}Benem\'{e}rita Universidad Aut\'{o}noma de Puebla, Puebla, Mexico
\item \Idef{org3}Bogolyubov Institute for Theoretical Physics, Kiev, Ukraine
\item \Idef{org4}Bose Institute, Department of Physics and Centre for Astroparticle Physics and Space Science (CAPSS), Kolkata, India
\item \Idef{org5}Budker Institute for Nuclear Physics, Novosibirsk, Russia
\item \Idef{org6}California Polytechnic State University, San Luis Obispo, CA, United States
\item \Idef{org7}Central China Normal University, Wuhan, China
\item \Idef{org8}Centre de Calcul de l'IN2P3, Villeurbanne, France
\item \Idef{org9}Centro de Aplicaciones Tecnol\'{o}gicas y Desarrollo Nuclear (CEADEN), Havana, Cuba
\item \Idef{org10}Centro de Investigaciones Energ\'{e}ticas Medioambientales y Tecnol\'{o}gicas (CIEMAT), Madrid, Spain
\item \Idef{org11}Centro de Investigaci\'{o}n y de Estudios Avanzados (CINVESTAV), Mexico City and M\'{e}rida, Mexico
\item \Idef{org12}Centro Fermi - Museo Storico della Fisica e Centro Studi e Ricerche ``Enrico Fermi'', Rome, Italy
\item \Idef{org13}Chicago State University, Chicago, USA
\item \Idef{org14}Commissariat \`{a} l'Energie Atomique, IRFU, Saclay, France
\item \Idef{org15}COMSATS Institute of Information Technology (CIIT), Islamabad, Pakistan
\item \Idef{org16}Departamento de F\'{\i}sica de Part\'{\i}culas and IGFAE, Universidad de Santiago de Compostela, Santiago de Compostela, Spain
\item \Idef{org17}Department of Physics and Technology, University of Bergen, Bergen, Norway
\item \Idef{org18}Department of Physics, Aligarh Muslim University, Aligarh, India
\item \Idef{org19}Department of Physics, Ohio State University, Columbus, OH, United States
\item \Idef{org20}Department of Physics, Sejong University, Seoul, South Korea
\item \Idef{org21}Department of Physics, University of Oslo, Oslo, Norway
\item \Idef{org22}Dipartimento di Fisica dell'Universit\`{a} 'La Sapienza' and Sezione INFN Rome, Italy
\item \Idef{org23}Dipartimento di Fisica dell'Universit\`{a} and Sezione INFN, Cagliari, Italy
\item \Idef{org24}Dipartimento di Fisica dell'Universit\`{a} and Sezione INFN, Trieste, Italy
\item \Idef{org25}Dipartimento di Fisica dell'Universit\`{a} and Sezione INFN, Turin, Italy
\item \Idef{org26}Dipartimento di Fisica e Astronomia dell'Universit\`{a} and Sezione INFN, Bologna, Italy
\item \Idef{org27}Dipartimento di Fisica e Astronomia dell'Universit\`{a} and Sezione INFN, Catania, Italy
\item \Idef{org28}Dipartimento di Fisica e Astronomia dell'Universit\`{a} and Sezione INFN, Padova, Italy
\item \Idef{org29}Dipartimento di Fisica `E.R.~Caianiello' dell'Universit\`{a} and Gruppo Collegato INFN, Salerno, Italy
\item \Idef{org30}Dipartimento di Scienze e Innovazione Tecnologica dell'Universit\`{a} del  Piemonte Orientale and Gruppo Collegato INFN, Alessandria, Italy
\item \Idef{org31}Dipartimento Interateneo di Fisica `M.~Merlin' and Sezione INFN, Bari, Italy
\item \Idef{org32}Division of Experimental High Energy Physics, University of Lund, Lund, Sweden
\item \Idef{org33}Eberhard Karls Universit\"{a}t T\"{u}bingen, T\"{u}bingen, Germany
\item \Idef{org34}European Organization for Nuclear Research (CERN), Geneva, Switzerland
\item \Idef{org35}Faculty of Engineering, Bergen University College, Bergen, Norway
\item \Idef{org36}Faculty of Mathematics, Physics and Informatics, Comenius University, Bratislava, Slovakia
\item \Idef{org37}Faculty of Nuclear Sciences and Physical Engineering, Czech Technical University in Prague, Prague, Czech Republic
\item \Idef{org38}Faculty of Science, P.J.~\v{S}af\'{a}rik University, Ko\v{s}ice, Slovakia
\item \Idef{org39}Frankfurt Institute for Advanced Studies, Johann Wolfgang Goethe-Universit\"{a}t Frankfurt, Frankfurt, Germany
\item \Idef{org40}Gangneung-Wonju National University, Gangneung, South Korea
\item \Idef{org41}Gauhati University, Department of Physics, Guwahati, India
\item \Idef{org42}Helsinki Institute of Physics (HIP), Helsinki, Finland
\item \Idef{org43}Hiroshima University, Hiroshima, Japan
\item \Idef{org44}Indian Institute of Technology Bombay (IIT), Mumbai, India
\item \Idef{org45}Indian Institute of Technology Indore, Indore (IITI), India
\item \Idef{org46}Institut de Physique Nucl\'eaire d'Orsay (IPNO), Universit\'e Paris-Sud, CNRS-IN2P3, Orsay, France
\item \Idef{org47}Institut f\"{u}r Informatik, Johann Wolfgang Goethe-Universit\"{a}t Frankfurt, Frankfurt, Germany
\item \Idef{org48}Institut f\"{u}r Kernphysik, Johann Wolfgang Goethe-Universit\"{a}t Frankfurt, Frankfurt, Germany
\item \Idef{org49}Institut f\"{u}r Kernphysik, Westf\"{a}lische Wilhelms-Universit\"{a}t M\"{u}nster, M\"{u}nster, Germany
\item \Idef{org50}Institut Pluridisciplinaire Hubert Curien (IPHC), Universit\'{e} de Strasbourg, CNRS-IN2P3, Strasbourg, France
\item \Idef{org51}Institute for Nuclear Research, Academy of Sciences, Moscow, Russia
\item \Idef{org52}Institute for Subatomic Physics of Utrecht University, Utrecht, Netherlands
\item \Idef{org53}Institute for Theoretical and Experimental Physics, Moscow, Russia
\item \Idef{org54}Institute of Experimental Physics, Slovak Academy of Sciences, Ko\v{s}ice, Slovakia
\item \Idef{org55}Institute of Physics, Academy of Sciences of the Czech Republic, Prague, Czech Republic
\item \Idef{org56}Institute of Physics, Bhubaneswar, India
\item \Idef{org57}Institute of Space Science (ISS), Bucharest, Romania
\item \Idef{org58}Instituto de Ciencias Nucleares, Universidad Nacional Aut\'{o}noma de M\'{e}xico, Mexico City, Mexico
\item \Idef{org59}Instituto de F\'{\i}sica, Universidad Nacional Aut\'{o}noma de M\'{e}xico, Mexico City, Mexico
\item \Idef{org60}iThemba LABS, National Research Foundation, Somerset West, South Africa
\item \Idef{org61}Joint Institute for Nuclear Research (JINR), Dubna, Russia
\item \Idef{org62}Korea Institute of Science and Technology Information, Daejeon, South Korea
\item \Idef{org63}KTO Karatay University, Konya, Turkey
\item \Idef{org64}Laboratoire de Physique Corpusculaire (LPC), Clermont Universit\'{e}, Universit\'{e} Blaise Pascal, CNRS--IN2P3, Clermont-Ferrand, France
\item \Idef{org65}Laboratoire de Physique Subatomique et de Cosmologie, Universit\'{e} Grenoble-Alpes, CNRS-IN2P3, Grenoble, France
\item \Idef{org66}Laboratori Nazionali di Frascati, INFN, Frascati, Italy
\item \Idef{org67}Laboratori Nazionali di Legnaro, INFN, Legnaro, Italy
\item \Idef{org68}Lawrence Berkeley National Laboratory, Berkeley, CA, United States
\item \Idef{org69}Lawrence Livermore National Laboratory, Livermore, CA, United States
\item \Idef{org70}Moscow Engineering Physics Institute, Moscow, Russia
\item \Idef{org71}National Centre for Nuclear Studies, Warsaw, Poland
\item \Idef{org72}National Institute for Physics and Nuclear Engineering, Bucharest, Romania
\item \Idef{org73}National Institute of Science Education and Research, Bhubaneswar, India
\item \Idef{org74}Niels Bohr Institute, University of Copenhagen, Copenhagen, Denmark
\item \Idef{org75}Nikhef, National Institute for Subatomic Physics, Amsterdam, Netherlands
\item \Idef{org76}Nuclear Physics Group, STFC Daresbury Laboratory, Daresbury, United Kingdom
\item \Idef{org77}Nuclear Physics Institute, Academy of Sciences of the Czech Republic, \v{R}e\v{z} u Prahy, Czech Republic
\item \Idef{org78}Oak Ridge National Laboratory, Oak Ridge, TN, United States
\item \Idef{org79}Petersburg Nuclear Physics Institute, Gatchina, Russia
\item \Idef{org80}Physics Department, Creighton University, Omaha, NE, United States
\item \Idef{org81}Physics Department, Panjab University, Chandigarh, India
\item \Idef{org82}Physics Department, University of Athens, Athens, Greece
\item \Idef{org83}Physics Department, University of Cape Town, Cape Town, South Africa
\item \Idef{org84}Physics Department, University of Jammu, Jammu, India
\item \Idef{org85}Physics Department, University of Rajasthan, Jaipur, India
\item \Idef{org86}Physik Department, Technische Universit\"{a}t M\"{u}nchen, Munich, Germany
\item \Idef{org87}Physikalisches Institut, Ruprecht-Karls-Universit\"{a}t Heidelberg, Heidelberg, Germany
\item \Idef{org88}Politecnico di Torino, Turin, Italy
\item \Idef{org89}Purdue University, West Lafayette, IN, United States
\item \Idef{org90}Pusan National University, Pusan, South Korea
\item \Idef{org91}Research Division and ExtreMe Matter Institute EMMI, GSI Helmholtzzentrum f\"ur Schwerionenforschung, Darmstadt, Germany
\item \Idef{org92}Rudjer Bo\v{s}kovi\'{c} Institute, Zagreb, Croatia
\item \Idef{org93}Russian Federal Nuclear Center (VNIIEF), Sarov, Russia
\item \Idef{org94}Russian Research Centre Kurchatov Institute, Moscow, Russia
\item \Idef{org95}Saha Institute of Nuclear Physics, Kolkata, India
\item \Idef{org96}School of Physics and Astronomy, University of Birmingham, Birmingham, United Kingdom
\item \Idef{org97}Secci\'{o}n F\'{\i}sica, Departamento de Ciencias, Pontificia Universidad Cat\'{o}lica del Per\'{u}, Lima, Peru
\item \Idef{org98}Sezione INFN, Bari, Italy
\item \Idef{org99}Sezione INFN, Bologna, Italy
\item \Idef{org100}Sezione INFN, Cagliari, Italy
\item \Idef{org101}Sezione INFN, Catania, Italy
\item \Idef{org102}Sezione INFN, Padova, Italy
\item \Idef{org103}Sezione INFN, Rome, Italy
\item \Idef{org104}Sezione INFN, Trieste, Italy
\item \Idef{org105}Sezione INFN, Turin, Italy
\item \Idef{org106}SSC IHEP of NRC Kurchatov institute, Protvino, Russia
\item \Idef{org107}SUBATECH, Ecole des Mines de Nantes, Universit\'{e} de Nantes, CNRS-IN2P3, Nantes, France
\item \Idef{org108}Suranaree University of Technology, Nakhon Ratchasima, Thailand
\item \Idef{org109}Technical University of Split FESB, Split, Croatia
\item \Idef{org110}The Henryk Niewodniczanski Institute of Nuclear Physics, Polish Academy of Sciences, Cracow, Poland
\item \Idef{org111}The University of Texas at Austin, Physics Department, Austin, TX, USA
\item \Idef{org112}Universidad Aut\'{o}noma de Sinaloa, Culiac\'{a}n, Mexico
\item \Idef{org113}Universidade de S\~{a}o Paulo (USP), S\~{a}o Paulo, Brazil
\item \Idef{org114}Universidade Estadual de Campinas (UNICAMP), Campinas, Brazil
\item \Idef{org115}University of Houston, Houston, TX, United States
\item \Idef{org116}University of Jyv\"{a}skyl\"{a}, Jyv\"{a}skyl\"{a}, Finland
\item \Idef{org117}University of Liverpool, Liverpool, United Kingdom
\item \Idef{org118}University of Tennessee, Knoxville, TN, United States
\item \Idef{org119}University of Tokyo, Tokyo, Japan
\item \Idef{org120}University of Tsukuba, Tsukuba, Japan
\item \Idef{org121}University of Zagreb, Zagreb, Croatia
\item \Idef{org122}Universit\'{e} de Lyon, Universit\'{e} Lyon 1, CNRS/IN2P3, IPN-Lyon, Villeurbanne, France
\item \Idef{org123}V.~Fock Institute for Physics, St. Petersburg State University, St. Petersburg, Russia
\item \Idef{org124}Variable Energy Cyclotron Centre, Kolkata, India
\item \Idef{org125}Vestfold University College, Tonsberg, Norway
\item \Idef{org126}Warsaw University of Technology, Warsaw, Poland
\item \Idef{org127}Wayne State University, Detroit, MI, United States
\item \Idef{org128}Wigner Research Centre for Physics, Hungarian Academy of Sciences, Budapest, Hungary
\item \Idef{org129}Yale University, New Haven, CT, United States
\item \Idef{org130}Yonsei University, Seoul, South Korea
\item \Idef{org131}Zentrum f\"{u}r Technologietransfer und Telekommunikation (ZTT), Fachhochschule Worms, Worms, Germany
\end{Authlist}
\endgroup

%% file: paperCDS.bbl
\begin{thebibliography}{99}

\bibitem{Satz:2000bn}
  H.~Satz,
  Rept.\ Prog.\ Phys.\  {\bf 63} (2000) 1511
  [hep-ph/0007069].
  
\bibitem{Bass:1998vz}
  S.~A.~Bass, M.~Gyulassy, H.~Stoecker and W.~Greiner,
  J.\ Phys.\ G {\bf 25} (1999) R1
  [hep-ph/9810281].
  
\bibitem{Shuryak:1984nq}
  E.~V.~Shuryak,
  Phys.\ Rept.\  {\bf 115} (1984) 151.
  
\bibitem{Cleymans:1985wb}
  J.~Cleymans, R.~V.~Gavai and E.~Suhonen,
  Phys.\ Rept.\  {\bf 130} (1986) 217.
  
\bibitem{Borsanyi:2010cj}
  S.~Borsanyi, G.~Endrodi, Z.~Fodor, A.~Jakovac, S.~D.~Katz, S.~Krieg, C.~Ratti and K.~K.~Szabo,
  JHEP {\bf 1011} (2010) 077
  [arXiv:1007.2580 [hep-lat]].

\bibitem{Bhattacharya:2014ara}
  T.~Bhattacharya, M.~I.~Buchoff, N.~H.~Christ, H.~-T.~Ding, R.~Gupta, C.~Jung, F.~Karsch and Z.~Lin {\it et al.},
  arXiv:1402.5175 [hep-lat].
  
\bibitem{Arsene:2004fa}
  I.~Arsene {\it et al.}  [BRAHMS Collaboration],
  Nucl.\ Phys.\ A {\bf 757} (2005) 1
  [nucl-ex/0410020].

\bibitem{Adcox:2004mh}
K. Adcox \textit{et al.} [PHENIX Collaboration],
Nucl. Phys. \textbf{A757} (2005) 184,
nucl-ex/0410003.
   
\bibitem{Back:2004je}
B. B. Back \textit{et al.} [PHOBOS Collaboration],
Nucl. Phys. \textbf{A757} (2005) 28,
nucl-ex/0410022.

\bibitem{Adams:2005dq}
J. Adams \textit{et al.} [STAR Collaboration],
Nucl. Phys. \textbf{A757} (2005) 102,
nucl-ex/0501009.

\bibitem{Aamodt:2010pb}
  KAamodt {\it et al.}  [ALICE Collaboration],
  Phys.\ Rev.\ Lett.\  {\bf 105} (2010) 252301
  [arXiv:1011.3916 [nucl-ex]].
 
\bibitem{Aamodt:2010cz}
  K.~Aamodt {\it et al.}  [ALICE Collaboration],
  Phys.\ Rev.\ Lett.\  {\bf 106} (2011) 032301
  [arXiv:1012.1657 [nucl-ex]].
 
\bibitem{Chatrchyan:2011pb}
  S.~Chatrchyan {\it et al.}  [CMS Collaboration],
  JHEP {\bf 1108} (2011) 141
  [arXiv:1107.4800 [nucl-ex]].

\bibitem{Aamodt:2011mr}
  K.~Aamodt {\it et al.}  [ALICE Collaboration],
  Phys.\ Lett.\ B {\bf 696} (2011) 328
  [arXiv:1012.4035 [nucl-ex]].
  
\bibitem{Aamodt:2010pa}
  K.~Aamodt {\it et al.}  [ALICE Collaboration],
  Phys.\ Rev.\ Lett.\  {\bf 105} (2010) 252302
  [arXiv:1011.3914 [nucl-ex]].

\bibitem{ATLAS:2011ah}
  G.~Aad {\it et al.}  [ATLAS Collaboration],
  Phys.\ Lett.\ B {\bf 707} (2012) 330
  [arXiv:1108.6018 [hep-ex]].

\bibitem{ATLAS:2012at}
  G.~Aad {\it et al.}  [ATLAS Collaboration],
  Phys.\ Rev.\ C {\bf 86} (2012) 014907
  [arXiv:1203.3087 [hep-ex]].
    
\bibitem{Chatrchyan:2012wg}
  S.~Chatrchyan {\it et al.}  [CMS Collaboration],
  Eur.\ Phys.\ J.\ C {\bf 72} (2012) 2012
  [arXiv:1201.3158 [nucl-ex]].
  
\bibitem{Chatrchyan:2012ta}
  S.~Chatrchyan {\it et al.}  [CMS Collaboration],
  Phys.\ Rev.\ C {\bf 87} (2013) 014902
  [arXiv:1204.1409 [nucl-ex]].
  
\bibitem{ALICE:2011ab}
  K.~Aamodt {\it et al.}  [ALICE Collaboration],
  Phys.\ Rev.\ Lett.\  {\bf 107} (2011) 032301
  [arXiv:1105.3865 [nucl-ex]].
  
\bibitem{Aad:2013xma}
  G.~Aad {\it et al.}  [ATLAS Collaboration],
  JHEP {\bf 1311} (2013) 183
  [arXiv:1305.2942 [hep-ex]].

\bibitem{Chatrchyan:2013kba}
  S.~Chatrchyan {\it et al.}  [CMS Collaboration],
  Phys.\ Rev.\ C {\bf 89} (2014) 044906
  [arXiv:1310.8651 [nucl-ex]].
  
\bibitem{Aamodt:2010jd}
  K.~Aamodt {\it et al.}  [ALICE Collaboration],
  Phys.\ Lett.\ B {\bf 696} (2011) 30
  [arXiv:1012.1004 [nucl-ex]].
  
\bibitem{Aad:2010bu}
  G.~Aad {\it et al.}  [ATLAS Collaboration],
  Phys.\ Rev.\ Lett.\  {\bf 105} (2010) 252303
  [arXiv:1011.6182 [hep-ex]].

\bibitem{Chatrchyan:2011sx}
  S.~Chatrchyan {\it et al.}  [CMS Collaboration],
  Phys.\ Rev.\ C {\bf 84} (2011) 024906
  [arXiv:1102.1957 [nucl-ex]].
  
\bibitem{Voloshin:1994mz} 
  S.~Voloshin and Y.~Zhang,
  Z.\ Phys.\ C {\bf 70}, 665 (1996)
  [hep-ph/9407282].

\bibitem{Poskanzer:1998yz} 
  A.~M.~Poskanzer and S.~A.~Voloshin,
  Phys.\ Rev.\ C {\bf 58}, 1671 (1998)
  [nucl-ex/9805001].

\bibitem{Bhalerao:2006tp}
  R.~S.~Bhalerao and J.~-Y.~Ollitrault,
  Phys.\ Lett.\ B {\bf 641} (2006) 260
  [nucl-th/0607009].

\bibitem{Alver:2008zza}
  B.~Alver, B.~B.~Back, M.~D.~Baker, M.~Ballintijn, D.~S.~Barton, R.~R.~Betts, R.~Bindel and W.~Busza {\it et al.},
  Phys.\ Rev.\ C {\bf 77} (2008) 014906
  [arXiv:0711.3724 [nucl-ex]].
  
\bibitem{Ollitrault:2009ie}
  J.~-Y.~Ollitrault, A.~M.~Poskanzer and S.~A.~Voloshin,
  Phys.\ Rev.\ C {\bf 80} (2009) 014904
  [arXiv:0904.2315 [nucl-ex]].
 
\bibitem{Alver:2010gr}
  B.~Alver and G.~Roland,
  Phys.\ Rev.\ C {\bf 81} (2010) 054905
   [Erratum-ibid.\ C {\bf 82} (2010) 039903]
  [arXiv:1003.0194 [nucl-th]].
  
\bibitem{Qiu:2011iv}
  Z.~Qiu and U.~W.~Heinz,
  Phys.\ Rev.\ C {\bf 84} (2011) 024911
  [arXiv:1104.0650 [nucl-th]].

\bibitem{Kovtun:2004de}
 P.~Kovtun, D.~T.~Son and A.~O.~Starinets,
  Phys.\ Rev.\ Lett.\  {\bf 94} (2005) 111601
  [hep-th/0405231].
 
\bibitem{Huovinen:2001cy}
  P.~Huovinen, P.~F.~Kolb, U.~W.~Heinz, P.~V.~Ruuskanen and S.~A.~Voloshin,
  Phys.\ Lett.\ B {\bf 503} (2001) 58
  [hep-ph/0101136].
  
\bibitem{Teaney:2000cw}
  D.~Teaney, J.~Lauret and E.~V.~Shuryak,
  Phys.\ Rev.\ Lett.\  {\bf 86} (2001) 4783
  [nucl-th/0011058].
  
\bibitem{Voloshin:1996nv} 
 S.~A.~Voloshin,
  Phys.\ Rev.\ C {\bf 55}, 1630 (1997)
  [nucl-th/9611038].

\bibitem{Shen:2011eg}
  C.~Shen, U.~Heinz, P.~Huovinen and H.~Song,
  Phys.\ Rev.\ C {\bf 84} (2011) 044903
  [arXiv:1105.3226 [nucl-th]].
  
\bibitem{Barrette:1997pt} 
  J.~Barrette {\it et al.}  [E877 Collaboration],
  Phys.\ Rev.\ C {\bf 56}, 3254 (1997)
  [nucl-ex/9707002].

\bibitem{Barrette:1998bz} 
  J.~Barrette {\it et al.}  [E877 Collaboration],
  Phys.\ Rev.\ C {\bf 59}, 884 (1999)
  [nucl-ex/9805006].

\bibitem{Appelshauser:1997dg} 
  H.~Appelshauser {\it et al.}  [NA49 Collaboration],
  Phys.\ Rev.\ Lett.\  {\bf 80}, 4136 (1998)
  [nucl-ex/9711001].

\bibitem{Alt:2003ab} 
  C.~Alt {\it et al.}  [NA49 Collaboration],
  Phys.\ Rev.\ C {\bf 68}, 034903 (2003)
  [nucl-ex/0303001].
  
\bibitem{Adams:2003am}
  J.~Adams {\it et al.}  [STAR Collaboration],
  Phys.\ Rev.\ Lett.\  {\bf 92} (2004) 052302
  [nucl-ex/0306007].
    
\bibitem{Abelev:2007qg}
  B.~Abelev {\it et al.}  [STAR Collaboration],
  Phys.\ Rev.\ C {\bf 75} (2007) 054906
  [nucl-ex/0701010].
 
\bibitem{Adams:2004bi}
  J.~Adams {\it et al.}  [STAR Collaboration],
  Phys.\ Rev.\ C {\bf 72} (2005) 014904
  [nucl-ex/0409033].
  
\bibitem{Adler:2003kt}
  S.~S.~Adler {\it et al.}  [PHENIX Collaboration],
  Phys.\ Rev.\ Lett.\  {\bf 91} (2003) 182301
  [nucl-ex/0305013].
  
\bibitem{Afanasiev:2007tv}
  S.~Afanasiev {\it et al.}  [PHENIX Collaboration],
  Phys.\ Rev.\ Lett.\  {\bf 99} (2007) 052301
  [nucl-ex/0703024 [NUCL-EX]].
  
\bibitem{Adare:2006ti}
  A.~Adare {\it et al.}  [PHENIX Collaboration],
  Phys.\ Rev.\ Lett.\  {\bf 98} (2007) 162301
  [nucl-ex/0608033].

\bibitem{Adare:2012vq}
  A.~Adare {\it et al.}  [PHENIX Collaboration],
  Phys.\ Rev.\ C {\bf 85} (2012) 064914
  [arXiv:1203.2644 [nucl-ex]].

\bibitem{Hirano:2005xf}
  T.~Hirano, U.~W.~Heinz, D.~Kharzeev, R.~Lacey and Y.~Nara,
  Phys.\ Lett.\ B {\bf 636} (2006) 299
  [nucl-th/0511046].
 
\bibitem{Biagi:1980ar}
  S.~F.~Biagi, M.~Bourquin, A.~J.~Britten, R.~M.~Brown, H.~Burckhart, A.~A.~Carter, J.~R.~Carter and C.~Dore {\it et al.},
  Nucl.\ Phys.\ B {\bf 186} (1981) 1.
  
\bibitem{Bass:1999tu}
  S.~A.~Bass, A.~Dumitru, M.~Bleicher, L.~Bravina, E.~Zabrodin, H.~Stoecker and W.~Greiner,
  Phys.\ Rev.\ C {\bf 60} (1999) 021902
  [nucl-th/9902062].

\bibitem{Dumitru:1999sf}
  A.~Dumitru, S.~A.~Bass, M.~Bleicher, H.~Stoecker and W.~Greiner,
  Phys.\ Lett.\ B {\bf 460} (1999) 411
  [nucl-th/9901046].

\bibitem{Bass:2000ib}
  S.~A.~Bass and A.~Dumitru,
  Phys.\ Rev.\ C {\bf 61} (2000) 064909
  [nucl-th/0001033].

\bibitem{Shor:1984ui}
  A.~Shor,
  Phys.\ Rev.\ Lett.\  {\bf 54} (1985) 1122.
 
\bibitem{Voloshin:2002wa} 
  S.~A.~Voloshin,
  Nucl.\ Phys.\ A {\bf 715}, 379 (2003)
  [nucl-ex/0210014].

\bibitem{Molnar:2003ff}
  D.~Molnar and S.~A.~Voloshin,
  Phys.\ Rev.\ Lett.\  {\bf 91} (2003) 092301
  [nucl-th/0302014].
  
\bibitem{Greco:2003mm}
  V.~Greco, C.~M.~Ko and P.~Levai,
  Phys.\ Rev.\ C {\bf 68} (2003) 034904
  [nucl-th/0305024].

\bibitem{Fries:2003kq}
  R.~J.~Fries, B.~Muller, C.~Nonaka and S.~A.~Bass,
  Phys.\ Rev.\ C {\bf 68} (2003) 044902
  [nucl-th/0306027].

\bibitem{Hwa:2003ic}
  R.~C.~Hwa and C.~B.~Yang,
  Phys.\ Rev.\ C {\bf 70} (2004) 024904
  [hep-ph/0312271].
  
\bibitem{Abelev:2012di}
  B.~Abelev {\it et al.}  [ALICE Collaboration],
  Phys.\ Lett.\ B {\bf 719} (2013) 18
  [arXiv:1205.5761 [nucl-ex]].
  
\bibitem{Carminati:2004fp}
  F.~Carminati {\it et al.}  [ALICE Collaboration],
  J.\ Phys.\ G {\bf 30} (2004) 1517.

\bibitem{Alessandro:2006yt}
  B.~Alessandro {\it et al.}  [ALICE Collaboration],
  J.\ Phys.\ G {\bf 32} (2006) 1295.

\bibitem{Aamodt:2008zz}
  K.~Aamodt {\it et al.}  [ALICE Collaboration],
  JINST {\bf 3} (2008) S08002.

\bibitem{Adler:2002pu}
  C.~Adler {\it et al.}  [STAR Collaboration],
  Phys.\ Rev.\ C {\bf 66} (2002) 034904
  [nucl-ex/0206001].
  
\bibitem{Ref:EventPlane} {
S.~Voloshin, A.~M.~Poskanzer, and R.~Snellings, in \emph{Relativistic Heavy Ion Physics}, Landolt-Bornstein Vol. 1 (Springer-Verlag, Berlin, 2010), pp. 5Ð54.}

\bibitem{Abelev:2013qoq}
  B.~Abelev {\it et al.}  [ALICE Collaboration],
  Phys.\ Rev.\ C {\bf 88} (2013) 044909
  [arXiv:1301.4361 [nucl-ex]].

\bibitem{Alme:2010ke}
  J.~Alme, Y.~Andres, H.~Appelshauser, S.~Bablok, N.~Bialas, R.~Bolgen, U.~Bonnes and R.~Bramm {\it et al.},
  Nucl.\ Instrum.\ Meth.\ A {\bf 622} (2010) 316
  [arXiv:1001.1950 [physics.ins-det]].

\bibitem{Abelev:2014ffa}
  B.~Abelev {\it et al.}  [ALICE Collaboration],
  arXiv:1402.4476 [nucl-ex].

\bibitem{Akindinov:2013tea}
  A.~Akindinov, A.~Alici, A.~Agostinelli, P.~Antonioli, S.~Arcelli, M.~Basile, F.~Bellini and G.~Cara Romeo {\it et al.},
  Eur.\ Phys.\ J.\ Plus {\bf 128} (2013) 44.
  
\bibitem{Abbas:2013taa}
  E.~Abbas {\it et al.}  [ALICE Collaboration],
  JINST {\bf 8} (2013) P10016
  [arXiv:1306.3130 [nucl-ex]].
      
\bibitem{Abelev:2013vea}
  B.~Abelev {\it et al.}  [ALICE Collaboration],
  Phys.\ Rev.\ C {\bf 88} (2013) 044910
  [arXiv:1303.0737 [hep-ex]].

\bibitem{Beringer:1900zz}
  J.~Beringer {\it et al.}  [Particle Data Group Collaboration],
  Phys.\ Rev.\ D {\bf 86} (2012) 010001.
  
\bibitem{Abelev:2013xaa}
  B.~Abelev {\it et al.}  [ALICE Collaboration],
  Phys.\ Rev.\ Lett.\  {\bf 111} (2013) 222301
  [arXiv:1307.5530 [nucl-ex]].
  
\bibitem{Ref:Armenteros} {
J.~Podolanski and R.~Armenteros, Phil. Mag. \textbf{45}, 13 (1954).}

\bibitem{ABELEV:2013zaa}
  B.~Abelev {\it et al.}  [ALICE Collaboration],
  Phys.\ Lett.\ B {\bf 728} (2014) 216
  [arXiv:1307.5543 [nucl-ex]].
  
\bibitem{Borghini:2004ra}
  N.~Borghini and J.~Y.~Ollitrault,
  Phys.\ Rev.\ C {\bf 70} (2004) 064905
  [nucl-th/0407041].
  
\bibitem{Abelev:2008ae}
  B.~Abelev {\it et al.}  [STAR Collaboration],
  Phys.\ Rev.\ C {\bf 77} (2008) 054901
  [arXiv:0801.3466 [nucl-ex]].

\bibitem{Brun:1994aa}
  R.~Brun, F.~Carminati and S.~Giani,
  CERN-W5013.

\bibitem{Abelev:2014uua}
  B.~Abelev {\it et al.}  [ALICE Collaboration],
  arXiv:1404.0495 [nucl-ex].
  
\bibitem{Song:2007fn}
  H.~Song and U.~W.~Heinz,
  Phys.\ Lett.\ B {\bf 658} (2008) 279
  [arXiv:0709.0742 [nucl-th]].

\bibitem{Song:2007ux}
  H.~Song and U.~W.~Heinz,
  Phys.\ Rev.\ C {\bf 77} (2008) 064901
  [arXiv:0712.3715 [nucl-th]].

\bibitem{Song:2008si}
  H.~Song and U.~W.~Heinz,
  Phys.\ Rev.\ C {\bf 78} (2008) 024902
  [arXiv:0805.1756 [nucl-th]].

\bibitem{Song:2010mg}
  H.~Song, S.~A.~Bass, U.~Heinz, T.~Hirano and C.~Shen,
  Phys.\ Rev.\ Lett.\  {\bf 106} (2011) 192301
   [Erratum-ibid.\  {\bf 109} (2012) 139904]
  [arXiv:1011.2783 [nucl-th]].

\bibitem{Song:2011hk}
  H.~Song, S.~A.~Bass, U.~Heinz, T.~Hirano and C.~Shen,
  Phys.\ Rev.\ C {\bf 83} (2011) 054910
   [Erratum-ibid.\ C {\bf 86} (2012) 059903]
  [arXiv:1101.4638 [nucl-th]].

\bibitem{Song:2013qma}
  H.~Song, S.~Bass and U.~W.~Heinz,
  arXiv:1311.0157 [nucl-th].

\bibitem{Drescher:2007ax}
  H.~-J.~Drescher and Y.~Nara,
  Phys.\ Rev.\ C {\bf 76} (2007) 041903
  [arXiv:0707.0249 [nucl-th]].

\bibitem{Bass:1998ca}
  S.~A.~Bass, M.~Belkacem, M.~Bleicher, M.~Brandstetter, L.~Bravina, C.~Ernst, L.~Gerland and M.~Hofmann {\it et al.},
  Prog.\ Part.\ Nucl.\ Phys.\  {\bf 41} (1998) 255
   [Prog.\ Part.\ Nucl.\ Phys.\  {\bf 41} (1998) 225]
  [nucl-th/9803035].
  
\bibitem{Bleicher:1999xi}
  M.~Bleicher, E.~Zabrodin, C.~Spieles, S.~A.~Bass, C.~Ernst, S.~Soff, L.~Bravina and M.~Belkacem {\it et al.},
  J.\ Phys.\ G {\bf 25} (1999) 1859
  [hep-ph/9909407].
        
\end{thebibliography}
